\PassOptionsToPackage{unicode}{hyperref}
\PassOptionsToPackage{hyphens}{url}
\PassOptionsToPackage{dvipsnames,svgnames,x11names}{xcolor}
\documentclass[
  12pt]{article}

\usepackage{amsmath,amssymb}
\usepackage{iftex}
\ifPDFTeX
  \usepackage[T1]{fontenc}
  \usepackage[utf8]{inputenc}
  \usepackage{textcomp} 
\else 
  \usepackage{unicode-math}
  \defaultfontfeatures{Scale=MatchLowercase}
  \defaultfontfeatures[\rmfamily]{Ligatures=TeX,Scale=1}
\fi
\usepackage{lmodern}
\ifPDFTeX\else  
\fi
\IfFileExists{upquote.sty}{\usepackage{upquote}}{}
\IfFileExists{microtype.sty}{
  \usepackage[]{microtype}
  \UseMicrotypeSet[protrusion]{basicmath} 
}{}
\makeatletter
\@ifundefined{KOMAClassName}{
  \IfFileExists{parskip.sty}{%
    \usepackage{parskip}
  }{
    \setlength{\parindent}{0pt}
    \setlength{\parskip}{6pt plus 2pt minus 1pt}}
}{
  \KOMAoptions{parskip=half}}
\makeatother
\usepackage{xcolor}
\makeatletter
\ifx\paragraph\undefined\else
  \let\oldparagraph\paragraph
  \renewcommand{\paragraph}{
    \@ifstar
      \xxxParagraphStar
      \xxxParagraphNoStar
  }
  \newcommand{\xxxParagraphStar}[1]{\oldparagraph*{#1}\mbox{}}
  \newcommand{\xxxParagraphNoStar}[1]{\oldparagraph{#1}\mbox{}}
\fi
\ifx\subparagraph\undefined\else
  \let\oldsubparagraph\subparagraph
  \renewcommand{\subparagraph}{
    \@ifstar
      \xxxSubParagraphStar
      \xxxSubParagraphNoStar
  }
  \newcommand{\xxxSubParagraphStar}[1]{\oldsubparagraph*{#1}\mbox{}}
  \newcommand{\xxxSubParagraphNoStar}[1]{\oldsubparagraph{#1}\mbox{}}
\fi
\makeatother

\usepackage{longtable,booktabs,array}
\usepackage{calc} 
\usepackage{etoolbox}
\makeatletter
\patchcmd\longtable{\par}{\if@noskipsec\mbox{}\fi\par}{}{}
\makeatother
\IfFileExists{footnotehyper.sty}{\usepackage{footnotehyper}}{\usepackage{footnote}}
\makesavenoteenv{longtable}
\usepackage{graphicx}
\makeatletter
\def\maxwidth{\ifdim\Gin@nat@width>\linewidth\linewidth\else\Gin@nat@width\fi}
\def\maxheight{\ifdim\Gin@nat@height>\textheight\textheight\else\Gin@nat@height\fi}
\makeatother
\setkeys{Gin}{width=\maxwidth,height=\maxheight,keepaspectratio}
\makeatletter
\def\fps@figure{htbp}
\makeatother

\makeatletter
\@ifpackageloaded{caption}{}{\usepackage{caption}}
\AtBeginDocument{%
\ifdefined\contentsname
  \renewcommand*\contentsname{Table of contents}
\else
  \newcommand\contentsname{Table of contents}
\fi
\ifdefined\listfigurename
  \renewcommand*\listfigurename{List of Figures}
\else
  \newcommand\listfigurename{List of Figures}
\fi
\ifdefined\listtablename
  \renewcommand*\listtablename{List of Tables}
\else
  \newcommand\listtablename{List of Tables}
\fi
\ifdefined\figurename
  \renewcommand*\figurename{Figure}
\else
  \newcommand\figurename{Figure}
\fi
\ifdefined\tablename
  \renewcommand*\tablename{Table}
\else
  \newcommand\tablename{Table}
\fi
}
\@ifpackageloaded{float}{}{\usepackage{float}}
\floatstyle{ruled}
\@ifundefined{c@chapter}{\newfloat{codelisting}{h}{lop}}{\newfloat{codelisting}{h}{lop}[chapter]}
\floatname{codelisting}{Listing}

\makeatother
\makeatletter
\@ifpackageloaded{caption}{}{\usepackage{caption}}
\@ifpackageloaded{subcaption}{}{\usepackage{subcaption}}
\makeatother

\ifLuaTeX
  \usepackage{selnolig}  
\fi
\usepackage[]{natbib}
\usepackage{bookmark}

\IfFileExists{xurl.sty}{\usepackage{xurl}}{} 
\hypersetup{
  pdftitle={Title},
  pdfauthor={Author 1; Author 2},
  pdfkeywords={3 to 6 keywords, that do not appear in the title},
  colorlinks=true,
  linkcolor={blue},
  filecolor={Maroon},
  citecolor={Blue},
  urlcolor={Blue},
  pdfcreator={LaTeX via pandoc}}

\newcommand{\anon}{1}

\usepackage[algoruled, linesnumbered]{algorithm2e}
\usepackage{bm}
\usepackage[font=footnotesize]{caption}
\usepackage{dsfont}
\usepackage{enumitem}
\usepackage[nodisplayskipstretch]{setspace}

\usepackage{siunitx}
\newcolumntype{T}{S[table-format=1.2, round-mode=places, round-precision=2]}

\usepackage{hyperref}

\newcommand{\forcond}{$i=1$ \KwTo $m$}
\newcommand{\xsub}[1]{{#1}^{\phantom{}}}
\newcommand{\xsup}[1]{{#1}{\rule{0pt}{12pt}}}

\begin{document}

\def\spacingset#1{\renewcommand{\baselinestretch}%
{#1}\small\normalsize} \spacingset{1}


\if1\anon
{
  \title{\vspace{-2ex} \bf A Mixed Self-Exciting Process to Model Epileptic Seizures}
  \author{Karen Kanaster \\
    Department of Biostatistics and Informatics, University of Colorado Anschutz \\
    Giovani L. Silva \\
    Departamento de Matemática - IST, Universidade de Lisboa \\
    Peter Müller \\
    Department of Statistics and Data Science, University of Texas at Austin \\
    Jacob Pellinen \\
    Department of Neurology, University of Colorado School of Medicine \\
    and \\
    Elizabeth Juarez-Colunga\thanks{
    Corresponding author. \textit{email address:} elizabeth.juarez-colunga@cuanschutz.edu} \\
    Department of Biostatistics and Informatics, University of Colorado Anschutz
    }
  \date{}
  \maketitle
} \fi

\if0\anon
{
  \bigskip
  \bigskip
  \bigskip
  \begin{center}
    {\LARGE\bf A Mixed Self-Exciting Process to Model Epileptic Seizures}
\end{center}
  \medskip
} \fi

\bigskip
\begin{abstract}
\noindent Epilepsy is a neurological disorder characterized by recurrent seizures affecting more than 70 million people worldwide. Often, an individual with epilepsy is more likely to experience subsequent seizures following an initial seizure, a process we call seizure clustering. Motivated by seizure diary data collected over three years from 407 individuals newly diagnosed with focal epilepsy in the Human Epilepsy Project (HEP), we propose a Bayesian mixed Hawkes process model that addresses seizure clustering and heterogeneity between individuals. In the Hawkes process, the intensity is accelerated each time an event occurs, through the composition of background and excitation intensity functions. The proposed model incorporates a Weibull baseline intensity to model a trend in background seizure rates over time, while the excitation process accounts for seizure clustering within individuals. We model heterogeneity among individuals by including covariates and random effects in both the background and excitation intensities. In the HEP study, the average time between primary and secondary seizures within an individual is 1.57 (95\% CrI: 1.43, 1.70) days, with an average of 2.20 (1.96, 2.47) seizures per cluster. We demonstrate that omitting random effects in the presence of heterogeneity leads to underestimation of the background intensity and overestimation of excitation rates.

\end{abstract}

\noindent%
\vfill

\newpage
\spacingset{1.8} 

\section{Introduction}\label{sec:intro}

Motivated by data from the Human Epilepsy Project (HEP), we develop a mixed Hawkes process model to address clinical and scientific research questions regarding seizure intensity and the clustering of seizure events in time. Specifically, we propose the use of random effects in both the background and excitation components of the Hawkes process intensity function to address heterogeneity in seizure trajectories between individuals. We demonstrate that failure to model such heterogeneity may lead to significantly biased inference on excitation rates, which directly impacts clinically relevant conclusions. 

Epilepsy is a neurological disorder characterized by recurrent seizures affecting more than 70 million people worldwide \citep{thijs_epilepsy_2019}. \textit{Clustering} of seizures occurs when an individual with epilepsy experiences additional seizures following an initial seizure. This has a significant impact on health and quality of life \citep{bauman_seizure_2021, jafarpour_seizure_2019}, as seizure clustering increases seizure-related morbidity and mortality compared to isolated time-limited seizures. Improving the understanding of seizure clustering can aid in monitoring, management, and treatment, as well as inform clinical trial design. In particular, it can help identify factors associated with the magnitude of clustering and determine the time window during which an individual with epilepsy is at the highest risk of seizure clustering. 

Evidence for seizure clustering has been documented across multiple data modalities and populations. Based on pre-surgical electroencephalogram (EEG) monitoring, individuals with drug-resistant epilepsy frequently exhibit pronounced seizure clustering over short time scales \citep{ferastraoaru_termination_2016, osorio_pharmacoresistant_2009}. Analyses of seizure diary data in broader populations suggest substantial heterogeneity in clustering patterns, with average inter-seizure intervals within clusters ranging from less than an hour to several days depending on individual seizure frequency \citep{chiang_individualizing_2020, haut_seizure_2006}. These findings indicate that seizure clustering varies considerably between individuals. We propose a mixed Hawkes process model that explicitly accounts for seizure clustering at the individual level while allowing for heterogeneity in seizure rates across individuals. 

A Hawkes process \citep{hawkes_spectra_1971, hawkes_cluster_1974} is a nonhomogeneous point process with the key feature that the intensity function is accelerated or excited each time an event occurs, which is achieved through the composition of background (immigrant) and excitation (offspring) intensity functions. Hawkes process models have been applied in a wide variety of areas, such as earthquakes \citep{ogata_statistical_1988}, terrorist activity \citep{white_terrorism_2013}, financial trading \citep{da_fonseca_hawkes_2014, omi_hawkes_2017}, social animal interactions \citep{ward_network_2022}, insurance claims \citep{lesage_hawkes_2022}, and sports injuries \citep{worrall_sports_2025}. Within epilepsy, EEG signals for a single subject have been modeled using a basic Hawkes process with constant background and offspring rates \citep{gerencser_real-time_2021}. Previous work incorporated covariates into the background intensity function to model disease networks \citep{choi_constructing_2015}, crime \citep{mohler_penalized_2018}, and smoking events \citep{engelhard_predicting_2018}. To model COVID-19 transmission, \citet{chiang_hawkes_2022} included covariates in both the background and offspring processes. \citet{kang_analyzing_2025} analyzed whale call data by combining covariates with a Gaussian process to account for additional variability in the background process. The present work is aimed at mixed modeling of seizure trajectories for multiple individuals, with background and offspring covariates, plus random effects in both intensities to account for unobserved heterogeneity. 

Practical challenges for Hawkes process models include the presence of aggregated count data and missing data. Due to dependence of the Hawkes process intensity on the complete event history, missingness or coarseness in the longitudinal process must be accounted for in order to estimate the model parameters. A number of methods have been proposed for the imputation of continuous event times for binned counts \citep{chen_estimating_2025, cheysson_spectral_2022, koyama_coarse-grained_2025, rizoiu_interval-censored_2022, shlomovich_parameter_2022, xu_learning_2017} and unobserved time intervals \citep{deutsch_abc_2021, le_multivariate_2018, linderman_bayesian_2017, shelton_hawkes_2018} in Hawkes process models. Previous methods that address aggregated events \citep{zhou_bayesian_2025} and missing intervals \citep{tucker_derek_handling_2019} through data augmentation are compatible with a Bayesian approach. We extend these methods to estimation of a more complex parameterization of a Hawkes process as a random effects model with a nonhomogeneous proportional baseline intensity. 

Our proposed approach is motivated by data collected over three or more years from 407 participants in the Human Epilepsy Project newly diagnosed with focal epilepsy. Individuals were expected to complete a seizure diary, recording the number of seizures occurring on a daily basis throughout their duration in the study. Seizure trajectory data among participants exhibit seizure clustering within individuals as well as high variability in seizure rates between individuals, including many with zero or few events following diagnosis and initiation of treatment. Challenges with applying a Hawkes process in this context include the aggregation of exact seizure event times into daily seizure counts, and the inherent missingness in the seizure diary, where seizures may have occurred but were not tracked. We address these challenges through multiple imputation, where under the assumption of non-informative missingness, the observed daily seizure counts are augmented by sampling unobserved continuous event times for both the binned counts and the unreported intervals. 

The present work aims to understand seizure occurrence among the HEP study participants through a novel mixed Hawkes process model. Our contribution is the introduction of background and offspring random effects to account for heterogeneity between independent event processes. Section \ref{sec:data} describes the motivating dataset from the Human Epilepsy Project. Section \ref{sec:methods} presents the mixed Hawkes process model and Markov Chain Monte Carlo (MCMC) methods. Section \ref{sec:app} discusses the application of the proposed model to the HEP study data, answering relevant clinical questions. Section \ref{sec:sim} contains two simulation studies; the first explores parameter estimation for varying levels of missing data and random effects variance, and the second investigates misspecification of the random effects structure. Section \ref{sec:disc} concludes with a discussion of limitations, extensions, and clinical relevance. 

\section{Human Epilepsy Project Data}\label{sec:data}

The first cohort of HEP enrolled 448 individuals newly diagnosed with focal epilepsy between 2012 and 2017 across 34 clinical centers worldwide. Eligible individuals were diagnosed between ages 12 and 60 and enrolled within four months of initiating medical treatment for seizures. Participants were asked to report the number of seizures that occurred each day using an electronic seizure diary \citep{fisher_tracking_2010} over a follow-up period of three or more years. Seizure trajectories were truncated to three years for the present work. Participants with no seizure diary entries (N=21), no properly tracked days (N=17), or no medication data (N=3) during the enrollment period were excluded from the present analysis, resulting in a dataset with 407 individuals. Additional details on study eligibility criteria, the electronic seizure diary entry process, and tracking protocol can be found in previous publications \citep{miller_long-term_2024, pellinen_focal_2020}. 

\begin{table}[!ht]
\centering{
\caption{Characteristics of Human Epilepsy Project participants. ``Pre-Tx" refers to the period between onset of seizures and initiation of treatment.}
\includegraphics[clip, trim=0.5cm 2cm 0.5cm 0, width=1.00\textwidth]{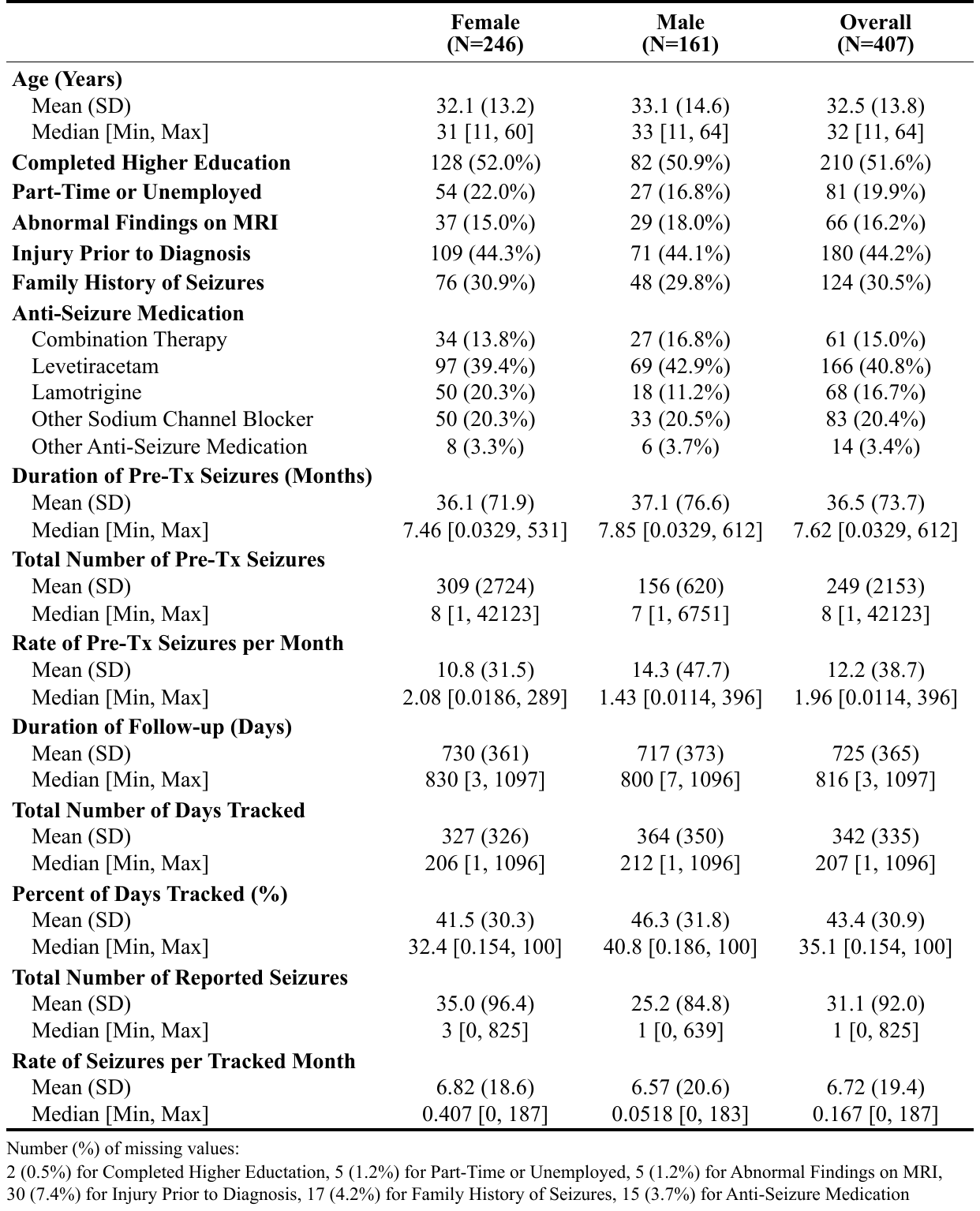}
\label{tab:hep_data}
}
\vspace{-2.5ex}
\end{table}

Data collected from the HEP study cohort include demographic and clinical variables at baseline, as well as anti-seizure medication records over time. The analysis cohort consisted of 246 (60.4\%) females and 161 (39.6\%) males, with a mean (SD) age of 32.5 (13.8) at enrollment (Table \ref{tab:hep_data}). At baseline, 61 (15.0\%) of participants were on combination therapy. The remaining on monotherapy consisted of 166 (40.8\%) on Levetiracetam, 68 (16.7\%) on Lamotrigine, 83 (24.0\%) on a sodium channel blocker other than Lamotrigine, and 14 (3.4\%) on any other anti-seizure medication. The mean (SD) duration between onset of seizures and study enrollment was 3.0 (6.1) years, with a median value of 7.6 months, and the median total number of pre-treatment seizures was 8. The average rate of seizures prior to treatment was 12.2 (38.7) per month, with a median of 2.0 seizures per month. The average duration of follow-up was 24 (12) months, with a range from 3 days to 36 months, where the end of follow-up was defined as the last tracked day. The average percent of days tracked over the follow-up period was 43.4\% (30.9\%), with a range from 0.15\% to 100\%, indicating a high level of missingness in the HEP study data, as well as high variability in tracking among participants. The average rate of reported seizures per tracked month was 6.7 (19.4), with a median value of 0.17 and a range from 0 to 187, indicating considerable heterogeneity in reported seizure rates between individuals. 

\begin{figure}[!ht]
\vspace{1ex}
\centering{
\includegraphics[width=1.00\textwidth]{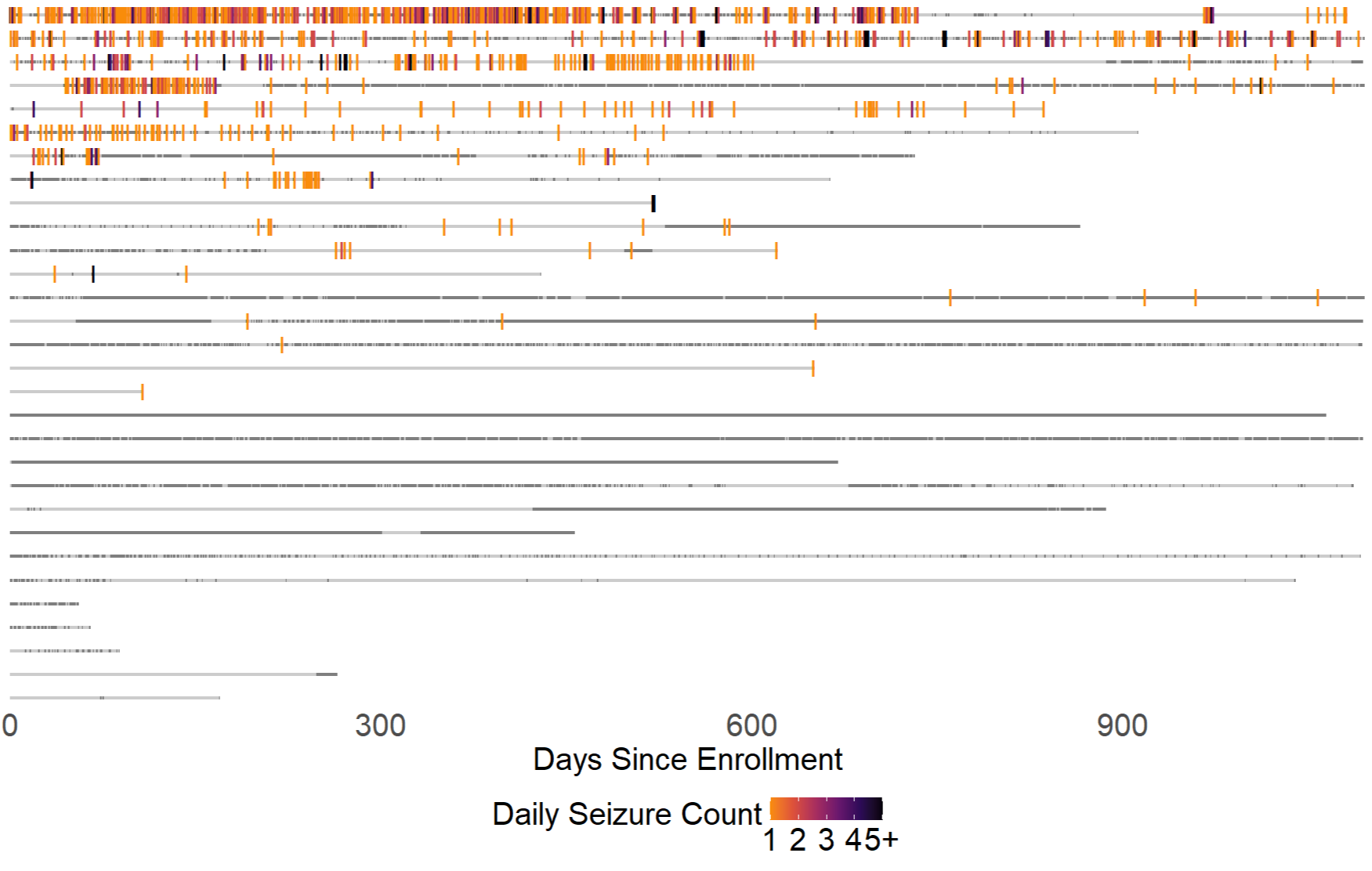}
\caption{Daily seizure counts for random sample of 30 individuals from HEP study. Darker tick marks correspond to higher seizure counts. Each horizontal line represents the follow-up period for one participant. Darker segments represent tracked periods while lighter segments indicate missing intervals. In the full HEP study data, maximum daily seizure count is 30, and seizure count exceeds 5 for 0.125\% of tracked days.}
\label{fig:hep_sample}
}
\vspace{-1ex}
\end{figure}

\begin{figure}[!ht]
\vspace{1ex}
\centering{
\includegraphics[width=0.80\textwidth]{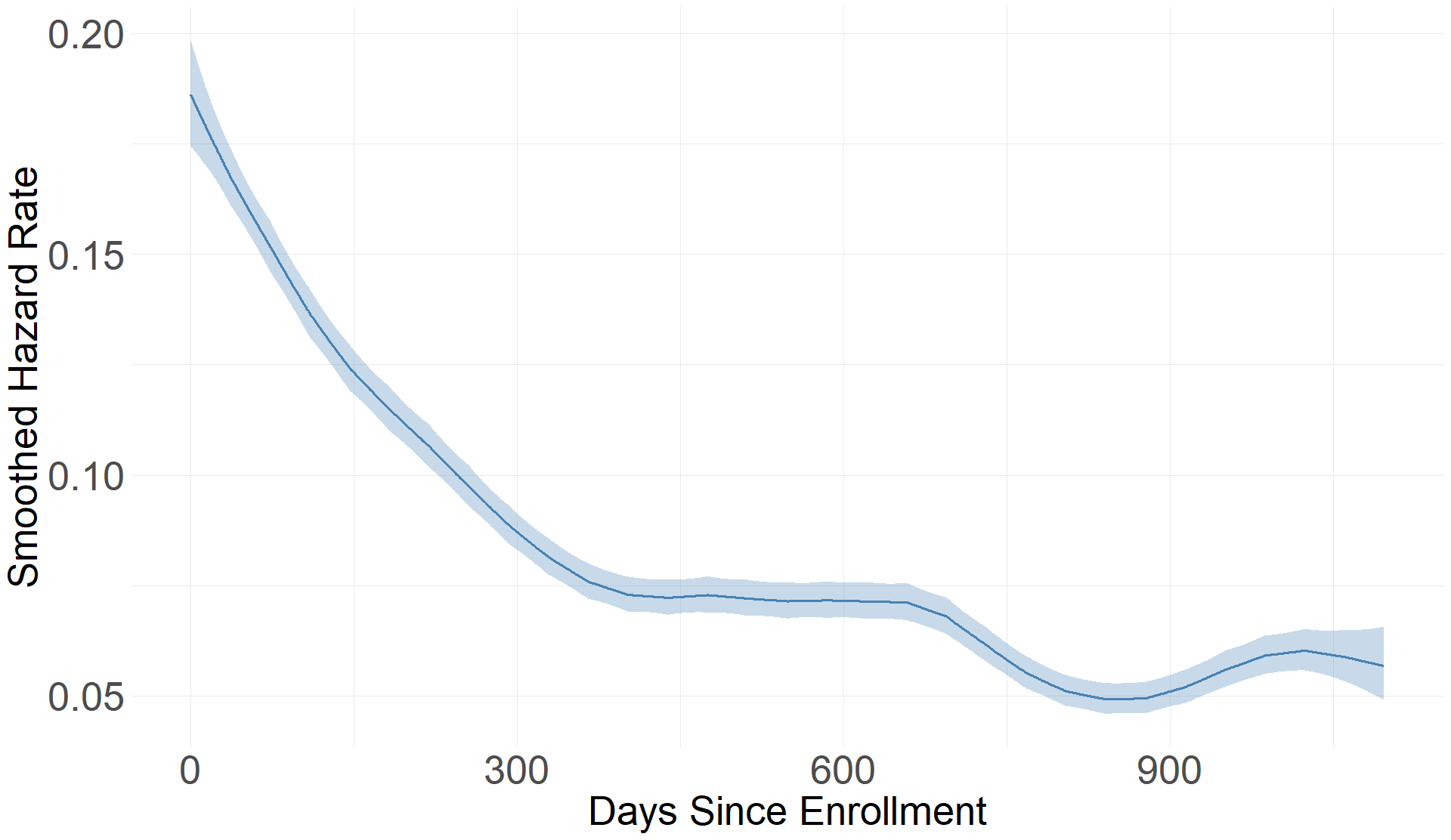}
\caption{Nonparametric smoothed hazard rate estimates with 95\% confidence intervals for HEP data. Event times were uniformly imputed from seizure counts within each day. Overall hazard of seizures is generally decreasing over the observation period, and is steeply declining during the first year after enrollment.}
\label{fig:hep_hazard}
}
\vspace{-1ex}
\end{figure}

Figure \ref{fig:hep_sample} shows daily seizure counts over time for a random sample of 30 participants from the HEP study data. This figure illustrates the high level of heterogeneity in missingness among individuals, as well as the high amount of unreported days overall. In addition, the high variability in seizure occurrence between individuals is visible, with some individuals reporting a high number of seizures during tracked periods and many individuals reporting very few or no seizures. Figure \ref{fig:hep_hazard} shows the nonparametric smoothed hazard rate estimates with 95\% confidence intervals for all seizure events reported in the HEP study data \citep{rebora_bshazard_2015-1}. The overall hazard of reported seizure events is generally decreasing over the observation period, and in particular is steeply declining during the first year of the study.

The HEP study provides one of the most extensive datasets to study epileptic seizures. Related questions of high interest to clinical collaborators include: What part of heterogeneity in the study population can be traced back to baseline characteristics, and what part is inherent to the disease? Can inference on seizure cluster sizes be biased by a failure to account for heterogeneity between individuals? After an initial seizure event, for how many days are individuals with epilepsy at risk for secondary seizure events? By how much can treatment after diagnosis be expected to lower background seizure rates, and for what demographic and clinical factors do these predictions need to be adjusted? In what sense are primary (background) and secondary (offspring) seizure events clinically different? In the following section, we develop model-based inference to address these questions. 

\section{Mixed Hawkes Process Model}\label{sec:methods}


Let $N(t)$, defined for $t \ge 0$, be a right-continuous counting process starting at $N(0)=0$. Additionally, let $\mathcal{H}_t = \mathcal{H}(t) = \{N(s) : 0 \le s < t;\, \bm{x}(s) : 0 \le s \le t\}$ denote the history of the counting process up to time $t$ and the covariate history up to and including time $t$. Then the conditional intensity function, representing the instantaneous rate of arrivals at time $t$, is defined as $\lambda^*(t) = \lambda(t \,|\, \mathcal{H}_t) = \lim_{\Delta t \to 0} Pr\{N(t + \Delta t) - N(t) = 1 \,|\, \mathcal{H}_t\} / \Delta t$. Suppose that we have a Hawkes process for $m$ individuals indexed by $i=1,...,m$, each with $n_i$ events. The general form of the conditional intensity function for individual $i$ with events $j=1,...,n_i$ is 
\begin{equation}
\lambda_i^*(t) = \mu_i(t) + \sum_{t_{ij} < t}\kappa_{ij} g_i(t - t_{ij}),  
\end{equation}

where $\mu_i(t)$ is the background (immigrant) arrival rate, $\kappa_{ij}$ is the self-excited jump size at the event time $t_{ij}$, and $g_i(\cdot)$ denotes the excitation (offspring) function. 

\subsection{Conditional Intensity}\label{sec:lambda_star}

Let $\{x_{ip} : p=0,...,P\}$ be a set of background covariates with corresponding coefficients $\{\beta_p \in \mathbb{R} : p=0,...,P\}$.  Similarly, let $\{z_{iq} : q=0,...,Q\}$ be a set of offspring covariates with corresponding coefficients $\{\zeta_q \in \mathbb{R} : q=0,...,Q\}$. Define $\eta_i = \exp\left\{\sum_{p=0}^P \beta_p x_{ip}\right\}$ as an exponentiated linear combination of the background process covariates. Similarly, define $\kappa_i = \exp\left\{\sum_{q=0}^Q \zeta_q z_{iq}\right\}$ based on the offspring process covariates. Here $\{x_{ip}\}$ and $\{z_{iq}\}$ are two (possibly overlapping) subsets of baseline covariates. Let the excitation function take the form of exponential decay, i.e., $g_i(t - t_{ij}) = \exp\{-\delta(t - t_{ij})\}$ with rate $\delta > 0$. This conveniently allows for recursive computation of the conditional intensity. In addition, it has an intuitive interpretation as the density function for the time elapsed between a parent event and its offspring. Then for our proposed model, a Hawkes process with a proportional intensity baseline, background and offspring process covariates, and independent random effects $\nu_i$ and $\omega_i$, the conditional intensity function for individual $i$ has the following form:
\vspace{1ex}
\begin{equation}\label{eq:lambda_star}
\lambda_i^*(t) = \nu_i \mu_0(t) \exp\left\{\sum_{p=0}^P \beta_p x_{ip}\right\} + \omega_i \exp\left\{\sum_{q=0}^Q \zeta_q z_{iq}\right\} \sum_{t_{ij} < t} \exp\{-\delta(t - t_{ij})\},
\end{equation}

where $\nu_i \sim \text{Gamma}(\phi, \phi)$, $\omega_i \sim \text{Gamma}(\xi, \xi)$, and $\delta, \phi, \xi > 0$. 

For our specific application, we use a Weibull baseline in the background intensity function, i.e., $\mu_0(t) = \alpha t^{\alpha-1}$ with $\alpha > 0$. While we show Weibull-specific formulas from this point forward, it is straightforward to extend our methods to other baseline functions. A Weibull baseline models a trend in the background rate that is decreasing with time for $\alpha < 1$, constant for $\alpha = 1$, and increasing for $\alpha > 1$. By expressing $\eta_i$ in terms of the background covariates, we allow the Weibull scale parameter to vary by individual. The background intensity is also modified by a latent multiplicative random effect to account for additional heterogenity between individuals. This frailty term $\nu_i$ is assumed to follow a Gamma distribution with a mean of one and precision $\phi > 0$ such that $\nu_i \sim \text{Gamma}(\phi, \phi)$.


Covariates are included in the offspring intensity through the jump size $\kappa_i = \exp\left\{\sum_{q=0}^Q \zeta_q z_{iq}\right\}$. The offspring covariate set $\{z_{iq}\}$ may or may not intersect with the background covariate set $\{x_{ip}\}$, allowing us to consider the influence of covariates separately between the processes. To account for additional heterogeneity between individuals in the excitation process, the offspring intensity is modified by a latent multiplicative random effect $\omega_i$ that is independent of the background random effect $\nu_i$. This offspring frailty term is similarly assumed to follow a Gamma distribution with a mean of one and precision $\xi > 0$, such that $\omega_i \sim \text{Gamma}(\xi, \xi)$.


\subsection{Branching Structure}\label{sec:branch_Y}

A Hawkes process can be constructed as a composition of Poisson cluster processes \citep{hawkes_cluster_1974, rasmussen_bayesian_2013}. In this view, immigrant arrivals occur according to a Poisson process with background rate $\mu_i(t)$. Each arrival at time $t_{ij}$ generates a new Poisson process with offspring rate $\kappa_{ij} g_i(t-t_{ij})$ for $t > t_{ij}$. 
With exponential decay, the offspring rate jumps to its highest point immediately following the event and decays exponentially. This cluster generation process occurs recursively, so that each event may then generate more offspring in a similar manner, resulting in the formation of temporal clusters. This gives rise to the latent branching structure of the Hawkes process, defined as 
\vspace{1ex}
\begin{equation}
Y_{ij} = \begin{cases}
   0 & \text{if } t_{ij} \text{ is an immigrant}, \\[-0.5ex]
   y & \text{if } t_{ij} \text{ is an offspring of } t_{iy} < t_{ij}.
\end{cases}
\end{equation}

Immigrant arrivals in $I_i$ are generated according to background intensity $\lambda_{I_i}(t) = \nu_i \eta_i \mu_0(t)$ and offspring in $O_{iy}$ are generated according to offspring intensity $\lambda_{O_{iy}}(t) = \omega_i \kappa_i g_i(t-t_{iy})$. The Hawkes process can therefore be viewed as a Poisson cluster process generated by
\vspace{1ex}
\begin{equation}\label{eq:lambda_Y}
\begin{aligned}
&\lambda_{I_i}(t) 
   = \nu_i \eta_i \alpha t^{\alpha-1} 
   &&\text{where } I_i = \{t_{ij}: Y_{ij} = 0\}, \\[1.5ex]
&\lambda_{O_{iy}}(t) 
   = \omega_i \kappa_i \exp\{-\delta(t - t_{iy})\}
   &&\text{where } O_{iy} = \{t_{ij}: Y_{ij} = y\}.
\end{aligned}
\end{equation}

The compensators, or the integrated intensity functions, then take the following form:
\vspace{1ex}
\begin{equation}\label{eq:Lambda_Y}
\begin{aligned}
&\Lambda_{I_i}(t) 
   = \nu_i \eta_i t^{\alpha} 
   &&\text{ where } I_i = \{t_{ij}: Y_{ij} = 0\}, \\[1.5ex]
&\Lambda_{O_{iy}}(t) 
   = \dfrac{\omega_i \kappa_i}{\delta} \left(1 - \exp\{-\delta(t - t_{iy})\}\right) 
   &&\text{ where } O_{iy} = \{t_{ij}: Y_{ij} = y\}.
\end{aligned}
\end{equation}

\vspace{0pt}
\smallskip
\textbf{Branching Ratio}. The branching ratio $r_i$ for an individual Hawkes process can be derived from the offspring intensity function as follows 
\citep{laub_elements_2021}: 
\vspace{0.5ex}
\begin{equation}\label{eq:branching_ratio}
r_i = \int_0^\infty \omega_i \kappa_i \exp(-\delta s)ds
= \dfrac{\omega_i \kappa_i}{\delta} \,.
\end{equation}

\vspace{-1ex}
\noindent This represents the expected number of offspring for an immigrant event. The expected total number of descendants for an immigrant event, where $k$ represents the generation index, is then $\sum_{k=1}^\infty r_i^k = r_i/(1-r_i)$ for a stationary process with $r_i < 1$. The expected cluster size, including the original immigrant event, is then $\sum_{k=0}^\infty r_i^k = 1/(1-r_i)$. The branching ratio can thus be interpreted as the expected proportion of offspring events in a cluster, or the probability of a Hawkes process event being generated from the self-excitation process. Note that if $r_i \ge 1$, the expected cluster size is infinite and the Hawkes process explodes \citep{laub_elements_2021, lim_simulation_2016}. Because the offspring random effects have a mean of one, we can also define a global branching ratio for a set of offspring covariate values. For a model with no offspring covariates, the global branching ratio is $r = \kappa/\delta$, where $\kappa = \exp(\zeta_0)$. 

\vspace{0pt}
\smallskip
\textbf{Conditional Likelihood}. The cluster process intensity pieces can be used to construct the joint likelihood for the mixed Hawkes process model. Suppose independent Hawkes processes are generated with parameters $\bm\theta = \{\alpha, \delta, \bm\beta, \bm\zeta, \phi, \xi\}$ according to the conditional intensity in Equation \eqref{eq:lambda_star}. For individuals $i = 1, ..., m$, each with event times $\bm{t}_i = \{t_{i1}, ..., t_{in_i}\}$ over the follow-up period $[0, T_i]$, the general form of the complete data joint density function \citep{cook_statistical_2007}, conditional on random effects $\bm\nu = \{\nu_1, ..., \nu_m\}$ and $\bm\omega = \{\omega_1, ..., \omega_m\}$, is
\vspace{1ex}
\begin{equation}\label{eq:like_star}
f(\bm{t} \,|\, \bm\nu, \bm\omega, \bm{\theta}) 
\propto \prod_{i=1}^m \left( \exp\{-\Lambda_i(T_i)\} 
  \prod_{j=1}^{n_i}\lambda_i^*(t_{ij})\right), 
  \text{ where } \Lambda_i(T_i) = \int_0^{T_i}\lambda_i^*(s)ds.   
\end{equation}

The cluster formation definition of the Hawkes process allows us to separate each set of events into independent Poisson processes. Thus, by conditioning on the latent branching structure, the density for individual $i$ can be restated as the product of density functions for immigrant set $I_i$ and offspring event sets $O_{iy}$. Let $\xsub{\gamma}_\phi$ and $\xsub{\gamma}_\xi$ denote the Gamma density functions for random effects $\nu_i$ and $\omega_i$, respectively. Then the joint density function for complete data $\bm{t}$ and random effects $\bm{\nu}$ and $\bm{\omega}$, conditional on branching structure $\bm{Y}$, becomes  

\vspace{-4ex}
{\allowdisplaybreaks
\begin{align}  
&p(\bm{t}, \bm\nu, \bm\omega \,|\, \bm{Y}, \bm{\theta})
\propto \prod_{i=1}^m \Bigg[ \xsub{\gamma}_\phi(\nu_i) \xsub{\gamma}_\xi(\omega_i) 
   p(I_i \,|\, \nu_i, \bm{Y}_i, \bm{\theta})
   \prod_{y=1}^{n_i} p(O_{iy} \,|\, \omega_i, \bm{Y}_i, \bm{\theta}) \Bigg] \\[2ex]
&\propto \prod_{i=1}^m \Bigg[ \xsub{\gamma}_\phi(\nu_i) \xsub{\gamma}_\xi(\omega_i) 
   \exp\left\{-\Lambda_{I_i}(T_i)\right\} \prod_{t_{ij} \in I_i} \lambda_{I_i}(t_{ij}) 
   \prod_{y=1}^{n_i} \Bigg( \exp\{-\Lambda_{O_{iy}}(T_i)\} 
   \prod_{t_{ij} \in O_{iy}} \hspace{-0.1em}
   \hspace{-0,1em} \lambda_{O_{iy}}(t_{ij}) \Bigg) \Bigg] \\[2ex]
&\propto \prod_{i=1}^m \Bigg[ 
   \dfrac{\phi^\phi}{\Gamma(\phi)} \nu_i^{\phi-1} 
   \exp(-\phi\nu_i) \dfrac{\xi^\xi}{\Gamma(\xi)} 
   \omega_i^{\xi-1} \exp(-\xi\omega_i) 
   \exp(-\nu_i \eta_i {T_i}^{\alpha})
   \prod_{t_{ij} \in I_i} \nu_i \eta_i \alpha t_{ij}^{\alpha-1} \notag \\[2ex]
&\qquad \times \prod_{y=1}^{n_i} 
   \Bigg( \exp\left\{ -\frac{\omega_i \kappa_i}{\delta} 
   \left( 1 - \exp\{-\delta(T_i - t_{iy})\} \right) \right\}
   \prod_{t_{ij} \in O_{iy}} \omega_i \kappa_i 
   \exp\{-\delta(t_{ij} - t_{iy})\} \Bigg) \Bigg].
\end{align}
}

\vspace{1ex}
\citet{rasmussen_bayesian_2013} found that Bayesian sampling methods that condition on the cluster representation of a Hawkes process are more computationally efficient than those based on the conditional intensity in Equation \eqref{eq:lambda_star}. This is because fewer points are involved in the estimation of each parameter and the dependence between parameters in separate processes is reduced or removed \citep{ross_bayesian_2021}. The Bayesian inference model is completed by specifying Gamma priors for the positive Hawkes process parameters and Inverse-Gamma priors for the exponentiated coefficients: 
$\alpha \sim \text{Gamma}(a_{\alpha}, b_{\alpha})$, 
$\delta \sim \text{Gamma}(\xsub{a}_{\delta}, \xsub{b}_{\delta})$, 
$\phi \sim \text{Gamma}(\xsub{a}_{\phi}, \xsub{b}_{\phi})$, 
$\xi \sim \text{Gamma}(\xsub{a}_{\xi}, \xsub{b}_{\xi})$, 
$\exp(\beta_p) \sim \text{Inverse-Gamma}(\xsub{a}_\beta, \xsub{b}_\beta)$, 
$\exp(\zeta_q) \sim \text{Inverse-Gamma}(\xsub{a}_\zeta, \xsub{b}_\zeta)$.

\subsection{Complete Data MCMC Algorithm}\label{sec:mcmc1}

Parameter estimation proceeds by iterating through the MCMC steps outlined below. To simplify notation, let $\bm\theta^* = \{\alpha, \delta, \bm\beta, \bm\zeta, \phi, \xi, \bm{\nu}, \bm{\omega}, \bm{Y}\}$ represent the extended set of parameters to be sampled, including the random effects and branching structure. In the first stage of the MCMC algorithm, the complete data joint posterior distribution 
\begin{equation}
\pi(\bm{\theta}^* |\, \bm{t})
\propto p(\bm{t} \,|\, \bm\nu, \bm\omega, \bm{Y}, \bm{\theta})
\,\xsub{\gamma}_\phi(\bm\nu) \,\xsub{\gamma}_\xi(\bm\omega) 
\,\pi(\bm{Y}) \,\pi(\bm\theta) \,
\end{equation}
is sampled under the assumption that the point process $\bm{t}$ is fully observed. Gibbs sampling is used to approximate the joint posterior, where each element is sequentially sampled from its full conditional distribution. The branching structure $\bm{Y}$ and random effects $\bm\nu$ and $\bm\omega$ can be sampled from known distributions. Building upon \citet{lim_simulation_2016}, the adaptive rejection sampling (ARS) algorithm \citep{gilks_adaptive_1992, hunziker_fast_2018} is used for the Hawkes process parameters $(\alpha, \delta, \phi, \xi)$. ARS usually allows for a better mixing Markov chain than a Metropolis-Hastings transition probability and is applicable when the full conditional posterior is log-concave. When covariates are strictly binary, conjugate distributions can be used to sample coefficients $\beta_p$ and $\zeta_q$ instead of ARS. The form of the coefficient samplers depends on the random effects parameterization (see section \ref{supp:mcmc1} for details). Derivations for the full conditional distributions are provided in Section \ref{supp:mcmc1} of the supplement. The transition probabilities of the complete data MCMC algorithm are summarized as follows: 


\begin{enumerate}[noitemsep]
\item Multinomial sampler for branching structure $Y_{ij}$, $i=1,...,m$, $j=1,...,n_i$. 
\item Gamma samplers for random effects $\nu_i$ and $\omega_i$, $i=1,...,m$. 
\item ARS for Hawkes process parameters $\alpha$, $\delta$, $\phi$, $\xi$. 
\item Inverse-Gamma samplers for coefficients $\beta_p$, $p=1,...,P$, and $\zeta_q$, $q=1,,,.,Q$. 
\end{enumerate}

\subsection{Latent Event Imputation}\label{sec:mcmc2}

\noindent The previous MCMC steps are based on the assumption of fully observed event trajectories $\bm{t}_i$ for individuals $i=1,...,m$. In the event of incomplete data, Bayesian estimation can proceed through data augmentation \citep{tanner_calculation_1987}. We consider point processes with two types of latent events, those reported in aggregate as counts and those that are unreported altogether. Let $\bm{t} = \{\bm{t}_{agg}, \bm{t}_{miss}\}$ denote the complete data, where $\bm{t}_{agg}$ consists of unobserved continous event times occurring during tracked intervals and reported as binned counts $\bm{N}$, and $\bm{t}_{miss}$ consists of unreported events occurring during untracked intervals. If aggregation is the only source of latent events, data augmentation proceeds by alternating between posterior sampling steps based on $p(\bm{\theta}^* \,|\, \bm{N},\bm{t}_{agg})$ and imputation steps based on $p(\bm{t}_{agg} \,|\, \bm{N}, \bm{\theta}^*)$ \citep{zhou_bayesian_2025}. If missing intervals are the only issue, we proceed by alternating between posterior sampling steps based on $p(\bm{\theta}^* \,|\, \bm{t})$ and imputation steps based on $p(\bm{t}_{miss} \,|\, \bm{\bm{t}_{obs}, \theta}^*)$, where $\bm{t}_{obs}$ consists of the observed continuous event times \citep{tucker_derek_handling_2019}. With both sources of latent event times, two imputation steps can be added to the MCMC algorithm, combining the two data augmentation approaches. 

First, impute continuous event times from the aggregated counts based on full conditional $p(\bm{t}_{agg} \,|\, \bm{t}_{miss}, \bm{N}, \bm{\theta}^*)$. Second, impute missing events across the unreported intervals based on $p(\bm{t}_{miss} \,|\, \bm{t}_{agg}, \bm{N}, \bm{\theta}^*)$. The two imputation steps combine to approximate the distribution $p(\bm{t}_{agg},\bm{t}_{miss} \,|\, \bm{N}, \bm{\theta}^*) = p(\bm{t} \,|\, \bm{N}, \bm{\theta}^*)$. The Gibbs sampler alternates between the imputation steps and the posterior steps from Section \ref{sec:mcmc1}, which collectively approximate the complete data joint posterior $p(\bm{\theta}^* \,|\, \bm{t}) = p(\bm{\theta}^* \,|\, \bm{N},\bm{t})$. This forms a data augmentation process with observed data $\bm{N}$ and latent data $\bm{t}$. The Rao-Blackwellized \citep{robert_raoblackwellisation_2021} average of distribution $p(\bm{\theta}^* \,|\, \bm{N},\bm{t})$, over many iterations of sampling the complete point process $\bm{t}$, converges to $p(\bm{\theta}^* \,|\, \bm{N})$. This is the joint posterior for all model parameters, given the observed counts $\bm{N}$. Details on the Metropolis-Hastings acceptance ratios for the two imputation steps are provided in Section \ref{supp:mcmc2} of the supplement. An issue arises when simulating missing event times over an untracked interval. Although it is not plausible for an individual to experience infinitely many events, it is still possible to sample an offspring random effect that results in a branching ratio above one, leading to an exploding process. To circumvent this problem, $r_i$ is capped at 0.99 within the missing data imputation step. 

\vspace{0pt}
\smallskip
\textbf{Model Selection}. The following criteria were considered for model selection: (1) Deviance information criterion (DIC) \citep{spiegelhalter_bayesian_2002}, (2) Watanabe-Akaike information criterion (WAIC) \citep{watanabe_asymptotic_2010-1}, (3) Pareto-smoothed importance sampling leave-one-out cross-validation (LOO) 
\citep{vehtari_practical_2017}, and (4) negative log pseudo-marginal likelihood ($-$LPML) \citep{geisser_predictive_1979}. As a result of the missing data imputation, the number of events is not constant across MCMC iterations, requiring the metrics to be computed at the individual level rather than by time point. Due to the complexity of the random effects structure and the flexibility of the proposed model, DIC and WAIC may not be reliable indicators of model fit. Cross-validation based methods such as LOO and $-$LPML are considered to be more stable and robust, particularly with the presence of weak priors and influential data points \citep{vehtari_practical_2017, vranckx_stability_2021}. 

\section{Application to HEP Study}\label{sec:app}

To prepare the HEP study data for analysis, anti-seizure medications started within the first 30 days of study enrollment were treated as the baseline values. Any missing baseline covariate values were filled in using the mode, with a maximum number imputed of 30 (7.4\%) per covariate. After discussion with our clinical collaborators, continuous covariates were discretized by grouping the participants into thirds. The Gibbs sampler algorithm in Section \ref{sec:methods} was implemented in R software version 4.5.1 \citep{r_core_team_r_2025} and compiled as a NIMBLE function \citep{valpine_programming_2017, valpine_nimble_2024} for faster run time. Four MCMC chains were constructed for each model in the analysis. The prior distributions were specified as follows: $\alpha \sim \text{Gamma}(2, 1)$, $\delta \sim \text{Gamma}(2, 1)$, $\phi \sim \text{Gamma}(2, 0.1)$, $\xi \sim \text{Gamma}(2, 0.1)$, $\exp(\beta_p) \sim \text{Inverse-Gamma}(0.001, 0.001)$, $\exp(\zeta_q) \sim \text{Inverse-Gamma}(0.001, 0.001)$. Initial values were sampled uniformly from the following intervals: $\alpha \in (0.5, 1.5)$, $\delta \in (0.5, 1.5)$, $\phi \in (10, 30)$, $\xi \in (10, 30)$, $\beta_0 \in (-3, -1)$, $\zeta_0 \in (-3, -1)$, $\beta_p \approx 0$ for $p > 0$, $\zeta_q \approx 0$ for $q > 0$. After discarding a burn-in period of 2000 iterations, we retained 10,000 iterations per chain for inference. Convergence of the MCMC chains was assessed through visual inspection of the trace plots and the R-hat convergence diagnostic \citep{vehtari_rank-normalization_2021}. 

\subsection{HEP Data Analysis}\label{sec:hep}

The proposed Hawkes process model, as defined by Equation \eqref{eq:lambda_star}, was applied to the HEP study data with the versions listed below, with random effects included in both processes. 

\begin{itemize}[noitemsep]
\item
M1: Model with full set of covariates and random effects in both processes. 
\item
M2: Model with selected covariates and random effects in both processes. Covariates retained from model M1 include basic demographics and those with significant effects. 
\item 
M3: Since there were no significant effects among the offspring covariates, M3 includes the selected covariates in the background only, and random effects in both processes. 
\end{itemize}

The model selection criteria in Table \ref{tab:hep_fit} consistently indicate that among models M1--M3, the full model (M1) provides the best fit to the HEP study data. However, to simplify the interpretation, results are reported for model M3 (Table \ref{tab:hep_par_M3}). Results for models M1 and M2 are shown in Tables \ref{tab:hep_par_M1} and \ref{tab:hep_par_M2}, respectively, of the supplement. 

\begin{table}[!ht]
\vspace{1ex}
\setlength{\tabcolsep}{7pt}
\centering{
\caption{Model selection criteria include deviance information criterion (DIC), Watanabe-Akaike information criterion (WAIC), Pareto-smoothed importance sampling leave-one-out cross-validation (LOO), and negative log pseudo-marginal likelihood ($-$LPML). DIC and WAIC displayed in thousands.}

\begin{tabular}{>{\centering\arraybackslash}p{0.30in}>{\raggedright\arraybackslash}p{1.70in}>{\raggedright\arraybackslash}p{0.75in}>{\raggedleft\arraybackslash}p{0.55in}>{\raggedleft\arraybackslash}p{0.55in}>{\raggedleft\arraybackslash}p{0.55in}>{\raggedleft\arraybackslash}p{0.55in}}
\toprule
\multicolumn{1}{c}{Model} & \multicolumn{1}{l}{Covariates} & \multicolumn{1}{l}{RE} & \multicolumn{1}{c}{DIC} & \multicolumn{1}{c}{WAIC} & \multicolumn{1}{c}{LOO} & \multicolumn{1}{c}{\textendash{}LPML}\\
\midrule
M1 & All in Both & Both & 72,186 & 68,057 & 177,485 & 89,974\\
M2 & Selected in Both & Both & 82,487 & 79,688 & 177,870 & 90,243\\
M3 & Selected in Background & Both & 85,767 & 82,266 & 177,888 & 90,425\\
M4 & Selected in Background & Background & 8,136 & 7,632 & 188,364 & 95,878\\
M5 & Selected in Background & Offspring & 14,235 & 12,586 & 200,617 & 101,681\\
M6 & Selected in Background & None & 14,880 & 14,477 & 200,913 & 101,667\\
\bottomrule
\end{tabular}

\label{tab:hep_fit}
}
\vspace{-2ex}
\end{table}

To assess the impact of the inclusion of random effects, we then compared the following to model M3, which has random effects in both processes. All models in this comparison have the same selected covariates in the background process and no offspring fixed effects. 

\begin{itemize}[noitemsep]
\item
M4: Model with selected background covariates and only background random effects. 
\item
M5: Model with selected background covariates and only offspring random effects. 
\item 
M6: Model with selected background covariates and no random effects. 
\end{itemize}

For models M3--M6, the model selection criteria are not consistent, but the cross-validation based metrics are expected to perform more reliably, as noted above. The model with both background and offspring random effects (M3) is supported by LOO and --LPML as having the preferred random effects structure for the HEP study data. 

\vspace{0pt}
\textbf{Hawkes Process}. Table \ref{tab:hep_par_M3} contains the posterior mean estimates and 95\% highest posterior density (HPD) intervals for the parameters and exponentiated coefficients in model M3. The estimated Weibull baseline shape parameter $\alpha$ is $0.872$ (95\% CrI: $0.844, 0.900$), indicating that the background seizure intensity drops by over 50\% during the first year and continues to decrease over the duration of the study, in agreement with the smoothed hazard in Figure \ref{fig:hep_hazard}. Since study participants were newly diagnosed at the time of enrollment, the declining background seizure rate is likely attributable to seizure control achieved for a portion of participants following introduction and adjustments to anti-seizure medication regimens. The estimated exponential decay rate $\delta$ is $0.638$ $(0.583, 0.694)$, corresponding to an expected elapsed time between parent and offspring seizures of $1.57$ $(1.43, 1.70)$ days. In other words, after an initial seizure event, individuals with epilepsy are at highest risk for a secondary seizure event for one to two days. The estimated background random effects variance $1/\phi$ is $4.652$ $(3.928, 5.418)$, indicating substantial heterogeneity in background intensity between individuals beyond that accounted for by the background process covariates. This implies that while part of the heterogeneity in the study population can be attributed to measured baseline characteristics, a significant portion may be inherent to the disease. The estimated offspring random effects variance $1/\xi$ is $0.299$ $(0.201, 0.404)$, which indicates a small amount of heterogeneity in offspring jump size among individuals. These results imply that the large amount of variability apparent in the HEP seizure count trajectories is largely driven by individual differences in the background seizure rate. 

\begin{table}[!ht]
\vspace{1ex}
\setlength{\tabcolsep}{5pt}
\centering{
\caption{Estimated posterior means and 95\% highest posterior density intervals for Hawkes process model M3 fit to HEP study data. Model includes selected covariates in background process only and random effects in both background and offspring processes.}

\begin{tabular}{clSSSS}\toprule
\multicolumn{1}{c}{} & \multicolumn{1}{c}{} & \multicolumn{1}{>{\centering\arraybackslash}p{0.70in}}{Posterior Mean} & \multicolumn{1}{>{\centering\arraybackslash}p{0.70in}}{95\% HPD Lower} & \multicolumn{1}{>{\centering\arraybackslash}p{0.70in}}{95\% HPD Upper} & \multicolumn{1}{>{\centering\arraybackslash}p{0.70in}}{R-Hat Statistic}\\
\midrule
\addlinespace[0.3em]
\multicolumn{6}{l}{\textbf{Hawkes Process Parameters}}\\
\hspace{1em}$\alpha$ & Weibull Shape & 0.872 & 0.844 & 0.900 & 1.003\\
\hspace{1em}$\delta$ & Exponential Decay Rate & 0.638 & 0.583 & 0.694 & 1.003\\
\hspace{1em}$1/\phi$ & Background RE Variance & 4.652 & 3.928 & 5.418 & 1.003\\
\hspace{1em}$1/\xi$ & Offspring RE Variance & 0.299 & 0.201 & 0.404 & 1.001\\
\addlinespace[0.3em]
\multicolumn{6}{l}{\textbf{Background Coefficients}}\\
\hspace{1em}$\exp(\beta_0)$ & Intercept & 0.028 & 0.013 & 0.045 & 1.003\\
\hspace{1em}$\exp(\beta_1)$ & Sex: Male & 0.678 & 0.355 & 1.031 & 1.002\\
\hspace{1em}$\exp(\beta_2)$ & Age: 25-39 & 2.110 & 0.833 & 3.638 & 1.000\\
\hspace{1em}$\exp(\beta_3)$ & Age: 40-64 & 3.300 & 1.406 & 5.584 & 1.004\\
\hspace{1em}$\exp(\beta_4)$ & Completed Higher Education & 0.574 & 0.275 & 0.925 & 1.000\\
\hspace{1em}$\exp(\beta_5)$ & Total Seizures Pre-Tx: 4-29 & 3.472 & 1.504 & 5.691 & 1.001\\
\hspace{1em}$\exp(\beta_6)$ & Total Seizures Pre-Tx: 30+ & 10.028 & 4.733 & 16.403 & 1.003\\
\hspace{1em}$\exp(\beta_7)$ & ASM: Combination Therapy & 3.775 & 1.537 & 6.495 & 1.000\\
\addlinespace[0.3em]
\multicolumn{6}{l}{\textbf{Offspring Coefficients}}\\
\hspace{1em}$\exp(\zeta_0)$ & Intercept & 0.348 & 0.310 & 0.386 & 1.005\\
\bottomrule
\end{tabular}

\label{tab:hep_par_M3}
}
\end{table}

\vspace{0pt}
\smallskip
\textbf{Covariates}. The model results also include demographic and clinical factors that may help identify individuals at higher overall risk of seizures. While not significant, the background intensity is on average 32.3\% lower for males, adjusting for all other background covariates. For participants in age groups $25-39$ and $40-64$, the adjusted background intensity is higher by a factor of 2.110 (0.833, 3.638) and 3.300 (1.406, 5.584), respectively, compared to those under 25 at baseline, indicating that seizure risk generally increases with age at diagnosis. The background intensity is on average 42.6\% (7.5\%, 72.5\%) lower for participants who had completed higher education at the time of enrollment, after adjustment, although this is unlikely to be a causal effect. For participants with 4$-$29 and 30+ total seizures prior to treatment initiation, the adjusted background intensity ratio is 3.472 (1.504, 5.691) and 10.028 (4.733, 16.403), respectively, relative to those with 1$-$3 pre-treatment seizures, reflecting that seizure control is harder to achieve in individuals with chronic issues or an explosive onset. The adjusted background intensity ratio for participants on combination therapy at baseline is 3.775 (1.537, 6.495), which makes sense since this includes participants whose seizures were not controlled by initial treatment with a single medication. No offspring covariates were included in model M3 because none were found to have significant effects on the offspring process in models M1 or M2. This suggests that primary and secondary seizure events may be clinically different, in that clinical characteristics influencing the background seizure intensity may not be predictive of clustering once a primary seizure event occurs, and that seizure clustering may depend more on latent physiological factors. 

\begin{table}[!ht]
\vspace{1ex}
\setlength{\tabcolsep}{3.5pt}
\centering{
\caption{Posterior medians and 95\% highest posterior density intervals for branching ratio and average cluster size of Hawkes models fit to HEP data. Models include random effects in both background and offspring processes, background random effects only, offspring random effects only, and no random effects. Selected covariates included in background process only.}

\begin{tabular}{lSSSTTT}
\toprule
\multicolumn{1}{p{1.0in}}{\centering  } & \multicolumn{3}{c}{Branching Ratio} & \multicolumn{3}{c}{Average Cluster Size} \\
\cmidrule(l{3pt}r{3pt}){2-4} \cmidrule(l{3pt}r{3pt}){5-7}
\multicolumn{1}{p{0.75in}}{Random Effects} & \multicolumn{1}{p{0.75in}}{\centering Posterior Median} & \multicolumn{1}{p{0.75in}}{\centering 95\% HPD Lower} & \multicolumn{1}{p{0.75in}}{\centering 95\% HPD Upper} & \multicolumn{1}{p{0.75in}}{\centering Posterior Median} & \multicolumn{1}{p{0.75in}}{\centering 95\% HPD Lower} & \multicolumn{1}{p{0.75in}}{\centering 95\% HPD Upper}\\
\midrule
M3: Both & 0.545 & 0.495 & 0.599 & 2.20 & 1.96 & 2.47\\
M4: Background & 0.815 & 0.791 & 0.839 & 5.41 & 4.72 & 6.13\\
M5: Offspring & 0.966 & 0.950 & 0.983 & 29.85 & 17.71 & 50.96\\
M6: None & 0.992 & 0.984 & 1.000 & 122.29 & 42.77 & 783.71\\
\bottomrule
\end{tabular}

\label{tab:hep_off}
}
\vspace{-1ex}
\end{table}

\vspace{0pt}
\smallskip
\textbf{Offspring Process}. Table \ref{tab:hep_off} shows the posterior median branching ratios and average cluster size estimates for the random effects scenarios. The parameter estimates for models M4--M6 are displayed in Tables \ref{tab:hep_par_M4}$-$\ref{tab:hep_par_M6} of the supplement. When random effects are included in both processes (M3), the estimated branching ratio is 0.545 (0.495, 0.599). This means that $45.5\%$ of the seizure events are background events, while $54.5\%$ are generated through excitation, corresponding to an average cluster size of 2.20 (1.96, 2.47) seizures. If random effects are accounted for only in the background process (M4), the estimated branching ratio is 0.815 (0.791, 0.839), corresponding to an average cluster size of 5.41 (4.72, 6.13) seizures. If random effects are incorporated only in the offspring process (M5), the estimated branching ratio is 0.966 (0.950, 0.983), corresponding to an average cluster size of $29.9$ $(17.7, 51.0)$ seizures. Finally, if no random effects are accounted for (M6), the estimated branching ratio is 0.992 (0.984, 1.000), with an average cluster size of 122 (42.8, 784) seizures. These results indicate that random effects have a greater impact on the model in the background than in the offspring process. In addition, when heterogeneity among individuals is not appropriately accounted for through random effects, the impact is apparent in the offspring component, where the branching ratio and cluster size estimates are overstated. 

\subsection{Posterior Predictive Checks}\label{sec:ppc}

To assess the appropriateness of the random effects structure in models M4--M6, we simulated 200 datasets from each posterior predictive distribution \citep{gelman_posterior_1996}. To make the simulated data comparable to the observed data, individual random effects were fixed at their sampled values rather than replicated from Gamma distributions \citep{gelman_bayesian_2013}. Conditioning on the random effects enabled us to align the simulated trajectories to the individuals in the original dataset. For each individual, we removed any untracked days in the HEP study data from the corresponding simulated trajectories and then computed the total number of reported seizure events per trajectory. 

\begin{figure}[!ht]
\centering{
\includegraphics[width=\textwidth]{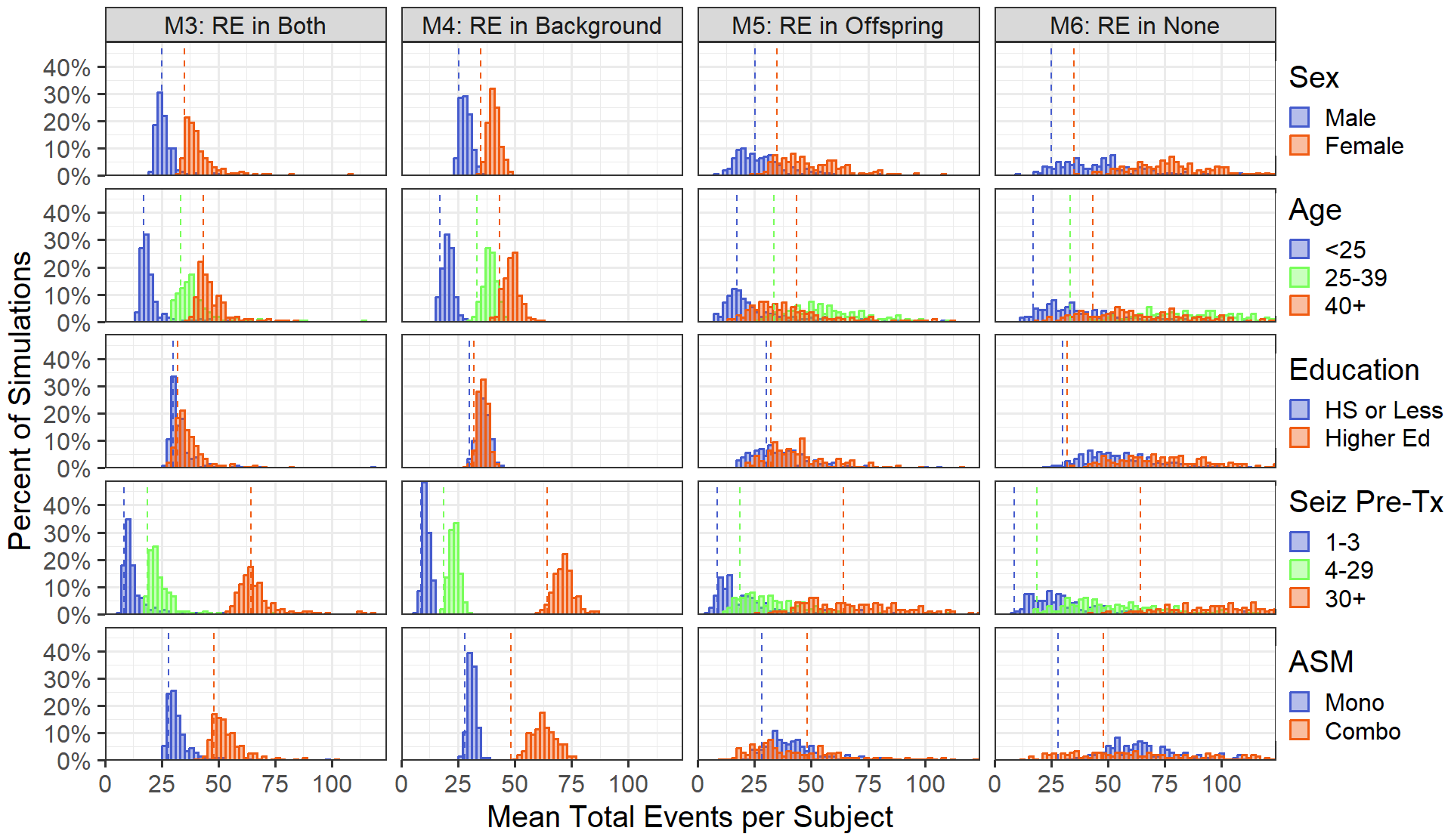}
\caption{Posterior predictive distributions of mean total number of reported seizure events per subject, with simulated events removed from untracked days in the HEP study data. Dashed vertical lines represent observed mean values by covariate group. Columns correspond to models including random effects in both background and offspring processes (M3), random effects in background process only (M4), random effects in offspring process only (M5), and no random effects (M6).}
\label{fig:ppc_global}
}
\end{figure}

Figure \ref{fig:ppc_global} displays histograms of the total simulated number of reported events per subject, averaged across individuals by covariate group. The dashed lines represent the mean total events per individual observed in the HEP study data. When the model accounts for both background and offspring random effects (M3), the posterior predictive distributions are consistent with the observed values. If random effects are included in the background only (M4), the simulated results tend to look high relative to the actual values. With only offspring random effects (M5), the histograms are consistent with the observed values but with wider spread, which indicates that removing the background random effects results in higher variability in the simulated trajectories. When no random effects are included (M6), the histograms become even flatter and wider, and also overestimate the actual values. 

\begin{figure}[!ht]
\centering{
\includegraphics[width=\textwidth]{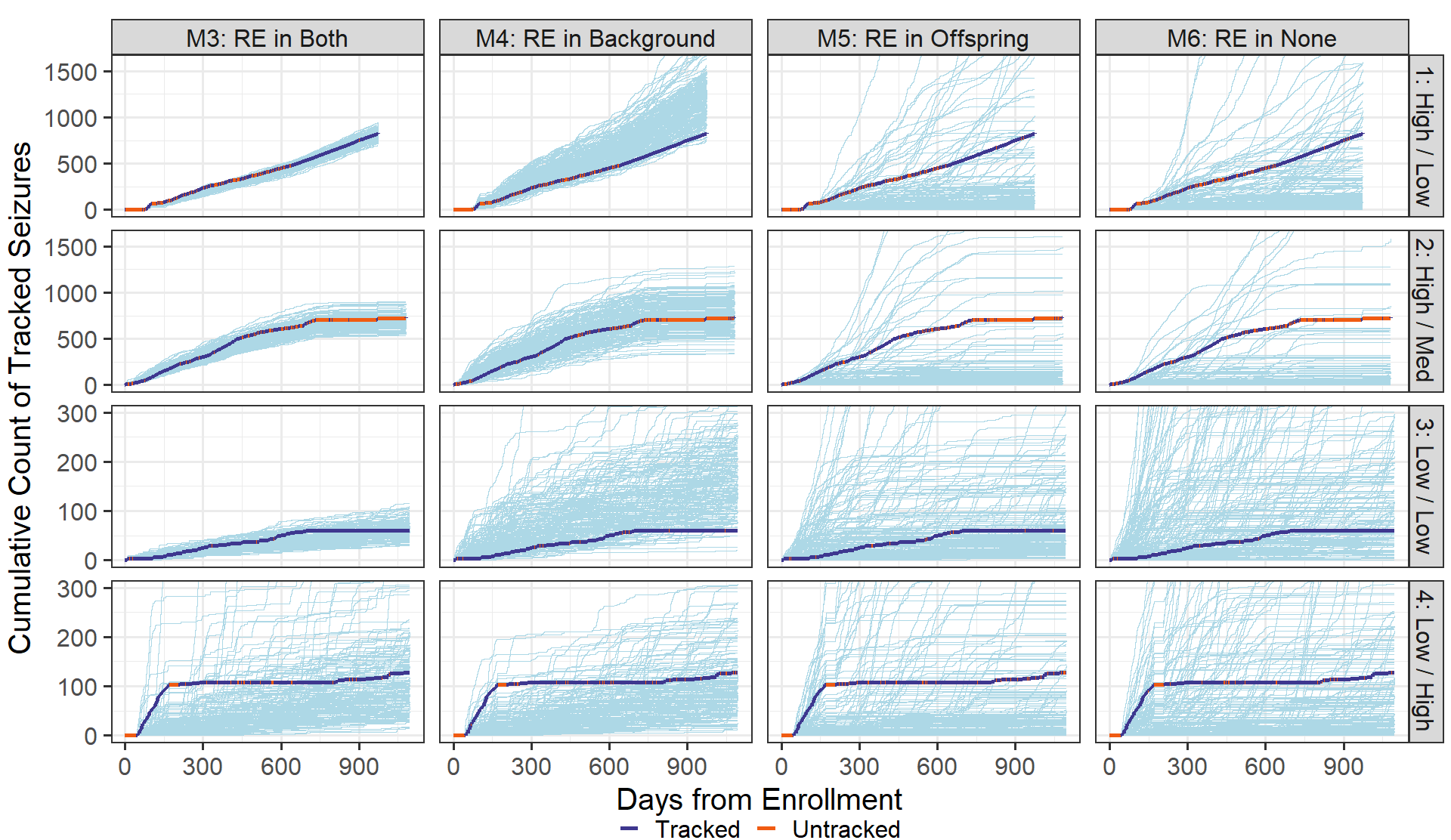}
\caption{Posterior predictive trajectories and observed trajectory for cumulative number of reported seizure events, with simulated counts removed from untracked days in the HEP study data. Columns correspond to models including random effects in both background and offspring processes (M3), random effects in background process only (M4), random effects in offspring process only (M5), and no random effects (M6). Rows correspond to example subjects with varying levels of background and offspring intensity. For example, the "High/Low" label for example 1 refers to a subject with high background rate and low offspring rate.}
\label{fig:ppc_individual}
}
\end{figure}

For an assessment at the individual level, we selected four subjects and performed posterior predictive checks based on their cumulative reported seizure count trajectories. Again, random effects were fixed at their sampled values and seizure events occurring on untracked days in the HEP study data were removed from the simulated trajectories. The individuals included in Figure \ref{fig:ppc_individual} were selected based on their estimated values of $\nu_i \eta_i$ and $\omega_i \kappa_i$ in the model with both random effects (M3). This allows us to assess the impact of subject-specific background and offspring rates, taking into account both fixed and random effects. For model M3, the simulated trajectories are consistent with the observed trajectories, with the exception of the low background, high offspring example (4). This example is unusual in that the slope is quite steep over the first 150 days but relatively flat afterwards, demonstrating that the model does not perform well in the presence of an abrupt change in seizure rate. 

For model M4, which contains random effects in the background process only, the estimated offspring rates are based on the average jump size across all individuals. For this reason, the simulated trajectories have good coverage for an individual with a medium jump size (example 2). However, if the predicted jump size in model M3 is low, the trajectories simulated based on model M4 will tend to be too high relative to the observed trajectory, as in examples 1 and 3. In the first three examples, removing the offspring random effect from the model results in higher variability in the simulated trajectories. 


In an analogous manner, when only offspring random effects are included (model M5), the simulated trajectories tend to be too low and coverage is poor when the predicted background rate in model M3 is high (examples 1 and 2). In addition, the offspring process variability appears inappropriately high for all four example subjects. This tendency for the variability to move into the offspring process is further exacerbated when random effects are removed from the model altogether (model M6). The variability of the simulated trajectories is too high, causing some individual processes to explode, particularly when the background rate estimated by model M3 is low (examples 3 and 4). These results indicate that consistency between the simulated and observed trajectories is highest when background random effects are accounted for in the model. Additionally, variability in the simulated trajectories is lowest when both random effects are included. Based on the posterior predictive checks, the model with both random effects (M3) is the most appropriate for the HEP study data. 

\section{Simulation Studies}\label{sec:sim}

We conducted two simulation studies to evaluate the performance of the proposed model on simulated data with practically plausible sample sizes and effects. \underline{Simulation Study I} examined the impacts of sample size, missing data, and random effects variance on the estimation of model parameters. \underline{Simulation Study II} investigated misspecification of the random effects structure in the presence of heterogeneity. Both simulation studies were set up to mimic the sample sizes and structure of the HEP study data. 

Simulation pseudocode can be found in Section \ref{supp:sim_alg} of the supplement. Tracking trajectories were constructed according to Algorithm \ref{alg:sim_track} by simulating transitions in tracking status based on a two-state discrete-time Markov chain. For states \{0 = missing, 1 = tracked\}, we define $p_i$ as the probability of individual $i$ moving from state 0 to 1, and $q_i$ as the state 1 to 0 probability, where $p_i$ and $q_i$ are generated from Beta distributions. The inputs required to generate the individual transition probabilities and tracking trajectories include dropout rate $\xsub r_T = 0.0008$, minimum follow-up $T_{min} = 3$ days, maximum follow-up $T_{max} = 1096$ days, average percent missing $\pi_0$, factors $\xsub w_1 = 2$ and $\xsub w_2 = 4$, and initial adjustment $\Delta \pi_1 = 0.1$. Factor $w_1 > 1$ increases the ``stickiness'' of tracking state runs beyond that of a binomial distribution, $w_2 > 1$ decreases the dispersion of the transition probabilities, and $\Delta \pi_1 > 0$ allows the initial tracked state probability to be higher than that of the long-run distribution. 

Seizure count trajectories were simulated from Equation \eqref{eq:lambda_star} using Algorithm \ref{alg:sim_event}, an extension of the exact simulation method of \citet{dassios_exact_2013}. For a stable Hawkes process, the branching ratio must be less than one. Branching ratios less than but close to one produce a finite but excessive number of events with a high computational cost. To ensure reasonable run times, simulated offspring random effects were restricted to cap the individual branching ratios at 0.90. The average number subject to capping, as a percentage of sample size, was 0.074\% for low variance, 5.4\% for medium variance, and 11.0\% for high variance scenarios. 

\subsection{Simulation Study I}\label{sec:sim_study1}

Scenarios for Simulation Study I consisted of all combinations of high $(m = 400)$ and low $(m = 100)$ sample size, high $(\pi_0 = 75\%)$, medium $(\pi_0 = 50\%)$, and low $(\pi_0 = 25\%)$ average percent missing, and high ($\phi=0.2$, $\xi=5$), medium ($\phi=1$, $\xi=10$), and low ($\phi=5$, $\xi=50$) random effects variance. The remaining model parameters were selected based on the HEP analysis: $\alpha = 0.9$, $\delta = 0.6$, $\beta_0 = -3.5$, $\beta_1 = -0.5$, $\beta_2 = 1.0$, $\xsub\zeta_0 = -1.1$, $\xsub\zeta_1 = -0.1$, $\xsub\zeta_2 = 0.1$. 300 datasets were simulated for each scenario and model fitting was performed in R version 4.5.1 (R Core Team 2025) using the Gibbs sampler outlined in Section \ref{sec:methods}. The prior distributions were specified with the same parameter values as in Section \ref{sec:app}. Initial values were generated by perturbing the true parameters by a uniformly sampled amount within 10\% of the true values. After initial convergence testing, we discarded a burn-in of 500 iterations and retained 2,000 iterations for inference from one MCMC chain per dataset. 

The selected simulation settings produced datasets comparable to the HEP study data, with an average follow-up time of 732 days across all scenarios and an average of 36.5 reported events per individual in scenarios with high random effects variance and 50\% average missing. Examples of simulated trajectories with 50\% average missing are shown in Figures \ref{fig:sim_plot_LM}--\ref{fig:sim_plot_HM} of the supplement. We can see that increasing the random effects variance results in higher heterogeneity in seizure rates between individuals. In particular, the high variance setting produces simulated trajectories with heterogeneity resembling that of the HEP study data (Figure \ref{fig:hep_sample}). The results of Simulation Study I are shown in Figure \ref{fig:sim_plot}. In addition, tables for all scenarios can be found in Section \ref{supp:sim} of the supplement. 

\begin{figure}[!ht]
\centering{
\includegraphics[width=\textwidth]{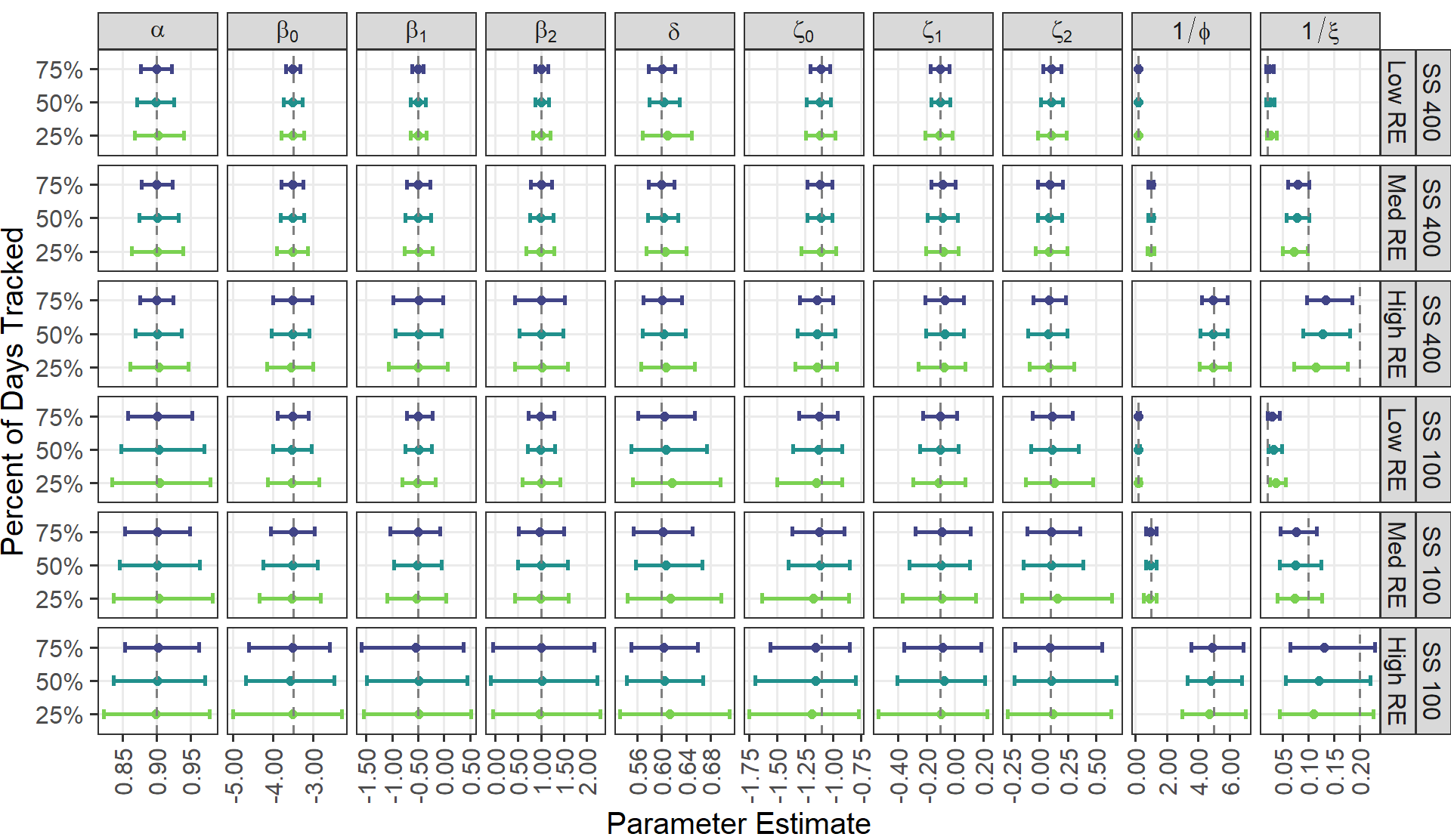}
\caption{Results of Simulation Study I for mixed Hawkes process model with covariates and random effects in both background and offspring processes. Scenarios vary by sample size (100 or 400 individuals), random effects variance (low: $\phi=5, \xi=50$; medium: $\phi=1, \xi=10$; high: $\phi=0.2, \xi=5$), and average percent of days tracked (25\%, 50\%, 75\%). Empirical means and 2.5th/97.5th percentiles based on 300 datasets per scenario. Dashed vertical lines represent true parameter values.}
\label{fig:sim_plot}
}
\end{figure}

For the background parameters ($\alpha$, $\beta_0$, $\beta_1$, $\beta_2$, $1/\phi$), bias is low for all scenarios, although coverage is consistently low for the Weibull shape $\alpha$, with values under 90\% for higher levels of missing data. For the offspring parameters $\delta$ and $\zeta_0$, bias is lower in magnitude for high sample size scenarios and increases with the average percent missing. Bias is positive for the exponential decay rate $\delta$ and negative for the offspring intercept $\zeta_0$, leading to underestimation of the offspring event rates when the percent tracked is low. Estimates of the offspring random effects variance $1/\xi$ show a small amount of positive bias when the true values are low, negative bias for medium variance, and substantial negative bias for high variance scenarios. Consequently, coverage is generally poor for the offspring variance parameter, with values around 50\% and 20\% under large sample size scenarios with medium and high variance, respectively. This underestimation of the offspring random effects variance is induced by the practical need to limit simulated branching ratios to 0.90. 

The variability of estimates is higher for smaller sample size and lower tracking scenarios, due to the increased uncertainty that comes with less observed data. Wider intervals are also evident for scenarios with higher random effects variance, where heterogeneity among individuals leads to more uncertainty in parameter estimation. Additionally, the imputation of aggregated events contributes to uncertainty. As a result, even with unbiased parameter recovery, covariate effects may not be detectable. This is particularly true for the offspring coefficients, which tend to be smaller in magnitude. For example, the intervals for background coefficients $\beta_1$ and $\beta_2$ only include zero for scenarios with high random effects variance and low sample size. Meanwhile, the intervals for offspring coefficients $\zeta_1$ and $\zeta_2$ include zero except for scenarios with low random effects variance and high sample size. 

\subsection{Simulation Study II}\label{sec:sim_study2}

For Simulation Study II, we simulated 300 datasets per scenario with 400 individuals, 50\% average untracked days, and high random effects variance ($\phi = 0.2$, $\xi = 5$) in order to resemble the HEP study data. All other simulation settings and parameter values were retained from Simulation Study I. Datasets were simulated from the mixed Hawkes process model with covariates and random effects in both the background and offspring processes. Models were then fit to the simulated datasets under four scenarios: random effects in both processes, background random effects only, offspring random effects only, and no random effects. This allows us to evaluate the results for model parameter (Table \ref{tab:sim_re}) and branching ratio (Table \ref{tab:sim_off}) estimates under varying misspecifications of the random effects structure. 

\begin{table}[!ht]
\setlength{\tabcolsep}{7pt}
\centering{
\caption{Results of Simulation Study II for data simulated from mixed Hawkes process model with random effects in both background and offspring processes. Models fit to simulated data under four scenarios: random effects in both processes, background random effects only, offspring random effects only, and no random effects. Datasets simulated with 400 individuals, 50\% average missing, and high random effects variance ($\phi=0.2, \xi=5$). Empirical means and 95\% HPD coverage based on 300 datasets per scenario.}

\centering
\begin{tabular}{cTSSSSSSSS}
\toprule
\multicolumn{2}{c}{ } & \multicolumn{2}{c}{Both} & \multicolumn{2}{c}{Background} & \multicolumn{2}{c}{Offspring} & \multicolumn{2}{c}{None} \\
\cmidrule(l{3pt}r{3pt}){3-4} \cmidrule(l{3pt}r{3pt}){5-6} \cmidrule(l{3pt}r{3pt}){7-8} \cmidrule(l{3pt}r{3pt}){9-10}
\multicolumn{1}{p{0.4cm}}{\centering } & \multicolumn{1}{p{1.2cm}}{\centering True Value} & \multicolumn{1}{p{1.2cm}}{\centering Emp Mean} & \multicolumn{1}{p{1.2cm}}{\centering HPD Cvg} & \multicolumn{1}{p{1.2cm}}{\centering Emp Mean} & \multicolumn{1}{p{1.2cm}}{\centering HPD Cvg} & \multicolumn{1}{p{1.2cm}}{\centering Emp Mean} & \multicolumn{1}{p{1.2cm}}{\centering HPD Cvg} & \multicolumn{1}{p{1.2cm}}{\centering Emp Mean} & \multicolumn{1}{p{1.2cm}}{\centering HPD Cvg}\\
\midrule
$\alpha$ & 0.9 & 0.900 & 0.893 & 0.901 & 0.883 & 0.853 & 0.433 & 0.838 & 0.297\\
$\beta_0$ & -3.5 & -3.479 & 0.933 & -3.618 & 0.907 & -3.992 & 0.233 & -4.019 & 0.237\\
$\beta_1$ & -0.5 & -0.534 & 0.940 & -0.479 & 0.897 & -0.218 & 0.207 & -0.187 & 0.193\\
$\beta_2$ & 1.0 & 0.976 & 0.950 & 1.001 & 0.940 & 0.481 & 0.057 & 0.428 & 0.063\\
$\delta$ & 0.6 & 0.602 & 0.943 & 0.560 & 0.250 & 0.291 & 0.000 & 0.271 & 0.000\\
$\zeta_0$ & -1.1 & -1.137 & 0.913 & -0.989 & 0.353 & -1.665 & 0.000 & -1.477 & 0.000\\
$\zeta_1$ & -0.1 & -0.076 & 0.920 & -0.075 & 0.317 & -0.067 & 0.910 & -0.053 & 0.283\\
$\zeta_2$ & 0.1 & 0.074 & 0.930 & 0.115 & 0.393 & 0.103 & 0.960 & 0.102 & 0.510\\
$1/\phi$ & 5.0 & 4.963 & 0.947 & 4.946 & 0.913 &  &  &  & \\
$1/\xi$ & 0.2 & 0.129 & 0.257 &  &  & 0.068 & 0.010 &  & \\
\bottomrule
\end{tabular}

\label{tab:sim_re}
}
\end{table}

As expected, when random effects are accounted for in both processes, the model parameters are estimated well, with the exception of the offspring random effects variance $1/\xi$. When only background random effects are included, empirical means and coverage are reasonable for background parameters $\alpha$ and $1/\phi$, as well as for background coefficients ($\beta_0$, $\beta_1$, $\beta_2$), but coverage is poor for all offspring parameters. With only offspring random effects, empirical means and coverage are only reasonable for offspring coefficients $\zeta_1$ and $\zeta_2$, whereas the remaining offspring parameters ($\delta$, $\zeta_0$, $1/\xi$) and all background parameters are estimated incorrectly. This demonstrates that failing to account for heterogenity in the background process may give rise to biased estimation in the offspring process as well. Finally, when no random effects are included in the model, estimation is poor for all parameters.

In scenarios that exclude background random effects, the Weibull shape parameter $\alpha$ and the Weibull scale group means $\eta_i$ are underestimated. This indicates that background events may be improperly treated as offspring events when the model does not account for heterogeneity in the background process. The exponential decay rate $\delta$ is underestimated in all misspecified scenarios, particularly when background random effects are removed, which causes the offspring intensity decay to stretch out over a longer period of time. The impact of model misspecification on the offspring process is shown in Table \ref{tab:sim_off}. For all offspring covariate combinations, the branching ratios increase across scenarios in the following order: both random effects, offspring only, background only, and none. The same pattern is evident for the cluster size estimates, which are notably elevated for the model with no random effects. These results demonstrate that variability unaccounted for in a misspecified model tends to move from the background intensity into the offspring process branching ratios. 


\begin{table}[!ht]
\setlength{\tabcolsep}{8.5pt}
\centering{
\caption{Results of Simulation Study II for data simulated from mixed Hawkes process model with two binary covariates and random effects in background and offspring processes. Results reported for combinations of binary offspring covariates $z_1$ and $z_2$. Models fit to simulated data under four scenarios: random effects in both processes, background random effects only, offspring random effects only, and no random effects. Datasets simulated with 400 individuals, 50\% average missing, and high random effects variance ($\phi=0.2, \xi=5$). True values are 0.555, 0.502, 0.613 for branching ratio and 2.25, 2.01, 2.58 for average cluster size. Means and percentiles based on medians of 300 datasets per scenario.}

\begin{tabular}{lSSSTTT}
\toprule
\multicolumn{1}{p{1.2in}}{\centering  } & \multicolumn{3}{c}{Branching Ratio} & \multicolumn{3}{c}{Average Cluster Size} \\
\cmidrule(l{3pt}r{3pt}){2-4} \cmidrule(l{3pt}r{3pt}){5-7}
\multicolumn{1}{p{1.2in}}{Random Effects Structure} & \multicolumn{1}{p{0.60in}}{\centering Mean of Medians} & \multicolumn{1}{p{0.60in}}{\centering $P_{2.5}$ of Medians} & \multicolumn{1}{p{0.60in}}{\centering $P_{97.5}$ of Medians} & \multicolumn{1}{p{0.60in}}{\centering Mean of Medians} & \multicolumn{1}{p{0.60in}}{\centering $P_{2.5}$ of Medians} & \multicolumn{1}{p{0.60in}}{\centering $P_{97.5}$ of Medians}\\
\midrule
\addlinespace[0.3em]
\multicolumn{7}{l}{\textbf{$\bm{z_1 = 0}$, $\bm{z_2 = 0}$}}\\
\hspace{1em}Both & 0.535 & 0.448 & 0.615 & 2.17 & 1.81 & 2.60\\
\hspace{1em}Background & 0.671 & 0.509 & 0.831 & 3.30 & 2.04 & 5.91\\
\hspace{1em}Offspring & 0.658 & 0.567 & 0.743 & 2.98 & 2.31 & 3.89\\
\hspace{1em}None & 0.851 & 0.728 & 0.936 & 7.72 & 3.67 & 15.58\\
\addlinespace[0.3em]
\multicolumn{7}{l}{\textbf{$\bm{z_1 = 1}$, $\bm{z_2 = 0}$}}\\
\hspace{1em}Both & 0.496 & 0.413 & 0.589 & 2.00 & 1.70 & 2.43\\
\hspace{1em}Background & 0.625 & 0.448 & 0.805 & 2.86 & 1.81 & 5.12\\
\hspace{1em}Offspring & 0.616 & 0.528 & 0.717 & 2.65 & 2.12 & 3.54\\
\hspace{1em}None & 0.807 & 0.695 & 0.904 & 5.62 & 3.28 & 10.45\\
\addlinespace[0.3em]
\multicolumn{7}{l}{\textbf{$\bm{z_1 = 0}$, $\bm{z_2 = 1}$}}\\
\hspace{1em}Both & 0.575 & 0.517 & 0.633 & 2.36 & 2.07 & 2.72\\
\hspace{1em}Background & 0.748 & 0.663 & 0.822 & 4.07 & 2.96 & 5.62\\
\hspace{1em}Offspring & 0.729 & 0.644 & 0.806 & 3.77 & 2.81 & 5.14\\
\hspace{1em}None & 0.941 & 0.905 & 0.966 & 18.33 & 10.49 & 29.54\\
\bottomrule
\end{tabular}
\label{tab:sim_off}
}
\vspace{5ex}
\end{table}

\section{Discussion}\label{sec:disc}

We propose a mixed Hawkes process model to characterize clustering and heterogeneity in epileptic seizures in a cohort of individuals newly diagnosed with focal epilepsy. Our model provides accurate estimation of seizure occurrence and covariate effects across individuals with variable seizure rates, while addressing the aggregation and missingness issues common in seizure diary data. We develop a Gibbs sampling method that allows for inference on data comprising multiple seizure count trajectories. We demonstrate through simulations that, for practically realistic sample sizes, parameters are estimated well under the correct model, even in scenarios with a high amount of missing data and high random effects variance. We also show that neglecting to account for heterogeneity through random effects may lead to poor estimation of model parameters and inappropriately elevated branching ratios. 

The application of the proposed model to the HEP study data demonstrates that the clustering of seizure events can be explained by a self-exciting process. The average time between parent and offspring events is 1.57 (1.43, 1.70) days and the average seizure cluster size is 2.20 (1.96, 2.47). In other words, after an initial seizure, the most likely time for a secondary seizure is between one and two days later, and only one additional seizure occurs on average. These estimates are consistent with prior clinical observations that seizure clustering typically occurs within a few hours to several days following an index seizure \citep{chiang_individualizing_2020, haut_seizure_2002, jafarpour_seizure_2019, mesraoua_seizure_2021}. Other studies based on EEG monitoring and long-term ambulatory recordings similarly report that increased seizure susceptibility following an initial seizure often dissipates over one to three days, though the exact window varies widely across individuals \citep{karoly_cycles_2021}. From a clinical standpoint, quantifying the expected duration of clustering and the number of seizures likely to occur may help patients and providers develop individualized care plans.

Background seizure intensity was found to be positively associated with age group, number of pre-treatment seizures, and combination therapy, but negatively associated with higher education. While education is unlikely to have a direct effect on seizures, it may be a proxy measure for factors such as medication adherence, access to healthcare, and communication with physicians. Additionally, covariate effects may be due to differences in seizure diary reporting rather than true differences in seizure occurrence, and job demands may be a factor. Notably, none of the examined covariates were associated with seizure recurrence through self-excitation, suggesting that clustering dynamics may be driven more by latent physiological factors than by the clinical characteristics considered. In contrast, covariates influencing the background seizure intensity may help identify individuals at higher overall risk of seizures but do not necessarily predict clustering once a seizure occurs. 

The simulation results emphasize an important statistical and clinical insight: unmodeled heterogeneity in background seizure risk can be misattributed to self-excited seizure activity, artificially exaggerating clustering strength and extending inferred post-seizure risk windows. Similar concerns have been raised implicitly in prior studies, where seizure clusters were identified but subject-specific background risk was not explicitly accounted for \citep{haut_seizure_2005, karoly_circadian_2018}. This finding is consistent with that of the simpler carryover effect model \citep{cigsar_assessing_2012} and cautions against interpreting elevated clustering metrics as evidence of seizure propagation without first addressing individual-level variability.

A few limitations of this work merit discussion. First, seizure diary data are known to suffer from underreporting, as individuals with epilepsy may not recall or recognize all seizures \citep{fisher_seizure_2012, hoppe_epilepsy_2007}. As a result, clustering, and seizure burden more generally, are likely to be underestimated. Future work incorporating prolonged EEG monitoring or wearable seizure tracking device data may help address this limitation. Second, missingness in seizure reporting may be informative, particularly if individuals with higher seizure burden are less likely to report consistently, as suggested by prior analyses \citep{miller_long-term_2024}. Extensions of the model to explicitly handle informative missingness represent an important direction for future research. Third, although for observed data the Hawkes process does not explode, estimation of individual branching ratios can yield values near or exceeding one. This study constrained branching ratios to remain below one during simulation and imputation to ensure model stability. Simulation results indicate that this constraint may underestimate the offspring random effects variance but does not substantially bias other parameters. However, a truncated random effects distribution may be a more appropriate assumption in future work. We also emphasize that the HEP study data is observational, and the analyses presented here are descriptive rather than causal. 

\smallskip
\textbf{Acknowledgments}. This research was done using services provided by the OSG Consortium \citep{osg_ospool_2006, osg_open_2015, pordes_open_2007, sfiligoi_pilot_2009} which is supported by the National Science Foundation awards \#2030508 and \#2323298. 

\smallskip
\textbf{Funding}. This research was supported by the National Institute Of Neurological Disorders And Stroke of the National Institutes of Health under Award Number \if1\anon{R01NS133040}\fi \if0\anon{XXXXXXXX}\fi. The content is solely the responsibility of the authors and does not necessarily represent the official views of the National Institutes of Health.

\smallskip
\textbf{Disclosure Statement}. The authors report there are no competing interests to declare.

\smallskip
\textbf{Data Availability Statement}. The data that support the findings of this study are available from The Human Epilepsy Project. Restrictions apply to the availability of these data, which were used under license for this study. Data are available from the corresponding author with the permission of The Human Epilepsy Project. 



\bibliography{refs.bib}

\appendix

\renewcommand{\thealgocf}{S\arabic{algocf}}
\renewcommand{\theequation}{S\arabic{equation}}
\renewcommand{\thefigure}{S\arabic{figure}}
\renewcommand{\thetable}{S\arabic{table}}

\setcounter{algocf}{0}
\setcounter{equation}{0}
\setcounter{figure}{0}
\setcounter{table}{0}

\clearpage
\begin{center}
{\large\bf A Mixed Self-Exciting Process to Model Epileptic Seizures: Supplementary Material}
\end{center}

\section{MCMC Algorithm}\label{supp:mcmc}

\subsection{Gibbs Sampler for Complete Data}\label{supp:mcmc1}

This section derives full conditional distributions for the Gibbs sampler outlined in Section \ref{sec:mcmc1}. Suppose independent Hawkes processes are generated according to Equation \eqref{eq:lambda_star} for individuals $i = 1, ..., m$ with event times $\bm{t}_i = \{t_{i1}, ..., t_{in_i}\}$ over the time period $[0, T_i]$. Let $\xsub{\gamma}_\phi$ and $\xsub{\gamma}_\xi$ denote the Gamma density functions for the random effects $\nu_i$ and $\omega_i$, respectively: 
\vspace{1ex}
\begin{align}
\nu_i \sim \text{Gamma}(\phi, \phi)
   &\implies \xsub\gamma_\phi (\nu_i) = 
    \dfrac{\phi^\phi}{\Gamma(\phi)} 
    \nu_i^{\phi-1} \exp(-\phi\nu_i) \\[1.5ex]
\omega_i \sim \text{Gamma}(\xi, \xi) 
   &\implies \xsub\gamma_\xi (\omega_i) = 
    \dfrac{\xi^\xi}{\Gamma(\xi)} 
    \omega_i^{\xi-1} \exp(-\xi\omega_i)
\end{align}

In addition, let $\gamma_\phi^c$ and $\gamma_\xi^c$ denote the scaled Gamma density functions for the random effects $\nu_i^c$ and $\omega_i^c$, respectively, under the centered parameterization: 
\vspace{1ex}
\begin{align}
\nu_i^c = \nu_i \eta_i 
\sim \text{Gamma}(\phi, \phi/\eta_i) 
   &\implies \gamma_\phi^c (\nu_i^c) = 
     \dfrac{(\phi/\eta_i)^\phi}{\Gamma(\phi)} 
     (\nu_i^c)^{\phi-1} 
     \, \exp\{-( \phi / \eta_i ) \, \nu_i^c\} \\[1.5ex]
\omega_i^c = \omega_i \kappa_i
\sim \text{Gamma}(\xi, \xi/\kappa_i) 
   &\implies \gamma_\xi^c (\omega_i^c) =
    \dfrac{(\xi/\kappa_i)^\xi}{\Gamma(\xi)} 
    (\omega_i^c)^{\xi-1} 
    \, \exp\{-( \xi / \kappa_i ) \, \omega_i^c\}
\end{align}

Recall that the Weibull scale $\eta_i$ is expressed in terms of the background covariates $\{x_{ip}\}_{p=0}^P$, while the individual jump size $\kappa_i$ is expressed in terms of the offspring covariates $\{z_{iq}\}_{q=0}^Q$: 
\vspace{1ex}
\begin{align}
&\eta_i = \exp\left\{\sum_{p=0}^P \beta_p x_{ip}\right\} 
       = \prod_{p=0}^P \exp(\beta_p x_{ip}) \\[1.5ex]
&\kappa_i = \exp\left\{\sum_{q=0}^Q \zeta_q z_{iq}\right\} 
          = \prod_{q=0}^Q \exp(\zeta_q z_{iq})
\end{align}

\noindent Conditional on the branching structure $\bm{Y}$, the complete data joint density function for the event times $\bm{t}$, background random effects $\bm\nu$, and offspring random effects $\bm\omega$ can be derived based on the intensity components in Equation \eqref{eq:lambda_Y} and the compensators in Equation \eqref{eq:Lambda_Y}.
\vspace{1ex}
\begin{align}  
&p(\bm{t}, \bm\nu, \bm\omega \,|\, \bm{Y}, \bm{\theta})
\propto \prod_{i=1}^m \Bigg[ \xsub{\gamma}_\phi(\nu_i)
   \xsub{\gamma}_\xi(\omega_i) 
   p(I_i \,|\, \nu_i, \bm{Y}_i, \bm{\theta})
   \prod_{y=1}^{n_i} p(O_{iy} \,|\, \omega_i, \bm{Y}_i, \bm{\theta}) \Bigg] \\[2ex]
&\propto \prod_{i=1}^m \Bigg[ \xsub{\gamma}_\phi(\nu_i)
   \xsub{\gamma}_\xi(\omega_i) 
   \exp\{-\Lambda_{I_i}(T_i)\} 
   \prod_{t_{ij} \in I_i} \lambda_{I_i}(t_{ij}) 
   \prod_{y=1}^{n_i} \Bigg( \exp\{-\Lambda_{O_{iy}}(T_i)\} 
   \hspace{-0.25em} \prod_{t_{ij} \in O_{iy}} \hspace{-0.25em}
   \lambda_{O_{iy}}(t_{ij}) \Bigg) \Bigg] \\[2ex]
&\propto \prod_{i=1}^m \Bigg[ 
   \dfrac{\phi^\phi}{\Gamma(\phi)} 
   \nu_i^{\phi-1} \exp(-\phi\nu_i) 
   \times \dfrac{\xi^\xi}{\Gamma(\xi)} 
   \omega_i^{\xi-1} \exp(-\xi\omega_i) 
   \times \exp(-\nu_i \eta_i {T_i}^\alpha) 
   \prod_{t_{ij} \in I_i} 
   \nu_i \eta_i \alpha t_{ij}^{\alpha-1} \notag \\[2ex]
&\qquad \times \prod_{y=1}^{n_i} 
   \Bigg( \exp\left\{ -\frac{\omega_i \kappa_i}{\delta} 
   \big( 1 - \exp\{-\delta(T_i - t_{iy})\} \big) \right\}
   \prod_{t_{ij} \in O_{iy}} \omega_i \kappa_i 
   \exp\{-\delta(t_{ij} - t_{iy})\} \Bigg) \Bigg] \\[2ex]
&\propto \xsup\phi^{m\phi} \times \xsup{[\Gamma(\phi)]}^{-m}
   \times \prod_{i=1}^m \xsup\nu_i^{\phi-1} 
   \times \exp\left\{-\phi \sum_{i=1}^m \nu_i \right\} 
   \notag \\[2ex]
&\qquad \times
   \xsup\xi^{m\xi} \times \xsup{[\Gamma(\xi)]}^{-m}
   \times \prod_{i=1}^m \xsup\omega_i^{\xi-1} 
   \times \exp\left\{-\xi \sum_{i=1}^m \omega_i \right\} 
   \notag \\[2ex]
&\qquad \times 
   \exp\left\{ -\sum_{i=1}^m \left( \nu_i {T_i}^\alpha
   \prod_{p=0}^P \exp(\beta_p x_{ip}) \right) \right\} 
   \times \prod_{i=1}^m \xsup\nu_i^{|I_i|}
   \times \xsup{\alpha}^{\sum_{i=1}^m {|I_i|}} \notag \\[2ex]
&\qquad \times 
   \exp\left\{ (\alpha-1)\sum_{i=1}^m
   \sum_{t_{ij} \in I_i}\log(t_{ij}) \right\} 
   \times \prod_{p=0}^P \exp\left\{ \beta_p 
   \sum_{i=1}^m |I_i|\, x_{ip} \right\} \notag \\[2ex]
&\qquad \times 
   \exp\left\{ {-\sum_{i=1}^m \sum_{y=1}^{n_i}
   \left[ \frac{\omega_i}{\delta} 
   \big( 1 - \exp\{-\delta \left( T_i - t_{iy} \right)\} \big)
   \prod_{q=0}^Q \exp(\zeta_q z_{iq}) \right]} \right\} 
   \times \prod_{i=1}^m 
   \xsup{\omega_i}^{\sum_{y=1}^{n_i} {|O_{iy}|}} \notag \\[2ex]
&\label{eq:joint_ncp} \qquad \times 
   \exp \left\{ -\delta \sum_{i=1}^m \sum_{y=1}^{n_i}
   \sum_{t_{ij} \in O_{iy}} (t_{ij} - t_{iy}) \right\}  
   \times \prod_{q=0}^Q \exp \left\{ \zeta_q \sum_{i=1}^m
   \sum_{y=1}^{n_i} |O_{iy}|\, z_{iq} \right\}
\end{align}  

\vspace{1ex}
Conditional on the branching structure $\bm{Y}$, the complete data joint density function for the event times $\bm{t}$, centered background random effects $\bm\nu^c$, and centered offspring random effects $\bm\omega^c$ can be derived in a similar manner.
\vspace{1ex}
\begin{align}  
&p(\bm{t}, \bm\nu^c, \bm\omega^c \,|\, \bm{Y}, \bm{\theta})
\propto \prod_{i=1}^m \Bigg[ {\xsub{\gamma}}^c_\phi(\nu_i^c) \,
   {\xsub{\gamma}}^c_\xi(\omega_i^c) \, 
   p(I_i \,|\, \nu_i^c, \bm{Y}_i, \bm{\theta})
   \prod_{y=1}^{n_i} p(O_{iy} \,|\, \omega_i^c, \bm{Y}_i, \bm{\theta}) \Bigg] \\[2ex]
&\propto \prod_{i=1}^m \Bigg[ {\xsub{\gamma}}^c_\phi(\nu_i^c) \,
   {\xsub{\gamma}}^c_\xi(\omega_i^c) \,
   \exp\{-\Lambda_{I_i}(T_i)\} 
   \hspace{-0.3em} \prod_{t_{ij} \in I_i} \hspace{-0.3em}
   \lambda_{I_i}(t_{ij}) 
   \prod_{y=1}^{n_i} \Bigg( \exp\{-\Lambda_{O_{iy}}(T_i)\} 
   \hspace{-0.6em} \prod_{t_{ij} \in O_{iy}} \hspace{-0.6em}
   \lambda_{O_{iy}}(t_{ij}) \Bigg) \Bigg] \\[2ex]
&\propto \prod_{i=1}^m \Bigg[ 
   \dfrac{\xsup{(\phi/\eta_i)}^\phi}{\Gamma(\phi)} 
   \xsup{(\nu_i^c)}^{\phi-1} 
   \exp\{-(\phi/\eta_i)\,\nu_i^c\} \notag \\[2ex]
&\qquad \times 
   \dfrac{\xsup{(\xi/\kappa_i)}^\xi}{\Gamma(\xi)} 
   \xsup{(\omega_i^c)}^{\xi-1} 
   \exp\{-(\xi/\kappa_i)\,\omega_i^c\}
   \times \exp(-\nu_i^c {T_i}^{\alpha})
   \prod_{t_{ij} \in I_i} 
   \nu_i^c \alpha t_{ij}^{\alpha-1} \notag \\[2ex]
&\qquad \times 
   \prod_{y=1}^{n_i} 
   \Bigg( \exp\Big\{ -\frac{\omega_i^c}{\delta} 
   \big( 1 - \exp\{-\delta(T_i - t_{iy})\} \big) \Big\}
   \prod_{t_{ij} \in O_{iy}} \omega_i^c 
   \exp\{-\delta(t_{ij} - t_{iy})\} \Bigg) \Bigg] \\[2ex]
&\propto \xsup\phi^{m\phi} \times \prod_{p=0}^P 
   \exp\left\{ -\phi \beta_p \sum_{i=1}^m x_{ip} \right\}
   \times \xsup{[\Gamma(\phi)]}^{-m} \notag \\[2ex]
&\qquad \times 
   \prod_{i=1}^m \xsup{(\nu_i^c)}^{\phi-1}
   \times \exp\left\{-\phi \sum_{i=1}^m 
   \left( \nu_i^c \prod_{p=0}^P 
   \exp(-\beta_p x_{ip}) \right) \right\} \notag \\[2ex]
&\qquad \times 
   \xsup\xi^{m\xi} \times \prod_{q=0}^Q 
   \exp\left\{ -\xi \zeta_q \sum_{i=1}^m z_{iq} \right\} 
   \times \xsup{[\Gamma(\xi)]}^{-m} \notag \\[2ex]
&\qquad \times 
   \prod_{i=1}^m \xsup{(\omega_i^c)}^{\xi-1} 
   \times \exp\left\{-\xi \sum_{i=1}^m \omega_i^c
   \prod_{q=0}^Q \exp(-\zeta_q z_{iq}) \right\} \notag \\[2ex]
&\qquad \times 
   \exp \left\{ -\sum_{i=1}^m \nu_i^c \, {T_i}^\alpha \right\} 
   \times \prod_{i=1}^m \xsup{(\nu_i^c)}^{|I_i|}
   \times \xsup{\alpha}^{\sum_{i=1}^m {|I_i|}}
   \times \exp\left\{ (\alpha-1)\sum_{i=1}^m
   \sum_{t_{ij} \in I_i}\log(t_{ij}) \right\} \notag \\[2ex]
&\qquad \times 
   \exp\left\{ -\sum_{i=1}^m \sum_{y=1}^{n_i}
   \frac{\omega_i^c}{\delta} 
   \big( 1 - \exp\{-\delta ( T_i - t_{iy}\} \big) \right\} 
   \notag \\[2ex]
&\label{eq:joint_cp} \qquad \times 
   \prod_{i=1}^m \xsup{(\omega_i^c)}^{\sum_{y=1}^{n_i} {|O_{iy}|}} 
   \times \exp \left\{ -\delta \sum_{i=1}^m \sum_{y=1}^{n_i}
   \sum_{t_{ij} \in O_{iy}} (t_{ij} - t_{iy}) \right\}  
\end{align}  

\vspace{1ex}
Derivations of the full conditional posterior distributions and the conditions for log-concavity follow below. To simplify notation, let $\bm\theta^* = \{\alpha, \delta, \bm\beta, \bm\zeta, \phi, \xi, \bm{\nu}, \bm{\omega}, \bm{Y}\}$ represent the extended set of parameters to be sampled, including the random effects and the branching structure. In addition, let $\ell(\theta) = \log\pi(\theta \,|\, \bm{t}, \bm\nu, \bm\omega, \bm{Y}, \bm{\theta}_{-\theta})$ denote the natural logarithm of the full conditional posterior for a parameter $\theta$ sampled using adaptive rejection sampling (ARS). 

\vspace{0ex}
\subsubsection{Multinomial Sampler for Branching Structure}

For each branching structure element $Y_{ij}$, assume a discrete uniform prior distribution over the set of possible parent indices $\{0, 1, ..., j-1\}$, so that $\pi(Y_{ij}) \propto 1$.  
\vspace{1ex}
\begin{align}  
\pi\left( Y_{ij} \,|\, \bm{t}, \bm\theta^*_{-Y_{ij}} \right) &\propto 
   \pi(Y_{ij}) \times \xsup{\big( \lambda_{I_i}(t_{ij}) \big)}
   ^{\mathds{1}(t_{ij} \in I_i)} 
   \times \prod_{y=1}^{j-1} \xsup{\left( \lambda_{O_{iy}}(t_{ij}) \right)}
   ^{\mathds{1}(t_{ij} \in O_{iy})} \\[2ex]
&\propto 
   \xsup{\left( \nu_i \eta_i \alpha t_{ij}^{\alpha-1} \right)}
   ^{\mathds{1}(Y_{ij} \,=\, 0)} 
   \times \prod_{y=1}^{j-1} 
   \xsup{\big( \omega_i \kappa_i 
   \exp\{-\delta(t_{ij} - t_{iy})\} \big)}
   ^{\mathds{1}(Y_{ij} \,=\, y)}
\end{align}

Since the sum of the above over all possible values of $Y_{ij}$ is the conditional intensity function from Equation \eqref{eq:lambda_star} evaluated at event time $t_{ij}$, $Y_{ij}$ can be sampled from a multinomial posterior distribution with one trial and the following outcome probabilities:
\vspace{1ex}
\begin{equation}
P\left( Y_{ij} = y \,|\, \bm{t}, \bm\theta^*_{-Y_{ij}} \right) =
\begin{cases}
\begin{aligned}
    &\hspace{2.5em} 
        \dfrac{\nu_i \eta_i \alpha t_{ij}^{\alpha-1}}
        {\lambda_i^*(t_{ij})}    
    &&\text{if } y = 0 \\[2ex] 
    &\dfrac{\omega_i \kappa_i 
        \exp\{-\delta(t_{ij} - t_{iy})\}}
        {\lambda_i^*(t_{ij})}
    &&\text{if } y = 1, 2, ..., j-1
\end{aligned}
\end{cases}
\end{equation}

\vspace{0ex}
\subsubsection{Gamma Sampler for Background Random Effects}

Under the non-centered parameterization, the background process random effects are assumed to follow a Gamma distribution with $\nu_i \sim \text{Gamma}(\phi, \phi)$ for $i=1,...,m$. Then each $\nu_i$ is sampled from a full conditional distribution that is also Gamma.
\vspace{1ex}
{
\allowdisplaybreaks
\begin{align}  
\pi\left( \nu_i \,|\, \bm{t}, \bm\theta^*_{-\nu_i} \right) 
&\propto \xsup\nu_i^{\phi-1} \times \exp(-\phi \nu_i) 
   \times \exp(-\nu_i \eta_i {T_i}^\alpha) 
   \times \xsup\nu_i^{|I_i|} \\[2ex]
&\propto \xsup\nu_i^{\phi \,+\, |I_i| \,-\, 1} 
   \times \exp\{-(\phi +  \eta_i {T_i}^\alpha) \nu_i\} \\[2ex]
\implies \nu_i \,|\, \bm{t}, \bm\theta^*_{-\nu_i}
&\sim \text{Gamma}\big(\phi + |I_i|,\, 
   \phi + \eta_i {T_i}^\alpha \big)
\end{align}
}

Under the centered parameterization, the background process random effects also follow a Gamma distribution, with $\nu_i^c = \nu_i \eta_i \sim \text{Gamma}(\phi, \phi/\eta_i)$ for $i=1,...,m$, so that the random effect is centered around the individual background fixed effect term $\eta_i$ rather than 1. The full conditional distribution for $\nu_i^c$, derived from equation \eqref{eq:joint_cp}, is also Gamma. 
\vspace{1ex}
\begin{align}  
\pi\left( \nu_i^c \,|\, \bm{t}, \bm\theta^*_{-\nu_i^c} \right) 
&\propto \xsup{(\nu_i^c)}^{\phi-1} 
   \times \exp\{-(\phi/\eta_i)\,\nu_i^c\} 
   \times \exp(-\nu_i^c {T_i}^\alpha) 
   \times \xsup{(\nu_i^c)}^{|I_i|} \\[2ex]
&\propto \xsup{(\nu_i^c)}^{\phi \,+\, |I_i| \,-\, 1} 
   \times \exp\{-( \phi / \eta_i + {T_i}^\alpha) \nu_i^c\} \\[2ex]
\implies \nu_i^c \,|\, \bm{t}, \bm\theta^*_{-\nu_i^c} 
&\sim \text{Gamma}\big(\phi + |I_i|,\, 
   \phi/\eta_i + {T_i}^\alpha \big)
\end{align}

\vspace{0ex}
\subsubsection{Gamma Sampler for Offspring Random Effects}

Under the non-centered parameterization, the offspring process random effects are assumed to follow a Gamma distribution with $\omega_i \sim \text{Gamma}(\xi, \xi)$ for $i=1,...,m$. Then each $\omega_i$ is sampled from a full conditional distribution that is also Gamma.
\vspace{1ex}
\begin{align}  
&\pi\left( \omega_i \,|\, \bm{t}, \bm\theta^*_{-\omega_i} \right) 
   \propto \xsup\omega_i^{\xi-1} \times \exp(-\xi \omega_i) 
   \times \xsup\omega_i^{\sum_{y=1}^{n_i} {|O_{iy}|}} \notag \\[2ex]
&\qquad\qquad\qquad \times 
   \exp\left\{ -\dfrac{\omega_i \kappa_i}{\delta} 
   \sum_{y=1}^{n_i} \big( 1 - \exp\{ -\delta (T_i - t_{iy}) \} 
   \big) \right\} \\[2ex]
&\qquad \propto 
   \xsup{\omega_i}^{\xi \,+\, \sum_{y=1}^{n_i} {|O_{iy}|} \,-\, 1} 
   \times \exp\left\{ -\left( \xi + \dfrac{\kappa_i}{\delta}
   \sum_{y=1}^{n_i} \big( 1 - \exp\{-\delta (T_i - t_{iy})\} 
   \big) \right) \omega_i \right\} \\[2ex]
&\implies \omega_i \,|\, \bm\theta^*_{-\omega_i} 
\sim \text{Gamma}\left(\xi + \sum_{y=1}^{n_i} {|O_{iy}|},\, 
   \xi + \dfrac{\kappa_i}{\delta} \sum_{y=1}^{n_i} 
   \big( 1 - \exp\{-\delta \left( T_i - t_{iy} \right)\} \big) \right)
\end{align}

Under the centered parameterization, the offspring process random effects also follow a Gamma distribution, with $\omega_i^c = \omega_i \kappa_i \sim \text{Gamma}(\xi, \xi/\kappa_i)$ for $i=1,...,m$. The random effect is centered around the individual offspring fixed effect term $\kappa_i$ rather than 1. The full conditional distribution for $\omega_i^c$, derived from equation \eqref{eq:joint_cp}, is also Gamma. 
\vspace{1ex}
\begin{align}  
&\pi\left( \omega_i^c \,|\, \bm{t}, \bm\theta^*_{-\omega_i^c} \right) \propto \xsup{(\omega_i^c)}^{\xi-1} 
   \times \exp\{-(\xi/\kappa_i)\,\omega_i^c\} 
   \times \xsup{(\omega_i^c)}^{\sum_{y=1}^{n_i} {|O_{iy}|}} 
   \notag \\[2ex]
&\qquad\qquad\qquad \times 
   \exp\left\{ -\frac{\omega_i^c}{\delta} \sum_{y=1}^{n_i} 
   \big( 1 - \exp\{ -\delta ( T_i - t_{iy} ) \} 
   \big) \right\} \\[3ex]
&\qquad \propto 
   \xsup{(\omega_i^c)}^{\xi \,+\, \sum_{y=1}^{n_i} {|O_{iy}|} \,-\, 1} 
   \times \exp\left\{ - \left( \dfrac{\xi}{\kappa_i} + 
   \dfrac{1}{\delta} \sum_{y=1}^{n_i} 
   \big( 1 - \exp\{ -\delta (T_i - t_{iy}) \} 
   \big) \right) \omega_i^c \right\} \\[3ex]
&\implies \omega_i^c \,|\, \bm{t}, \bm\theta^*_{-\omega_i^c} 
   \sim \text{Gamma}\left(\xi + \sum_{y=1}^{n_i} {|O_{iy}|},\, 
   \dfrac{\xi}{\kappa_i} + \dfrac{1}{\delta} \sum_{y=1}^{n_i} 
   \big( 1 - \exp\{ -\delta \left( T_i - t_{iy} \right) \} \big) \right)
\end{align}

\vspace{0ex}
\subsubsection{ARS for Weibull Shape Parameter}

Given the prior distribution $\pi(\alpha) \sim \text{Gamma}(a_\alpha, b_\alpha)$, let $\ell(\alpha)$ denote the natural logarithm of the full conditional posterior distribution. Since $\ell''(\alpha) < 0$ as long as at least one immigrant event has occurred, the posterior density is log-concave and ARS can be applied.
\vspace{1ex}
\begin{align}
&\pi(\alpha \,|\, \bm{t}, \bm{\theta}^*_{-\alpha}) 
   \propto \xsup\alpha^{a_\alpha-1} 
   \times \exp(-b_\alpha \alpha)
   \times \exp\left\{ -\sum_{i=1}^m \nu_i \eta_i {T_i}^\alpha \right\} 
   \notag \\[2ex]
&\qquad\qquad\qquad \times \xsup{\alpha}^{\sum_{i=1}^m {|I_i|}} 
   \times \exp\left\{ \alpha\sum_{i=1}^m 
   \sum_{t_{ij} \in I_i}\log(t_{ij}) \right\} \\[3ex]
&\ell(\alpha) \propto 
   \left(a_\alpha + \sum_{i=1}^m {|I_i|} - 1\right)
   \log\alpha - b_\alpha\alpha 
   - \sum_{i=1}^m \nu_i \eta_i {T_i}^\alpha
   + \alpha\sum_{i=1}^m\sum_{t_{ij} \in I_i}\log(t_{ij}) \\[3ex]
&\ell'(\alpha) = 
   \frac{a_\alpha + \sum_{i=1}^m {|I_i|} - 1}{\alpha} - b_\alpha 
   - \sum_{i=1}^m \nu_i \eta_i {T_i}^\alpha\log T_i   
   + \sum_{i=1}^m\sum_{t_{ij} \in I_i}\log(t_{ij}) \\[3ex]
&\ell''(\alpha) = 
   - \frac{a_\alpha + \sum_{i=1}^m {|I_i|} - 1}{\alpha^2} 
   - \sum_{i=1}^m \nu_i \eta_i {T_i}^\alpha\log^2 T_i
\end{align}

\vspace{0ex}
\subsubsection{ARS for Exponential Decay Parameter}

Given the prior distribution $\pi(\delta) \sim \text{Gamma}(\xsub{a}_\delta, \xsub{b}_\delta)$, let $\ell(\delta)$ denote the natural logarithm of the full conditional posterior distribution. We demonstrate that the posterior density for $\delta$ is log-concave and thus ARS can be used for values of the prior shape parameter $\xsub{a}_\delta > 1$. 
\vspace{1ex}
\begin{align}
&\pi(\delta \,|\, \bm{t}, \bm{\theta}^*_{-\delta})
   \propto \xsup\delta^{\xsub{a}_\delta-1} 
   \times \exp(-\xsub{b}_\delta \delta) 
   \times \exp\left\{ -\delta \sum_{i=1}^m \sum_{y=1}^{n_i} 
   \sum_{t_{ij} \in O_{iy}} (t_{ij} - t_{iy}) \right\} \\[2ex]
&\qquad\quad \times \exp\left\{ -\sum_{i=1}^m \sum_{y=1}^{n_i} 
   \dfrac{\omega_i \kappa_i}{\delta} 
   \big( 1 - \exp\{ -\delta(T_i - t_{iy}) \} \big) \right\} \\[2ex]
&\ell(\delta) \propto (\xsub{a}_\delta-1)\log\delta 
   - \xsub{b}_\delta\delta - \delta \sum_{i=1}^m \sum_{y=1}^{n_i} 
   \sum_{t_{ij} \in O_{iy}} (t_{ij} - t_{iy}) \notag \\[2ex]
&\qquad\quad - \sum_{i=1}^m \sum_{y=1}^{n_i} 
   \dfrac{\omega_i \kappa_i}{\delta} 
   \big( 1 - \exp\{ -\delta(T_i - t_{iy}) \} \big) \\[2ex]
&\ell'(\delta) = \dfrac{\xsub{a}_\delta-1}{\delta} 
   - \xsub{b}_\delta - \sum_{i=1}^m \sum_{y=1}^{n_i} 
   \sum_{t_{ij} \in O_{iy}} (t_{ij} - t_{iy}) \notag \\[2ex]
&\qquad\quad + \sum_{i=1}^m \sum_{y=1}^{n_i} 
   \left[ \dfrac{\omega_i \kappa_i}{\delta^2} 
   \big( 1 - \exp\{ -\delta(T_i - t_{iy}) \} \big) 
   - \dfrac{\omega_i \kappa_i}{\delta}  (T_i - t_{iy}) 
   \exp\{ -\delta(T_i - t_{iy}) \} \right] \\[2ex]
&\ell''(\delta) = -\dfrac{\xsub{a}_\delta - 1}{\delta^2}  
   + \sum_{i=1}^m \sum_{y=1}^{n_i} 
   \Big[ -\dfrac{2 \omega_i \kappa_i}{\delta^3} 
   \big( 1 - \exp\{ -\delta(T_i - t_{iy}) \} \big) \notag \\[2ex]
&\qquad\quad + \dfrac{2 \omega_i \kappa_i}{\delta^2} 
   (T_i - t_{iy}) \exp\{ -\delta(T_i - t_{iy}) \} 
   + \dfrac{\omega_i \kappa_i}{\delta} (T_i - t_{iy})^2 
   \exp\{ -\delta(T_i - t_{iy}) \} \Big] \\[2ex]
&\qquad = -\dfrac{\xsub{a}_\delta - 1}{\delta^2}  
   \hspace{-0.1em} - \hspace{-0.1em}  
   \sum_{i=1}^m \sum_{y=1}^{n_i} \dfrac{\omega_i \kappa_i}{\delta^3} 
   \big[ 2 \hspace{-0.1em} - \hspace{-0.1em}
   \big( \delta^2 (T_i \hspace{-0.1em} - \hspace{-0.1em} t_{iy})^2
   \hspace{-0.1em} + \hspace{-0.1em} 2\delta (T_i 
   \hspace{-0.1em} - \hspace{-0.1em} t_{iy}) + 
   \hspace{-0.1em} 2 \hspace{-0.1em} \big)    
   \exp\{ -\delta(T_i \hspace{-0.2em} - \hspace{-0.2em} t_{iy}) 
   \} \big] \\[2ex]
&\qquad = -\dfrac{\xsub{a}_\delta - 1}{\delta^2}  
   - \sum_{i=1}^m \sum_{y=1}^{n_i} 
   \dfrac{\omega_i \kappa_i}{\delta^3} 
   \left[ 2 - \left( \left[ \delta (T_i - t_{iy}) 
   + 1 \right]^2 + 1 \right)
   \exp\{ -\delta(T_i - t_{iy}) \} \right] 
\end{align}

Following \citet{lim_simulation_2018}, let $f_{iy}(\delta) = 2 - \left( \left[ \delta (T_i - t_{iy}) + 1 \right]^2 + 1 \right) \exp\{-\delta(T_i - t_{iy})\}$ and note that $\lim_{\delta \downarrow 0} f_{iy}(\delta) = 2 - \left( (0 + 1)^2 + 1 \right) \exp(0) = 0$. Next, we show that the first derivative of $f_{iy}(\delta)$ is positive for $\delta > 0$, and therefore the function is monotonically increasing:  
\vspace{1ex}
\begin{align}
f'_{iy}(\delta) 
&= \left( \left[ \delta (T_i - t_{iy}) + 1 \right]^2 
   - 2\left[\delta (T_i - t_{iy}) + 1 \right] + 1 \right) 
     (T_i - t_{iy}) \exp\{-\delta(T_i - t_{iy})\} \\[2ex]
&= \Big( \left[ \delta (T_i - t_{iy}) + 1 \right] - 1 \Big)^2
     (T_i - t_{iy}) \exp\{ -\delta(T_i - t_{iy}) \} \\[2ex]
&= \delta^2 (T_i - t_{iy})^3 \exp\{ -\delta(T_i - t_{iy }) \}
\end{align}

Since $T_i > t_{iy}$, the first derivative $f'_{iy}(\delta)$ is positive and $f_{iy}(\delta)$ is increasing from 0, so it must be positive for $\delta > 0$ and for all $i = 1, ..., m$; $y = 1, ..., n_i$. Thus, we have shown that
\vspace{1ex}
\begin{equation}
\ell''(\delta) = -\dfrac{\xsub{a}_\delta - 1}{\delta^2}  
    - \sum_{i=1}^m \sum_{y=1}^{n_i} 
    \dfrac{\omega_i \kappa_i}{\delta^3} 
    f_{iy}(\delta)
\end{equation}

is negative as long as the Gamma prior shape parameter $\xsub{a}_\delta > 1$. This means that the full conditional posterior for $\delta$ is log-concave and ARS applies under the condition $\xsub{a}_\delta > 1$.  

\vspace{0pt}
\subsubsection{ARS for Background Random Effects Precision}

Given the prior distribution $\pi(\phi) \sim \text{Gamma}(\xsub{a}_\phi, \xsub{b}_\phi)$, let $\ell(\phi)$ denote the natural logarithm of the full conditional posterior distribution. $\ell(\phi)$ and its first two derivatives are shown below, where $\psi^{(0)}(\cdot)$ and $\psi^{(1)}(\cdot)$ denote the digamma and trigamma functions, respectively. Using a known property of the trigamma function, we show that the full conditional posterior density is log-concave and ARS can be used as long as the number of individuals $m \ge 2$.
\vspace{1ex}
{
\allowdisplaybreaks
\begin{align}
&\pi(\phi \,|\, \bm{t}, \bm{\theta}^*_{-\phi}) \propto 
   \xsup\phi^{a_\phi-1} \times \exp(-\xsub{b}_\phi \phi) 
   \times \xsup\phi^{m\phi} \times \xsup{[\Gamma(\phi)]}^{-m} 
   \times \prod_{i=1}^m \xsup\nu_i^{\phi-1} 
   \times \exp\left\{ -\phi \sum_{i=1}^m \nu_i \right\} \\[2ex]
&\quad \propto \xsup\phi^{a_\phi-1} \times \exp(-\xsub{b}_\phi \phi) 
   \times \xsup\phi^{m\phi} \times \xsup{[\Gamma(\phi)]}^{-m} 
   \times \exp\left\{ \phi \sum_{i=1}^m \log\nu_i \right\} 
   \times \exp\left\{ -\phi \sum_{i=1}^m \nu_i \right\} \\[2ex]
&\ell(\phi) \propto (\xsub{a}_\phi-1)\log\phi - \xsub{b}_\phi\phi 
   + m\phi\log\phi - m\log\Gamma(\phi)
   + \phi \sum_{i=1}^m \log\nu_i -\phi \sum_{i=1}^m \nu_i \\[2ex]
&\ell'(\phi) = \dfrac{\xsub{a}_\phi-1}{\phi} - \xsub{b}_\phi
   + m\log\phi + m - m\psi^{(0)}(\phi)
   + \sum_{i=1}^m \log\nu_i - \sum_{i=1}^m \nu_i \\[2ex]
&\ell''(\phi) = \dfrac{1 - \xsub{a}_\phi}{\phi^2} 
   + \dfrac{m}{\phi} - m\psi^{(1)}(\phi)
\end{align}
}

Next, we use the standard lower bound for the trigamma function to identify the conditions under which $\ell''(\phi) < 0$:  
\begin{equation}
\label{eq:trigamma}
\psi^{(1)}(x) > \dfrac{1}{x} + \dfrac{1}{2x^2}
\end{equation}

Suppose $\xsub{a}_\phi > 1 - m/2$, where $m$ represents the number of individual processes in the dataset.  Then we can derive the following inequality based on Equation \eqref{eq:trigamma}:  
\vspace{1ex}
\begin{align}
\xsub{a}_\phi > 1 - \frac{m}{2}  
&\implies 1 - \xsub{a}_\phi < \dfrac{m}{2} \\[1.5ex]
&\implies \dfrac{m}{\phi} + \dfrac{1 - \xsub{a}_\phi}{\phi^2} 
   < \dfrac{m}{\phi} + \dfrac{m}{2\phi^2} \\[1.5ex]
&\implies \dfrac{m}{\phi} + \dfrac{1 - \xsub{a}_\phi}{\phi^2} 
   < m\psi^{(1)}(\phi) \\[1.5ex]
&\implies \dfrac{1 - \xsub{a}_\phi}{\phi^2} 
   + \dfrac{m}{\phi} - m\psi^{(1)}(\phi) < 0 \\[1.5ex]
&\implies \ell''(\phi) < 0
\end{align}

Since $\ell''(\phi) < 0$, the full conditional posterior for $\phi$ is log-concave and ARS applies under the condition $\xsub{a}_\phi > 1 - m/2$. Since $\xsub{a}_\phi$ is a Gamma prior shape parameter, it must be positive.  Thus, for a dataset consisting of $m \ge 2$ individual processes, $\ell''(\phi) < 0$ for all $\xsub{a}_\phi > 0$. Under the centered parameterization of random effects, the full conditional distribution is equivalent up to a proportionality constant, so the same formulas apply. 

\vspace{0pt}
\subsubsection{ARS for Offspring Random Effects Precision}

Given the prior distribution $\pi(\xi) \sim \text{Gamma}(\xsub{a}_\xi, \xsub{b}_\xi)$, let $\ell(\xi)$ denote the natural logarithm of the full conditional posterior distribution. Since the background and offspring random effects are both assumed to be generated from Gamma distributions with a mean of 1, the full conditionals of the precision parameters $\phi$ and $\xi$ likewise follow the same form. This means that all formulas in the previous section can also be applied to the offspring precision sampler by substituting $\xi$ for $\phi$, $\omega_i$ for $\nu_i$, and $\kappa_i$ for $\nu_i$. ARS is applicable to $\xi$ for all $\xsub{a}_\xi > 0$ under the condition $m \ge 2$, where $m$ denotes the number of individual processes.
\vspace{1ex}
\begin{align}
&\pi(\xi \,|\, \bm{t}, \bm{\theta}^*_{-\xi}) \propto 
   \xsup\xi^{a_\xi-1} \times \exp(-\xsub{b}_\xi \xi) 
   \times \xsup\xi^{m\xi} \times \xsup{[\Gamma(\xi)]}^{-m} 
   \times \prod_{i=1}^m \xsup\omega_i^{\xi-1} 
   \times \exp\left\{ -\xi \sum_{i=1}^m \omega_i \right\} \\[2ex]
&\quad \propto 
   \xsup\xi^{a_\xi-1} \times \exp(-\xsub{b}_\xi \xi) 
   \times \xsup\xi^{m\xi} \times \xsup{[\Gamma(\xi)]}^{-m} 
   \times \exp\left\{ \xi \sum_{i=1}^m \log\omega_i \right\} 
   \times \exp\left\{ -\xi \sum_{i=1}^m \omega_i \right\} \\[2ex]
&\ell(\xi) \propto (\xsub{a}_\xi-1)\log\xi - \xsub{b}_\xi\xi 
   + m\xi\log\xi - m\log\Gamma(\xi)
   + \xi \sum_{i=1}^m \log\omega_i 
   - \xi \sum_{i=1}^m \omega_i \\[2ex]
&\ell'(\xi) = \dfrac{\xsub{a}_\xi-1}{\xi} - \xsub{b}_\xi 
   + m\log\xi + m - m\psi^{(0)}(\xi)
   + \sum_{i=1}^m \log\omega_i - \sum_{i=1}^m \omega_i \\[2ex]
&\ell''(\xi) = \dfrac{1 - \xsub{a}_\xi}{\xi^2} 
   + \dfrac{m}{\xi} - m\psi^{(1)}(\xi)
\end{align}

\vspace{0pt}
\subsubsection{Samplers for Background Coefficients}

The sampler used for the background process coefficients depends on the variable type. In the general case, including continuous and mean-centered binary covariates, we assume the normal prior $\pi(\beta_p) \sim \text{Normal}(0, \sigma_\beta^2)$. Under the non-centered parameterization of background random effects, the full conditional log-posterior $\ell(\beta_p)$ and its first two derivatives are below. Since $\ell''(\beta_p) < 0$ for all $\beta_p \in \mathbb{R}$, the posterior is log-concave and ARS applies. 
\vspace{1ex}
{\allowdisplaybreaks
\begin{align}
&\pi\left( \beta_p \,|\, \bm{t}, \bm\theta^*_{-\beta_p} \right) 
   \propto \exp \left\{-\dfrac{\beta_p^2}{2\sigma_\beta^2} \right\} 
   \times \exp \left\{ \beta_p \sum_{i=1}^m |I_i| \, x_{ip} \right\} 
   \notag \\[2ex]
&\qquad\qquad\qquad \times 
   \exp \left\{-\sum_{i=1}^m \left( \nu_i {T_i}^\alpha 
   \prod_{r=0}^P \exp(\beta_r x_{ir}) \right) \right\} 
   \\[2ex]
&\ell(\beta_p) \propto 
   -\dfrac{\beta_p^2}{2\sigma_\beta^2} 
   + \beta_p \sum_{i=1}^m |I_i|\, x_{ip} - \sum_{i=1}^m 
   \Bigg( \nu_i {T_i}^\alpha \exp(\beta_p x_{ip}) 
   \prod_{\substack{r=0 \\ r \neq p}}^P 
   \exp(\beta_r x_{ir}) \Bigg) \\[2ex]
&\ell'(\beta_p) = 
   -\dfrac{\beta_p}{\sigma_\beta^2} 
   + \sum_{i=1}^m |I_i|\, x_{ip} - \sum_{i=1}^m 
   \Bigg( \nu_i {T_i}^\alpha x_{ip} \exp(\beta_p x_{ip}) 
   \prod_{\substack{r=0 \\ r \neq p}}^P 
   \exp(\beta_r x_{ir}) \Bigg) \\[2ex]
&\ell''(\beta_p) =
   -\dfrac{1}{\sigma_\beta^2} - \sum_{i=1}^m 
   \Bigg( \nu_i {T_i}^\alpha x_{ip}^2 \exp(\beta_p x_{ip}) 
   \prod_{\substack{r=0 \\ r \neq p}}^P 
   \exp(\beta_r x_{ir}) \Bigg) 
\end{align}
}

Under the centered parameterization, $\nu_i^c = \nu_i\eta_i$ and the background coefficient terms move into the Gamma distribution of background random effects. Since $\phi > 0$, $\ell''(\beta_p) < 0$ for all $\beta_p \in \mathbb{R}$, so the full conditional posterior distribution is log-concave and ARS can be used.  
\vspace{1ex}
\begin{align}
&\pi(\beta_p \,|\, \bm{t}, \bm\theta^*_{-\beta_p}) \propto 
   \exp\left\{-\dfrac{\beta_p^2}{2\sigma_\beta^2} \right\} 
   \times \exp\left\{-\phi \beta_p \sum_{i=1}^m x_{ip} \right\}
   \notag \\[2ex]
&\qquad\qquad\qquad \times 
   \exp\left\{-\phi \sum_{i=1}^m  
   \left( \nu_i^c \prod_{r=0}^P \exp(-\beta_r x_{ir}) 
   \right) \right\} \\[2ex]
&\ell(\beta_p) \propto -\dfrac{\beta_p^2}{2\sigma_\beta^2}
   - \phi \beta_p \sum_{i=1}^m x_{ip}
   - \phi \sum_{i=1}^m \Bigg( \nu_i^c \exp(-\beta_p x_{ip}) 
   \prod_{\substack{r=0 \\ r \neq p}}^P \exp(-\beta_r x_{ir}) 
   \Bigg) \\[2ex]
&\ell'(\beta_p) = -\dfrac{\beta_p}{\sigma_\beta^2}
   - \phi \sum_{i=1}^m x_{ip}
   + \phi \sum_{i=1}^m \Bigg( \nu_i^c x_{ip} \exp(-\beta_p x_{ip}) 
   \prod_{\substack{r=0 \\ r \neq p}}^P \exp(-\beta_r x_{ir}) 
   \Bigg) \\[2ex]
&\ell''(\beta_p) = -\dfrac{1}{\sigma_\beta^2} 
   - \phi \sum_{i=1}^m \Bigg( \nu_i^c x_{ip}^2 \exp(-\beta_p x_{ip}) 
   \prod_{\substack{r=0 \\ r \neq p}}^P \exp(-\beta_r x_{ir}) \Bigg)
\end{align}

For strictly binary background covariates, we derive conjugate samplers for the exponentiated coefficients. Under the non-centered parameterization of background random effects, the prior distribution $\exp(\beta_p) \sim \text{Gamma}(a_\xsub\beta, b_\xsub\beta)$ results in a Gamma full conditional posterior. Because we are conditioning on $\bm{\theta}^*_{-\beta_p}$ and $x_{ip}$ is binary, any terms with $x_{ip} = 0$ are constant with respect to $\exp(\beta_p)$ and can be ignored, and $\exp(\beta_p)$ can be factored out of the following:  

\vspace{-5ex}
\begin{equation}
\exp \Bigg( \prod_{r=0}^P \exp(\beta_r x_{ir}) \Bigg)
= \exp \Bigg( \exp(\beta_p x_{ip}) 
 \prod_{\substack{r=0 \\ r \neq p}}^P 
 \exp(\beta_r x_{ir}) \Bigg)
 \propto \exp \Bigg( x_{ip} \exp(\beta_p) 
 \prod_{\substack{r=0 \\ r \neq p}}^P 
 \exp(\beta_r x_{ir}) \Bigg)
\end{equation}

This simplification allows us to derive the following Gamma background coefficient sampler:  
\addvspace{-5ex}
\begin{align}  
&\pi\left(\exp(\beta_p) \,\big|\, 
   \bm{t}, \bm\theta^*_{-\beta_p}\right)
   \propto \exp\{ \beta_p(\xsub{a}_\beta-1) \} 
   \times \exp\{ -\xsub{b}_\beta \exp(\beta_p) \} \notag \\[2ex]
&\qquad\qquad\qquad \times
   \exp\left\{ \beta_p \sum_{i=1}^m |I_i|\, x_{ip} \right\}
   \times \exp\left\{ -\sum_{i=1}^m 
   \left( \nu_i {T_i}^\alpha \prod_{r=0}^P 
   \exp(\beta_r x_{ir}) \right) \right\} \\[2ex]
&\qquad \propto \exp\{\beta_p(\xsub{a}_\beta-1)\} 
   \times \exp\{ -\xsub{b}_\beta \exp(\beta_p) \}
   \times \exp\left\{ \beta_p 
   \sum_{i=1}^m |I_i|\, x_{ip} \right\} \notag \\[2ex]
&\qquad\qquad\qquad \times \exp\Bigg\{-\exp(\beta_p) 
   \sum_{i=1}^m \Bigg( x_{ip} \nu_i {T_i}^\alpha 
   \prod_{\substack{r=0 \\ r \neq p}}^P 
   \exp(\beta_r x_{ir}) \Bigg) \Bigg\} \\[2ex]
&\qquad \propto 
   \exp\left\{ \beta_p \left(\xsub{a}_\beta + 
   \sum_{i=1}^m |I_i|\,x_{ip} - 1\right) \right\} \notag \\[2ex]
&\qquad\qquad\qquad \times 
   \exp\Bigg\{ -\Bigg( \xsub{b}_\beta + 
   \sum_{i=1}^m \Bigg[ x_{ip} \nu_i {T_i}^\alpha 
   \prod_{\substack{r=0 \\ r \neq p}}^P 
   \exp(\beta_r x_{ir}) \Bigg] \Bigg) 
   \exp(\beta_p) \Bigg\} \\[2ex]
&\hspace{-1.1em} \implies 
   \exp(\beta_p) \,\big|\, \bm{t}, \bm\theta^*_{-\beta_p} 
   \sim \text{Gamma} \Bigg(\xsub{a}_\beta + 
   \sum_{i=1}^m |I_i|\, x_{ip},\, 
   \xsub{b}_\beta + \sum_{i=1}^m 
   \Bigg\{ x_{ip}\nu_i {T_i}^\alpha 
   \prod_{\substack{r=0 \\ r \neq p}}^P 
   \exp(\beta_r x_{ir}) \Bigg\} \Bigg)
\end{align}

However, with centered background random effects, the Inverse-Gamma distribution is more suitable. This is because the prior distribution $\xsup \exp(\beta_p) \sim \text{IG}(a_\xsub\beta, b_\xsub\beta)$ results in an Inverse-Gamma full conditional posterior distribution with positive shape and scale parameters.
\vspace{-4ex}
{
\allowdisplaybreaks
\begin{align}  
&\pi\left( \exp(\beta_p) \,\big|\, 
   \bm{t}, \bm\theta^*_{-\beta_p} \right)
   \propto \exp\{ \beta_p(-\xsub{a}_\beta-1) \} 
   \times \exp\{ -\xsub{b}_\beta \exp(-\beta_p) \} \notag \\[2ex]
&\qquad\qquad\qquad \times 
   \exp\left\{ -\phi\beta_p \sum_{i=1}^m x_{ip} \right\}
   \times \exp\left\{ -\phi\sum_{i=1}^m 
   \left( \nu_i^c \prod_{r=0}^P \exp(-\beta_r x_{ir}) 
   \right) \right\} \\[2ex]
&\qquad \propto \exp\{ \beta_p(-\xsub{a}_\beta-1) \} 
   \times \exp\{ -\xsub{b}_\beta \exp(-\beta_p) \}
   \times \exp\left\{ -\phi\beta_p \sum_{i=1}^m x_{ip} \right\} 
   \notag \\[2ex]
&\qquad\qquad\qquad \times 
   \exp\Bigg\{ -\phi \exp(-\beta_p) 
   \sum_{i=1}^m \Bigg[ x_{ip} \nu_i^c 
   \prod_{\substack{r=0 \\ r \neq p}}^P 
   \exp(-\beta_r x_{ir}) \Bigg] \Bigg\} \\[2ex]
&\qquad \propto \exp\Bigg\{\beta_p 
   \Bigg(-\xsub{a}_\beta - \phi 
   \sum_{i=1}^m x_{ip} - 1\Bigg) \Bigg\} \notag \\[2ex]
&\qquad\qquad\qquad \times 
   \exp\Bigg\{-\Bigg( \xsub{b}_\beta + \phi 
   \sum_{i=1}^m \Bigg[ x_{ip} \nu_i^c 
   \prod_{\substack{r=0 \\ r \neq p}}^P 
   \exp(-\beta_r x_{ir}) \Bigg] \Bigg) 
   \exp(-\beta_p) \Bigg\} \\[2ex]
&\implies 
   \exp(\beta_p) \,\big|\, \bm{t}, \bm\theta^*_{-\beta_p} 
   \sim \text{IG} \Bigg(\xsub{a}_\beta + 
   \phi \sum_{i=1}^m x_{ip},\, \xsub{b}_\beta + 
   \phi \sum_{i=1}^m \Bigg\{ x_{ip} \nu_i^c   
   \prod_{\substack{r=0 \\ r \neq p}}^P 
   \exp(-\beta_r x_{ir}) \Bigg\} \Bigg)
\end{align}
}

\vspace{0pt}
\subsubsection{Samplers for Offspring Coefficients} 

The sampler for the offspring process coefficients depends on the variable type. In the general case, including continuous covariates, we assume the normal prior $\pi(\zeta_q) \sim \text{Normal}(0, \sigma_\zeta^2)$. Under the non-centered parameterization of offspring random effects, the full conditional log-posterior $\ell(\zeta_q)$ and its first two derivatives are shown below. Since $T_i > t_{iy}$ for $i = 1,...,m$; $y = 1, ..., n_i$, $\ell''(\zeta_q) < 0$ for all $\zeta_q \in \mathbb{R}$. Thus, the full conditional posterior is always log-concave and ARS can be used. Let $d_{iy} = 1 - \exp\{-\delta(T_i - t_{iy})\}$ to simplify notation. 
\vspace{2ex}
\begin{align}
& \pi(\zeta_q \,|\, \bm{t}, \bm\theta^*_{-\zeta_q}) 
   \propto \exp\left\{ -\dfrac{\zeta_q^2}{2\sigma_\zeta^2} \right\} 
   \times \exp\left\{\zeta_q \sum_{i=1}^m 
   \sum_{y=1}^{n_i} |O_{iy}|\, z_{iq} \right\} \notag \\[2ex]
&\qquad\qquad\qquad 
   \times \exp\bigg\{ -\sum_{i=1}^m \sum_{y=1}^{n_i} 
   \bigg[ \dfrac{\omega_i}{\delta} \, d_{iy}
   \prod_{r=0}^Q \exp(\zeta_r z_{ir}) \bigg] \bigg\} \\[4ex]
&\ell(\zeta_q) \propto -\dfrac{\zeta_q^2}{2\sigma_\zeta^2}
   + \zeta_q \sum_{i=1}^m \sum_{y=1}^{n_i} |O_{iy}|\, z_{iq} 
   - \sum_{i=1}^m \sum_{y=1}^{n_i} \bigg[ \dfrac{\omega_i}{\delta} 
   \, d_{iy} \exp(\zeta_q z_{iq}) 
   \prod_{\substack{r=0 \\ r \neq q}}^Q \exp(\zeta_r z_{ir}) \bigg]
   \\[3ex]
&\ell'(\zeta_q) = -\dfrac{\zeta_q}{\sigma_\zeta^2}
   + \sum_{i=1}^m \sum_{y=1}^{n_i} |O_{iy}|\, z_{iq} 
   - \sum_{i=1}^m \sum_{y=1}^{n_i} \bigg[ \dfrac{\omega_i}{\delta} 
   \, d_{iy} z_{iq} \exp(\zeta_q z_{iq}) 
   \prod_{\substack{r=0 \\ r \neq q}}^Q \exp(\zeta_r z_{ir}) \bigg] 
   \\[3ex]
&\ell''(\zeta_q) = -\dfrac{1}{\sigma_\zeta^2}
   - \sum_{i=1}^m \sum_{y=1}^{n_i} \bigg[ \dfrac{\omega_i}{\delta} 
   \big( 1 - \exp\{-\delta(T_i - t_{iy})\} \big) 
   \, z_{iq}^2 \exp(\zeta_q z_{iq}) 
   \prod_{\substack{r=0 \\ r \neq q}}^Q \exp(\zeta_r z_{ir}) \bigg] 
\end{align}

Under the centered parameterization, $\omega_i^c = \omega_i\kappa_i$, so the offspring coefficient terms move into the Gamma distribution of offspring random effects. Since the shape parameter $\xi > 0$, $\ell''(\zeta_q) < 0$ for all $\zeta_q \in \mathbb{R}$, so the full conditional distribution is log-concave and ARS applies.  
\addvspace{-5ex}
\begin{align}
&\pi(\zeta_q \,|\, \bm{t}, \bm\theta^*_{-\zeta_q}) \propto 
   \exp\left\{ -\frac{\zeta_q^2}{2\sigma_\zeta^2} \right\} 
   \times \exp\left\{ -\xi \zeta_q \sum_{i=1}^m z_{iq} \right\} 
   \notag \\[2ex]
&\qquad\qquad\qquad \times 
   \exp\left\{ -\xi \sum_{i=1}^m \left( \omega_i^c 
   \prod_{r=0}^Q \exp(-\zeta_r z_{ir}) \right) \right\} \\[2ex]
&\ell(\zeta_q) \propto - \dfrac{\zeta_q^2}{2\sigma_\zeta^2}
   - \xi \zeta_q \sum_{i=1}^m z_{iq} - \xi \sum_{i=1}^m 
   \Bigg( \omega_i^c \exp(-\zeta_q z_{iq}) 
   \prod_{\substack{r=0 \\ r \neq q}}^Q 
   \exp(-\zeta_r z_{ir}) \Bigg) \\[2ex]
&\ell'(\zeta_q) = - \dfrac{\zeta_q}{\sigma_\zeta^2}
   - \xi \sum_{i=1}^m z_{iq} + \xi \sum_{i=1}^m 
   \Bigg( \omega_i^c z_{iq} \exp(-\zeta_q z_{iq}) 
   \prod_{\substack{r=0 \\ r \neq q}}^Q 
   \exp(-\zeta_r z_{ir}) \Bigg) \\[2ex]
&\ell''(\zeta_q) = - \dfrac{1}{\sigma_\zeta^2} - \xi 
   \sum_{i=1}^m \Bigg( \omega_i^c z_{iq}^2 \exp(-\zeta_q z_{iq}) 
   \prod_{\substack{r=0 \\ r \neq q}}^Q \exp(-\zeta_q z_{iq}) \Bigg)
\end{align}

For strictly binary offspring covariates, we construct conjugate samplers for the exponentiated coefficients. Under the non-centered parameterization of offspring random effects, the prior distribution $\exp(\zeta_q) \sim \text{Gamma}(a_\xsub\zeta, b_\xsub\zeta)$ results in a Gamma full conditional posterior. Because we are conditioning on $\bm{\theta}^*_{-\zeta_q}$ and $z_{iq}$ is binary, any terms with $z_{iq} = 0$ are constant with respect to $\exp(\zeta_q)$ and can be ignored, and $\exp(\zeta_q)$ can be factored out of the following:

\vspace{-5ex}
\begin{equation}
\exp \Bigg( \prod_{r=0}^Q \exp(\zeta_r z_{ir}) \Bigg)
= \exp \Bigg( \exp(\zeta_q z_{iq}) 
  \prod_{\substack{r=0 \\ r \neq q}}^Q 
  \exp(\zeta_r z_{ir}) \Bigg)
  \propto \exp \Bigg( z_{iq} \exp(\zeta_q)  
  \prod_{\substack{r=0 \\ r \neq q}}^Q 
  \exp(\zeta_r z_{ir}) \Bigg)
\end{equation}

\vspace{-2ex}
This simplification allows us to derive the following Gamma offspring coefficient sampler: 
{\allowdisplaybreaks
\begin{align}  
&\pi\big( \exp(\zeta_q) \,\big|\, \bm{t}, \bm\theta^*_{-\zeta_q} \big)
   \propto \exp\{ \zeta_q(\xsub{a}_\zeta-1) \} 
   \times \exp\{ -\xsub{b}_\zeta \exp(\zeta_q) \} \notag \\[2ex]
&\qquad\qquad\qquad \times \exp\Bigg\{ \zeta_q \sum_{i=1}^m 
   \sum_{y=1}^{n_i} |O_{iy}| \, z_{iq} \Bigg\}
   \times \exp\Bigg\{ -\sum_{i=1}^m \sum_{y=1}^{n_i} 
   \Bigg[ \dfrac{\omega_i}{\delta} \, d_{iy}
   \prod_{r=0}^Q \exp(\zeta_r z_{ir}) \Bigg] \Bigg\} \\[1.5ex]
&\qquad \propto \exp\{ \zeta_q(\xsub{a}_\zeta-1) \} 
   \times \exp\{ -\xsub{b}_\zeta \exp(\zeta_q) \} 
   \times \exp\Bigg\{ \zeta_q \sum_{i=1}^m 
   \sum_{y=1}^{n_i} |O_{iy}| \, z_{iq} \Bigg\} \notag \\[1.5ex]
&\qquad\qquad\qquad \times 
   \exp\Bigg\{ -\exp(\zeta_q) \sum_{i=1}^m \sum_{y=1}^{n_i} 
   \Bigg[ \dfrac{\omega_i z_{iq}}{\delta} \, d_{iy}
   \prod_{\substack{r=0 \\ r \neq q}}^Q 
   \exp(\zeta_r z_{ir}) \Bigg] \Bigg\} \\[1.5ex]
&\qquad \propto 
   \exp\Bigg\{ \zeta_q \left(\xsub{a}_\zeta + 
   \sum_{i=1}^m \sum_{y=1}^{n_i} |O_{iy}| 
   \, z_{iq} - 1\right) \Bigg\} \notag \\[1.5ex]
&\qquad\qquad\qquad \times 
   \exp\Bigg\{ -\Bigg( \xsub{b}_\zeta + 
   \sum_{i=1}^m \sum_{y=1}^{n_i} 
   \Bigg[ \dfrac{\omega_i z_{iq} }{\delta} \, d_{iy}
   \prod_{\substack{r=0 \\ r \neq q}}^Q 
   \exp(\zeta_r z_{ir}) \Bigg] \Bigg) \, 
   \exp(\zeta_q) \Bigg\} \\[1.5ex]
&\qquad \implies 
   \exp(\zeta_q) \,\big|\, \bm{t}, \bm\theta^*_{-{\zeta_q}} 
   \sim \text{Gamma}\Bigg( \xsub{a}_\zeta + 
   \sum_{i=1}^m \sum_{y=1}^{n_i} |O_{iy}| \, z_{iq}, \notag \\[1.5ex]
&\qquad\qquad\qquad 
   \xsub{b}_\zeta + \sum_{i=1}^m \sum_{y=1}^{n_i} 
   \Bigg\{\dfrac{\omega_i z_{iq}}{\delta}
   \big( 1 - \exp\{-\delta(T_i - t_{iy})\} \big) 
   \prod_{\substack{r=0 \\ r \neq q}}^Q 
   \exp(\zeta_r z_{ir}) \Bigg\} \Bigg)
\end{align}
}

However, with centered offspring random effects, the Inverse-Gamma distribution is more suitable. This is because the prior distribution $\exp(\zeta_q) \sim \text{IG}(a_\xsub\zeta, b_\xsub\zeta)$ results in an Inverse-Gamma full conditional posterior distribution with positive shape and scale parameters.  
\vspace{-4ex}
{\allowdisplaybreaks
\begin{align}  
&\pi\left(\exp(\zeta_q) \,\big|\, 
   \bm{t}, \bm\theta^*_{-\zeta_q}\right)
   \propto \exp\{ \zeta_q(-\xsub{a}_\zeta-1) \} 
   \times \exp\{ -\xsub{b}_\zeta \exp(-\zeta_q) \} \notag \\[1.5ex]
&\qquad\qquad\qquad \times 
   \exp\Bigg\{ -\xi\zeta_q \sum_{i=1}^m z_{iq} \Bigg\}
   \times \exp\Bigg\{ -\xi\sum_{i=1}^m 
   \Bigg( \omega_i^c \prod_{r=0}^Q 
   \exp(-\zeta_q x_{ir}) \Bigg) \Bigg\} \\[1.5ex]
&\qquad \propto \exp\{ \zeta_q(-\xsub{a}_\zeta-1) \} 
   \times \exp\{ -\xsub{b}_\zeta \exp(-\zeta_q) \}
   \exp\Bigg\{ -\xi\zeta_q \sum_{i=1}^m z_{iq} \Bigg\} \notag \\[1.5ex]
&\qquad\qquad\qquad \times 
   \exp\Bigg\{ -\xi \exp(-\zeta_q) 
   \sum_{i=1}^m \Bigg[ z_{iq} \, \omega_i^c 
   \prod_{\substack{r=0 \\ r \neq q}}^Q 
   \exp(-\zeta_q z_{ir}) \Bigg] \Bigg\} \\[1.5ex]
&\qquad \propto 
   \exp\Bigg\{ \zeta_q \left(-\xsub{a}_\zeta 
  - \xi \sum_{i=1}^m z_{iq} - 1\right) \Bigg\} \notag \\[1.5ex]
&\qquad\qquad\qquad \times 
   \exp\Bigg\{ -\Bigg( \xsub{b}_\zeta + 
   \xi \sum_{i=1}^m \Bigg[ z_{iq} \, \omega_i^c 
   \prod_{\substack{r=0 \\ r \neq q}}^Q 
   \exp(-\zeta_q z_{ir}) \Bigg] \Bigg) 
   \exp(-\zeta_q) \Bigg\} \\[1.5ex]
&\implies \exp(\zeta_q) \big|\, \bm{t}, \bm\theta^*_{-\zeta_q} 
   \sim \text{IG}\Bigg( \xsub{a}_\zeta + \xi 
   \sum_{i=1}^m z_{iq},\, \xsub{b}_\zeta + \xi 
   \sum_{i=1}^m \Bigg\{ z_{iq} \, \omega_i^c 
   \prod_{\substack{r=0 \\ r \neq q}}^Q 
   \exp(-\zeta_r z_{ir}) \Bigg\} \Bigg)
\end{align}
}

\vspace{0pt}
\subsection{Data Augmentation Steps}\label{supp:mcmc2}

\noindent The previous steps of the MCMC algorithm are based on the assumption of fully observed event trajectories $\bm{t}_i$ for individuals $i=1,...,m$. In the event of incomplete data, Bayesian estimation can proceed through data augmentation. Let $\bm{t} = \{\bm{t}_{agg}, \bm{t}_{miss}\}$ represent the complete data, where $\bm{t}_{agg}$ consists of the unobserved continous event times occurring during tracked intervals and reported in aggregate as counts, and $\bm{t}_{miss}$ consists of the missing events occurring during untracked intervals. Each follow up period $[0, T_i]$ can be partitioned into $u_i$ untracked intervals $(U_{i1}, U_{i2}, ..., U_{i u_i})$ and 
$v_i$ tracked bins $(V_{i1}, V_{i2}, ..., V_{i v_i})$ with observed counts $\bm{N}_i = (N_{i1}, N_{i2}, ..., N_{i v_i})$.

\vspace{0pt}
\subsubsection{Imputation of Event Times from Aggregated Counts}

For tracked intervals, continuous event times are sampled from 
the aggregate counts based on $p(\bm{t}_{agg} \,|\, \bm{t}_{miss}, \bm{N}, \bm{\theta}^*)$, extending the approach given by \citet{zhou_bayesian_2025}. The imputation process takes place through Gibbs sampling, by stepping through $p(t_{ij} \,|\, \bm{t}_{-ij}, \bm{N}, \bm{\theta}^*)$ for each latent continuous event time $t_{ij} \in \bm{t}_{agg}$. Since each latent event time must be consistent with the observed binned count data $\bm{N}$, a point $t_{ij}$ occurring during binned interval $V_{i\ell}$ is updated by proposing an event time $t^\prime_{ij}$ from a continuous uniform distibution with support restricted to $V_{i\ell}$. The proposal distribution may be further limited by the branching structure, as the proposed event time must take place after any parent event and prior to any offspring events. Suppose $t_{ij} \in V_{i\ell}$ and $Y_{ij} = y$, and let $t_{i0} = 0$ denote the starting time of the process. Then $t_{ij}'$ is proposed from a uniform distribution over $V_{i\ell} \cap (t_{iy}, \text{min}\{t: t \in O_{ij}\})$. The Metropolis-Hastings acceptance probability is shown below. Because it is conditional on all other event times in $\bm{t}_{-ij}$, only terms from the likelihood involving the specific time point $t_{ij}$ are retained. This includes either an immigrant term if $Y_{ij} = 0$ or a term for parent event $t_{iy}$ if $Y_{ij} > 0$, as well as terms for offspring events $t \in O_{ij}$. 
\vspace{-4ex}
\begin{align}
A_{t_{ij}} &= \min \left\{ 1, \,
   \dfrac{\pi(t_{ij}' \,|\, \bm{t}_{-ij}, \bm{N}, \bm{\theta}^*)}
         {\pi(t_{ij}  \,|\, \bm{t}_{-ij}, \bm{N}, \bm{\theta}^*)} \right\} \\[4ex]
&= \min \left\{1, \,
   \dfrac{\xsup{t'_{ij}}^{(\alpha-1) 
   \mathds{1}(t_{ij} \,\in\, I_i)} 
      \times \exp\left\{ -\dfrac{\omega_i \kappa_i}{\delta} 
      \left( 1 - \exp\{ -\delta(T_i - t_{ij}') \} 
      \right) \right\}} 
   {\xsup{t_{ij}}^{(\alpha-1) 
   \mathds{1}(t_{ij} \,\in\, I_i)} 
      \times \, \exp\left\{ -\dfrac{\omega_i \kappa_i}{\delta} 
      \left( 1 - \exp\{ -\delta(T_i - t_{ij}) \} 
      \right) \right\}} \right. \notag \\[3ex]
&\qquad\qquad\qquad \times \left.
   \dfrac{\exp\{ -\delta(t_{ij}' - t_{iy}) 
      \mathds{1}(t_{ij} \,\in\, O_{iy}) \} 
      \times \displaystyle\prod_{t \in O_{ij}} 
      \exp\{ -\delta(t - t_{ij}') \}} 
   {\exp\{ -\delta(t_{ij} - t_{iy}) 
      \mathds{1}(t_{ij} \,\in\, O_{iy}) \} 
      \times \displaystyle\prod_{t \in O_{ij}} 
      \exp\{ -\delta(t - t_{ij}) \}} \right\} \\[3ex]
&= \min \Bigg\{1, \, 
   \left( \dfrac{t'_{ij}}{t_{ij}} \right)^{(\alpha-1)
     \mathds{1}(Y_{ij} \,=\, 0)} 
     \times \exp\left\{-\delta \left(t_{ij} - t_{ij}' \right)
            \big(|O_{ij}| \,-\, \mathds{1}(Y_{ij} \,>\, 0) 
            \big) \right\} \notag \\[1ex]
&\qquad\qquad\qquad \times
   \exp \left\{ \dfrac{\omega_i \kappa_i}{\delta} 
     \left( \exp\{ -\delta(T_i - t_{ij}') \} 
          - \exp\{ -\delta(T_i - t_{ij}) \} \right) \right\} \Bigg\}
\end{align}

\vspace{0pt}
\subsubsection{Imputation of Event Times over Untracked Intervals}

Continuous event times are imputed over untracked intervals based on $p(\bm{t}_{miss} \,|\, \bm{t}_{agg}, \bm{N}, \bm{\theta}^*)$, extending the approach from \citet{tucker_derek_handling_2019}. Let \{$U_{i\ell} : i=1,...,m; \ell=1,...,u_i\}$ represent the set of all missing intervals, where $\bm{t}_{miss,i\ell} = \{t_{ij} \in U_{i\ell}\}$ consists of the unreported events occuring during the interval $U_{i\ell}$, while $\bm{t}_{miss,-i\ell} = \{t_{ij} \in \bm{t}_{miss} \setminus \bm{t}_{miss,i\ell}\}$ consists of the unreported events occurring during all other missing intervals. The imputation process takes place through Gibbs sampling based on $p(\bm{t}_{miss,i\ell} \,|\, \bm{t}_{miss,-i\ell}, \bm{t}_{agg}, \bm{N}, \bm{\theta}^*)$ for each untracked interval $U_{i\ell}$ with left endpoint $L(U_{i\ell})$ and right endpoint $R(U_{i\ell})$. Let $\bm{t}_{miss,i\ell}$ and $\bm{t}^\prime_{miss,i\ell}$ consist of the current and proposed event times occurring during $U_{i\ell}$, such that $\bm{t} = \{\bm{t}_{miss,i\ell}, \bm{t}_{miss,-i\ell}, \bm{t}_{agg}\}$ and $\bm{t}^\prime = \{\bm{t}^\prime_{miss,i\ell}, \bm{t}_{miss,-i\ell}, \bm{t}_{agg}\}$ consist of the current and proposed complete event time data, respectively. The full conditional posterior distribution for the event times occurring during the interval $U_{i\ell}$ is then
\vspace{1ex}
\begin{equation}
p(\bm{t}_{miss,i\ell} \,|\, 
   \bm{t}_{miss,-i\ell}, \bm{t}_{agg}, \bm{\theta}^*) 
=\dfrac{f(\bm{t}_i \,|\, \bm\theta^*) 
   f(\bm{t}_{-i} \,|\, \bm\theta^*)}
   {f(\bm{t}_{miss,-i\ell}, \bm{t}_{agg} \,|\, 
   \bm\theta^*)}\,,
\end{equation}

where $f(\cdot \,|\, \bm\theta^*)$ refers to the Hawkes process density function in Equation \eqref{eq:like_star}, which is based on the conditional intensity function in equation \eqref{eq:lambda_star} and does not rely upon the branching structure $\bm{Y}$. A set of missing points is simulated over $U_{i\ell}$ from the proposal distribution $q$, which conditions on $\mathcal{H}_{L(U_{i\ell})}$, the history of events occurring prior to $U_{i\ell}$ for individual $i$. Similarly, $\mathcal{H}_{R(U_{i\ell})}$ contains the history of events up to and including the interval $U_{i\ell}$. To simulate a proposed set of missing points, Algorithm \ref{alg:sim_event} is adapted by initializing the offspring intensity $\lambda_1$ to the value at $L(U_{i\ell})$, which is computed based on the history $\mathcal{H}_{L(U_{i\ell})}$.
\addvspace{-3ex}
\begin{equation}
q(\bm{t}_{miss,i\ell} \,|\, 
   \mathcal{H}_{L(U_{i\ell})}, \bm{\theta}^*) 
=\dfrac{f(\mathcal{H}_{R(U_{i\ell})} \,|\, \bm{\theta}^*)}
   {f(\mathcal{H}_{L(U_{i\ell})} \,|\, \bm{\theta}^*)}
\end{equation}

\addvspace{2ex}
The Metropolis-Hastings acceptance probability for the missing interval $U_{i\ell}$ is derived below. It can be simplified by noting that $t_{ij} = t'_{ij}$ for $t_{ij} < L(U_{i\ell})$ and $t_{ij} > R(U_{i\ell})$. The remaining terms include the conditional intensity for points occurring after $U_{i\ell}$ and the integrated offspring intensity for event times occurring within $U_{i\ell}$. Since the simulated event times do not account for the branching structure, the missing event augmentation is the final step in the MCMC algorithm. Once imputation is complete across all untracked intervals, the next iteration of the MCMC algorithm begins by sampling the branching structure conditional on the updated complete event time data $\bm{t}$. 
\vspace{2ex}
{\allowdisplaybreaks
\begin{align}
&A_{U_{i\ell}} 
= \min \left\{1,\, 
 \dfrac{p(\bm{t}_{{miss}, i\ell}' \,|\, 
        \bm{t}_{{miss}, -i\ell},\, 
        \bm{t}_{agg},\, \bm\theta^*)}
       {p(\bm{t}_{{miss}, i\ell}  \,|\, 
        \bm{t}_{{miss}, -i\ell},\, 
        \bm{t}_{agg},\,\bm\theta^*)} 
 \times 
 \dfrac{q(\bm{t}_{{miss}, i\ell} \,|\,
        \mathcal{H}_{L(U_{i\ell})}, \bm\theta^*)}
        {q(\bm{t}_{{miss}, i\ell}' \,|\, \mathcal{H}_{L(U_{i\ell})}, \bm\theta^*)} 
        \right\} \\[4ex]
&\hspace{1.7em} = \min\left\{1,\, 
 \dfrac{f(\bm{t}'_i \,|\, \bm\theta^*)}
       {f(\bm{t}_i \,|\, \bm\theta^*)} 
 \times \dfrac{f(\mathcal{H}_{R(U_{i\ell})}  \,|\, \bm\theta^*)}
              {f(\mathcal{H}_{R(U_{i\ell})}' \,|\, \bm\theta^*)} \right\} \\[4ex]
&= \min\left\{1,\, 
 \dfrac{\displaystyle 
        \left( \prod_{t_{ij} > R(U_{i\ell})} \hspace{-1em}
        \lambda_i(t_{ij} \,|\, \mathcal{H}'_{ij}) \right)
        \exp\left\{ -\Lambda_i \big( T_i \,|\, 
        \bm{t}'_i \big) \right\} \,
        \exp\left\{ -\Lambda_i \left( R(U_{i\ell}) \,|\, \mathcal{H}_{R(U_{i\ell})} \right) \right\}}
       {\displaystyle 
        \left( \prod_{t_{ij} > R(U_{i\ell})} \hspace{-1em}
        \lambda_i(t_{ij} \,|\, \mathcal{H}_{ij}) \right)
        \exp\left\{ -\Lambda_i \big( T_i \,|\, 
        \bm{t}_i \big) \right\} \, 
        \exp\left\{ -\Lambda_i \left( R(U_{i\ell}) \,|\, \mathcal{H}'_{R(U_{i\ell})} \right) \right\}} 
        \right\} \\[6ex]
&= \min\left\{1,\, 
 \dfrac{\displaystyle 
        \prod_{t_{ij} > R(U_{i\ell})} \hspace{-1em}
        \lambda_i(t_{ij} \,|\, \mathcal{H}'_{ij}) 
        \hspace{-0.5em} \prod_{t'_{ij} \in U_{i\ell}} \hspace{-0.5em} 
        \exp\left\{ -\omega_i \kappa_i G_i 
        \big( T_i - t'_{ij} \big) \right\}  
        \hspace{-0.5em} \prod_{t_{ij} \in U_{i\ell}} \hspace{-0.5em}
        \exp\left\{ -\omega_i \kappa_i G_i 
        \big( R(U_{i\ell}) - t_{ij} \big) \right\}}
       {\displaystyle 
        \prod_{t_{ij} > R(U_{i\ell})} \hspace{-1em}
        \lambda_i(t_{ij} \,|\, \mathcal{H}_{ij}) 
         \hspace{-0.5em} \prod_{t_{ij} \in U_{i\ell}} \hspace{-0.5em}  
        \exp\left\{ -\omega_i \kappa_i G_i 
        \big( T_i - t_{ij} \big) \right\} 
        \hspace{-0.5em} \prod_{t'_{ij} \in U_{i\ell}} \hspace{-0.5em} 
        \exp\left\{-\omega_i \kappa_i G_i 
        \left( R(U_{i\ell}) - t'_{ij} \right) \right\}} 
        \right\} \\[2ex]
&= \min\left\{1,\, 
 \dfrac{\displaystyle 
        \prod_{t_{ij} > R(U_{i\ell})} \hspace{-1em}
        \lambda_i(t_{ij} \,|\, \mathcal{H}'_{ij}) 
        \hspace{-0.5em} \prod_{t'_{ij} \in U_{i\ell}} \hspace{-0.5em}
        \exp\left\{ \dfrac{\omega_i \kappa_i}{\delta} 
        \bigg( \exp\left\{ -\delta \big( T_i - t'_{ij} \big) \right\} - 
        \exp\left\{ -\delta \big( R(U_{i\ell}) - t'_{ij} \big) \right\} \bigg) \right\}}
       {\displaystyle \prod_{t_{ij} > R(U_{i\ell})} \hspace{-1em}
        \lambda_i(t_{ij} \,|\, \mathcal{H}_{ij}) 
        \hspace{-0.5em} \prod_{t_{ij} \in U_{i\ell}} \hspace{-0.5em}
        \exp\left\{ \dfrac{\omega_i \kappa_i}{\delta} 
        \bigg( \exp\left\{ -\delta \big( T_i - t_{ij} \big) \right\} - 
        \exp\left\{ -\delta \big( R(U_{i\ell}) - t_{ij} \big) \right\} \bigg) \right\}} \right\}
\end{align}
}

\clearpage
\vspace{0pt}
\section{Additional HEP Analysis Results}
\label{supp:app}

\begin{table}[!ht]
\setlength{\tabcolsep}{4.75pt}
\centering{
\caption{Estimated posterior means and 95\% highest posterior density intervals for parameters of Hawkes process model M1 fit to HEP study data. Model includes full set of covariates in background and offspring processes and random effects in both processes.}

\begin{tabular}{clSSSS}\toprule
\multicolumn{1}{c}{} & \multicolumn{1}{c}{} & \multicolumn{1}{>{\centering\arraybackslash}p{0.70in}}{Posterior Mean} & \multicolumn{1}{>{\centering\arraybackslash}p{0.70in}}{95\% HPD Lower} & \multicolumn{1}{>{\centering\arraybackslash}p{0.70in}}{95\% HPD Upper} & \multicolumn{1}{>{\centering\arraybackslash}p{0.70in}}{R-Hat Statistic}\\
\midrule
\addlinespace[0.3em]
\multicolumn{6}{l}{\textbf{Hawkes Process Parameters}}\\
\hspace{1em}$\alpha$ & Weibull Shape & 0.876 & 0.849 & 0.901 & 1.002\\
\hspace{1em}$\delta$ & Exponential Decay Rate & 0.648 & 0.590 & 0.707 & 1.028\\
\hspace{1em}$1/\phi$ & Background RE Variance & 4.686 & 3.947 & 5.461 & 1.001\\
\hspace{1em}$1/\xi$ & Offspring RE Variance & 0.344 & 0.228 & 0.466 & 1.004\\
\addlinespace[0.3em]
\multicolumn{6}{l}{\textbf{Background Coefficients}}\\
\hspace{1em}$\exp(\beta_0)$ & Intercept & 0.040 & 0.011 & 0.079 & 1.000\\
\hspace{1em}$\exp(\beta_1)$ & Sex: Male & 0.633 & 0.335 & 0.981 & 1.001\\
\hspace{1em}$\exp(\beta_2)$ & Age: 25-39 & 2.120 & 0.784 & 3.790 & 1.000\\
\hspace{1em}$\exp(\beta_3)$ & Age: 40-64 & 2.952 & 1.073 & 5.218 & 1.000\\
\hspace{1em}$\exp(\beta_4)$ & Completed Higher Education & 0.600 & 0.280 & 0.987 & 1.000\\
\hspace{1em}$\exp(\beta_5)$ & Part-Time or Unemployed & 1.703 & 0.771 & 2.874 & 1.000\\
\hspace{1em}$\exp(\beta_6)$ & Abnormal Findings on MRI & 1.195 & 0.493 & 2.056 & 1.001\\
\hspace{1em}$\exp(\beta_7)$ & Injury Prior to Diagnosis & 0.973 & 0.519 & 1.482 & 1.003\\
\hspace{1em}$\exp(\beta_8)$ & Family History of Seizures & 1.033 & 0.507 & 1.624 & 1.000\\
\hspace{1em}$\exp(\beta_9)$ & Total Seizures Pre-Tx: 4-29 & 3.176 & 1.341 & 5.340 & 1.000\\
\hspace{1em}$\exp(\beta_{10})$ & Total Seizures Pre-Tx: 30+ & 9.128 & 3.979 & 15.051 & 1.001\\
\hspace{1em}$\exp(\beta_{11})$ & ASM: Combination Therapy & 2.870 & 0.870 & 5.458 & 1.000\\
\hspace{1em}$\exp(\beta_{12})$ & ASM: Levetiracetam & 0.857 & 0.335 & 1.482 & 1.009\\
\hspace{1em}$\exp(\beta_{13})$ & ASM: Lamotrigine & 0.609 & 0.185 & 1.164 & 1.001\\
\hspace{1em}$\exp(\beta_{14})$ & ASM: Other Non-SCB & 0.526 & 0.079 & 1.236 & 1.001\\
\addlinespace[0.3em]
\multicolumn{6}{l}{\textbf{Offspring Coefficients}}\\
\hspace{1em}$\exp(\zeta_0)$ & Intercept & 0.284 & 0.189 & 0.381 & 1.000\\
\hspace{1em}$\exp(\zeta_1)$ & Sex: Male & 0.905 & 0.725 & 1.091 & 1.000\\
\hspace{1em}$\exp(\zeta_2)$ & Age: 25-39 & 1.077 & 0.808 & 1.368 & 1.000\\
\hspace{1em}$\exp(\zeta_3)$ & Age: 40-64 & 1.049 & 0.778 & 1.354 & 1.000\\
\hspace{1em}$\exp(\zeta_4)$ & Completed Higher Education & 1.020 & 0.799 & 1.255 & 1.000\\
\hspace{1em}$\exp(\zeta_5)$ & Part-Time or Unemployed & 1.055 & 0.821 & 1.299 & 1.000\\
\hspace{1em}$\exp(\zeta_6)$ & Abnormal Findings on MRI & 1.033 & 0.798 & 1.284 & 1.000\\
\hspace{1em}$\exp(\zeta_7)$ & Injury Prior to Diagnosis & 0.958 & 0.778 & 1.151 & 1.000\\
\hspace{1em}$\exp(\zeta_8)$ & Family History of Seizures & 1.040 & 0.836 & 1.251 & 1.000\\
\hspace{1em}$\exp(\zeta_9)$ & Total Seizures Pre-Tx: 4-29 & 1.127 & 0.831 & 1.441 & 1.000\\
\hspace{1em}$\exp(\zeta_{10})$ & Total Seizures Pre-Tx: 30+ & 1.154 & 0.880 & 1.456 & 1.001\\
\hspace{1em}$\exp(\zeta_{11})$ & ASM: Combination Therapy & 0.956 & 0.648 & 1.273 & 1.001\\
\hspace{1em}$\exp(\zeta_{12})$ & ASM: Levetiracetam & 1.195 & 0.885 & 1.509 & 1.001\\
\hspace{1em}$\exp(\zeta_{13})$ & ASM: Lamotrigine & 1.058 & 0.734 & 1.405 & 1.001\\
\hspace{1em}$\exp(\zeta_{14})$ & ASM: Other Non-SCB & 1.500 & 0.895 & 2.211 & 1.001\\
\bottomrule
\end{tabular}

\label{tab:hep_par_M1}
}
\end{table}

\begin{table}[!ht]
\setlength{\tabcolsep}{5pt}
\centering{
\caption{Estimated posterior means and 95\% highest posterior density intervals for parameters of Hawkes process model M2 fit to HEP study data. Model includes selected covariates in background and offspring processes and random effects in both processes.}

\begin{tabular}{clSSSS}\toprule
\multicolumn{1}{c}{} & \multicolumn{1}{c}{} & \multicolumn{1}{>{\centering\arraybackslash}p{0.70in}}{Posterior Mean} & \multicolumn{1}{>{\centering\arraybackslash}p{0.70in}}{95\% HPD Lower} & \multicolumn{1}{>{\centering\arraybackslash}p{0.70in}}{95\% HPD Upper} & \multicolumn{1}{>{\centering\arraybackslash}p{0.70in}}{R-Hat Statistic}\\
\midrule
\addlinespace[0.3em]
\multicolumn{6}{l}{\textbf{Hawkes Process Parameters}}\\
\hspace{1em}$\alpha$ & Weibull Shape & 0.874 & 0.849 & 0.899 & 1.005\\
\hspace{1em}$\delta$ & Exponential Decay Rate & 0.640 & 0.591 & 0.693 & 1.001\\
\hspace{1em}$1/\phi$ & Background RE Variance & 4.662 & 3.939 & 5.414 & 1.000\\
\hspace{1em}$1/\xi$ & Offspring RE Variance & 0.319 & 0.218 & 0.434 & 1.000\\
\addlinespace[0.3em]
\multicolumn{6}{l}{\textbf{Background Coefficients}}\\
\hspace{1em}$\exp(\beta_0)$ & Intercept & 0.028 & 0.013 & 0.047 & 1.000\\
\hspace{1em}$\exp(\beta_1)$ & Sex: Male & 0.687 & 0.365 & 1.054 & 1.001\\
\hspace{1em}$\exp(\beta_2)$ & Age: 25-39 & 2.034 & 0.811 & 3.530 & 1.000\\
\hspace{1em}$\exp(\beta_3)$ & Age: 40-64 & 3.183 & 1.347 & 5.334 & 1.005\\
\hspace{1em}$\exp(\beta_4)$ & Completed Higher Education & 0.579 & 0.283 & 0.927 & 1.004\\
\hspace{1em}$\exp(\beta_5)$ & Total Seizures Pre-Tx: 4-29 & 3.380 & 1.515 & 5.625 & 1.000\\
\hspace{1em}$\exp(\beta_6)$ & Total Seizures Pre-Tx: 30+ & 9.768 & 4.503 & 15.824 & 1.002\\
\hspace{1em}$\exp(\beta_7)$ & ASM: Combination Therapy & 3.873 & 1.567 & 6.726 & 1.000\\
\addlinespace[0.3em]
\multicolumn{6}{l}{\textbf{Offspring Coefficients}}\\
\hspace{1em}$\exp(\zeta_0)$ & Intercept & 0.327 & 0.242 & 0.413 & 1.002\\
\hspace{1em}$\exp(\zeta_1)$ & Sex: Male & 0.923 & 0.754 & 1.100 & 1.002\\
\hspace{1em}$\exp(\zeta_2)$ & Age: 25-39 & 1.039 & 0.790 & 1.314 & 1.000\\
\hspace{1em}$\exp(\zeta_3)$ & Age: 40-64 & 1.032 & 0.774 & 1.300 & 1.001\\
\hspace{1em}$\exp(\zeta_4)$ & Completed Higher Education & 1.033 & 0.820 & 1.254 & 1.001\\
\hspace{1em}$\exp(\zeta_5)$ & Total Seizures Pre-Tx: 4-29 & 1.112 & 0.843 & 1.402 & 1.002\\
\hspace{1em}$\exp(\zeta_6)$ & Total Seizures Pre-Tx: 30+ & 1.127 & 0.877 & 1.404 & 1.002\\
\hspace{1em}$\exp(\zeta_7)$ & ASM: Combination Therapy & 0.844 & 0.634 & 1.059 & 1.001\\
\bottomrule
\end{tabular}

\label{tab:hep_par_M2}
}
\end{table}

\begin{table}[!ht]
\setlength{\tabcolsep}{5pt}
\centering{
\caption{Estimated posterior means and 95\% highest posterior density intervals for parameters of Hawkes process model M4 fit to HEP study data. Model includes selected covariates in background process only and random effects in background process only.}

\begin{tabular}{clSSSS}\toprule
\multicolumn{1}{c}{} & \multicolumn{1}{c}{} & \multicolumn{1}{>{\centering\arraybackslash}p{0.70in}}{Posterior Mean} & \multicolumn{1}{>{\centering\arraybackslash}p{0.70in}}{95\% HPD Lower} & \multicolumn{1}{>{\centering\arraybackslash}p{0.70in}}{95\% HPD Upper} & \multicolumn{1}{>{\centering\arraybackslash}p{0.70in}}{R-Hat Statistic}\\
\midrule
\addlinespace[0.3em]
\multicolumn{6}{l}{\textbf{Hawkes Process Parameters}}\\
\hspace{1em}$\alpha$ & Weibull Shape & 0.826 & 0.791 & 0.860 & 1.004\\
\hspace{1em}$\delta$ & Exponential Decay Rate & 0.336 & 0.303 & 0.371 & 1.012\\
\hspace{1em}$1/\phi$ & Background RE Variance & 3.757 & 3.150 & 4.408 & 1.001\\
\addlinespace[0.3em]
\multicolumn{6}{l}{\textbf{Background Coefficients}}\\
\hspace{1em}$\exp(\beta_0)$ & Intercept & 0.023 & 0.012 & 0.036 & 1.002\\
\hspace{1em}$\exp(\beta_1)$ & Sex: Male & 0.692 & 0.407 & 1.016 & 1.001\\
\hspace{1em}$\exp(\beta_2)$ & Age: 25-39 & 2.165 & 0.931 & 3.586 & 1.000\\
\hspace{1em}$\exp(\beta_3)$ & Age: 40-64 & 2.737 & 1.295 & 4.438 & 1.000\\
\hspace{1em}$\exp(\beta_4)$ & Completed Higher Education & 0.542 & 0.286 & 0.837 & 1.001\\
\hspace{1em}$\exp(\beta_5)$ & Total Seizures Pre-Tx: 4-29 & 2.933 & 1.469 & 4.575 & 1.000\\
\hspace{1em}$\exp(\beta_6)$ & Total Seizures Pre-Tx: 30+ & 8.795 & 4.562 & 13.685 & 1.000\\
\hspace{1em}$\exp(\beta_7)$ & ASM: Combination Therapy & 2.519 & 1.155 & 4.106 & 1.000\\
\addlinespace[0.3em]
\multicolumn{6}{l}{\textbf{Offspring Coefficients}}\\
\hspace{1em}$\exp(\zeta_0)$ & Intercept & 0.273 & 0.251 & 0.297 & 1.014\\
\bottomrule
\end{tabular}

\label{tab:hep_par_M4}
}
\end{table}

\clearpage
\vspace*{\fill}
\begin{table}[!ht]
\setlength{\tabcolsep}{5pt}
\centering{
\caption{Estimated posterior means and 95\% highest posterior density intervals for parameters of Hawkes process model M5 fit to HEP study data. Model includes selected covariates in background process only and random effects in offspring process only.}

\begin{tabular}{clSSSS}\toprule
\multicolumn{1}{c}{} & \multicolumn{1}{c}{} & \multicolumn{1}{>{\centering\arraybackslash}p{0.70in}}{Posterior Mean} & \multicolumn{1}{>{\centering\arraybackslash}p{0.70in}}{95\% HPD Lower} & \multicolumn{1}{>{\centering\arraybackslash}p{0.70in}}{95\% HPD Upper} & \multicolumn{1}{>{\centering\arraybackslash}p{0.70in}}{R-Hat Statistic}\\
\midrule
\addlinespace[0.3em]
\multicolumn{6}{l}{\textbf{Hawkes Process Parameters}}\\
\hspace{1em}$\alpha$ & Weibull Shape & 0.481 & 0.441 & 0.523 & 1.004\\
\hspace{1em}$\delta$ & Exponential Decay Rate & 0.133 & 0.125 & 0.140 & 1.037\\
\hspace{1em}$1/\xi$ & Offspring RE Variance & 0.004 & 0.003 & 0.006 & 1.000\\
\addlinespace[0.3em]
\multicolumn{6}{l}{\textbf{Background Coefficients}}\\
\hspace{1em}$\exp(\beta_0)$ & Intercept & 0.067 & 0.044 & 0.091 & 1.001\\
\hspace{1em}$\exp(\beta_1)$ & Sex: Male & 0.601 & 0.487 & 0.717 & 1.001\\
\hspace{1em}$\exp(\beta_2)$ & Age: 25-39 & 2.416 & 1.864 & 3.024 & 1.000\\
\hspace{1em}$\exp(\beta_3)$ & Age: 40-64 & 1.613 & 1.206 & 2.043 & 1.000\\
\hspace{1em}$\exp(\beta_4)$ & Completed Higher Education & 0.560 & 0.452 & 0.678 & 1.000\\
\hspace{1em}$\exp(\beta_5)$ & Total Seizures Pre-Tx: 4-29 & 1.739 & 1.312 & 2.186 & 1.000\\
\hspace{1em}$\exp(\beta_6)$ & Total Seizures Pre-Tx: 30+ & 3.574 & 2.786 & 4.429 & 1.000\\
\hspace{1em}$\exp(\beta_7)$ & ASM: Combination Therapy & 1.101 & 0.839 & 1.369 & 1.000\\
\addlinespace[0.3em]
\multicolumn{6}{l}{\textbf{Offspring Coefficients}}\\
\hspace{1em}$\exp(\zeta_0)$ & Intercept & 0.128 & 0.121 & 0.136 & 1.034\\
\bottomrule
\end{tabular}

\label{tab:hep_par_M5}
}
\end{table}
\vspace*{\fill}

\clearpage
\vspace*{\fill}
\begin{table}[!ht]
\setlength{\tabcolsep}{5pt}
\centering{
\caption{Estimated posterior means and 95\% highest posterior density intervals for parameters of Hawkes process model M6 fit to HEP study data. Model includes selected covariates in background process only and no random effects.}

\begin{tabular}{clSSSS}\toprule
\multicolumn{1}{c}{} & \multicolumn{1}{c}{} & \multicolumn{1}{>{\centering\arraybackslash}p{0.70in}}{Posterior Mean} & \multicolumn{1}{>{\centering\arraybackslash}p{0.70in}}{95\% HPD Lower} & \multicolumn{1}{>{\centering\arraybackslash}p{0.70in}}{95\% HPD Upper} & \multicolumn{1}{>{\centering\arraybackslash}p{0.70in}}{R-Hat Statistic}\\
\midrule
\addlinespace[0.3em]
\multicolumn{6}{l}{\textbf{Hawkes Process Parameters}}\\
\hspace{1em}$\alpha$ & Weibull Shape & 0.474 & 0.434 & 0.515 & 1.000\\
\hspace{1em}$\delta$ & Exponential Decay Rate & 0.132 & 0.125 & 0.140 & 1.004\\
\addlinespace[0.3em]
\multicolumn{6}{l}{\textbf{Background Coefficients}}\\
\hspace{1em}$\exp(\beta_0)$ & Intercept & 0.068 & 0.045 & 0.092 & 1.001\\
\hspace{1em}$\exp(\beta_1)$ & Sex: Male & 0.612 & 0.498 & 0.730 & 1.000\\
\hspace{1em}$\exp(\beta_2)$ & Age: 25-39 & 2.352 & 1.830 & 2.928 & 1.000\\
\hspace{1em}$\exp(\beta_3)$ & Age: 40-64 & 1.600 & 1.203 & 2.032 & 1.000\\
\hspace{1em}$\exp(\beta_4)$ & Completed Higher Education & 0.573 & 0.464 & 0.691 & 1.000\\
\hspace{1em}$\exp(\beta_5)$ & Total Seizures Pre-Tx: 4-29 & 1.726 & 1.318 & 2.163 & 1.000\\
\hspace{1em}$\exp(\beta_6)$ & Total Seizures Pre-Tx: 30+ & 3.505 & 2.777 & 4.336 & 1.001\\
\hspace{1em}$\exp(\beta_7)$ & ASM: Combination Therapy & 1.104 & 0.840 & 1.371 & 1.000\\
\addlinespace[0.3em]
\multicolumn{6}{l}{\textbf{Offspring Coefficients}}\\
\hspace{1em}$\exp(\zeta_0)$ & Intercept & 0.131 & 0.124 & 0.139 & 1.004\\
\bottomrule
\end{tabular}

\label{tab:hep_par_M6}
}
\end{table}
\vspace*{\fill}

\clearpage
\vspace{0pt}
\section{Simulation Algorithms and Examples}
\label{supp:sim_alg}

To simulate a dataset comprising $m$ aggregated seizure event trajectories from the proposed Hawkes process model, the exact simulation method of \citet{dassios_exact_2013} is extended to incorporate covariates, random effects, and a Weibull baseline intensity function \citep{bender_generating_2005}. The corresponding tracking trajectories are constructed by simulating transitions in tracking status based on a two-state discrete-time Markov chain. For states \{0 = missing, 1 = tracked\}, we define $p_i$ as the probability of individual $i$ moving from state 0 to 1, and $q_i$ as the probability of moving from state 1 to 0. The inputs required to generate the transition probabilities and tracking trajectories include dropout rate $r_T$, minimum follow-up time $T_{min}$, maximum follow-up time $T_{max}$, average percent of bins missing $\pi_0$, transition probability factors $w_1$ and $w_2$, and initial probability adjustment $\Delta \pi_1$. Note that $w_1 > 1$ increases the ``stickiness'' of the tracking state runs beyond that of a binomial distribution, $w_2 > 1$ reduces the dispersion of the transition probabilities, and $\Delta \pi_1 > 0$ allows the probability of an individual starting in the tracked state to be higher than that of the long-run distribution. Assuming that the aggregation bin width is equal to one time unit, the simulation of tracking trajectories proceeds according to Algorithm \ref{alg:sim_track}. Then, given values for the mixed Hawkes process model parameters $\alpha$, $\delta$, $\bm\eta$, $\bm\kappa$, $\phi$, and $\xi$, the simulation of event trajectories follows according to Algorithm \ref{alg:sim_event}, where $br_{max}$ represents the maximum allowable value for the individual branching ratios.

\vspace{2ex}
\begin{algorithm}
\setstretch{1.20}
\DontPrintSemicolon
\caption{Simulation of Tracking Trajectories}
\label{alg:sim_track}
\KwIn{$\xsub r_T, T_{min}, T_{max}, \xsub\pi_0, \xsub w_1, \xsub w_2, \Delta \xsub\pi_1$}
$\xsub\pi_1 \gets 1 - \xsub\pi_0$\;
Mean of transition probability:\;
    $\quad \bar{p} \gets \xsub\pi_1/\xsub w_1$\;
    $\quad \bar{q} \gets \xsub\pi_0/\xsub w_1$\;
Variance of transition probability:\;
    $\quad v_p \gets \bar{p}/\xsub w_2$\;
    $\quad v_q \gets \bar{q}/\xsub w_2$\;
Shape parameters for $0-1$ probability:\;
    $\quad c_p \gets \bar{p}(1-\bar{p})/v_p-1$\;
    $\quad a_p \gets \bar{p}c_p$\;
    $\quad b_p \gets (1-\bar{p})c_p$\;
Shape parameters for $1-0$ probability:\;
    $\quad c_q \gets \bar{q}(1-\bar{q})/v_q-1$\;
    $\quad a_q \gets \bar{q}c_q$\;
    $\quad b_q \gets (1-\bar{q})c_q$\;
\For{\forcond}{
Generate discrete follow-up times:\;
    $\quad X_i \sim \text{Exp}(\xsub r_T)$\;
    $\quad T_i \gets (\lfloor X_i \rfloor + T_{min}) \wedge T_{max}$\;
Generate transition probabilities:\;
    $\quad p_i \sim \text{Beta}(a_p, b_p)$\;
    $\quad q_i \sim \text{Beta}(a_q, b_q)$\;
Initialize tracking state $\tau_{ij}$:\;
    $\quad j \gets 1$\;
    $\quad \tau_{ij} \sim \text{Bern}(\pi_1 + \Delta\pi_1)$\;
\While{$j < T_i$}{
    \eIf{$\tau_{ij} = 0$}
        {$\tau_{i,j+1} \sim \text{Bern}(p_i)$\;}
        {$\tau_{i,j+1} \sim \text{Bern}(1-q_i)$\;}
  $j \gets j+1$
}
\KwRet{$\bm\tau_i$}
}
\end{algorithm}

\begin{algorithm}
\setstretch{1.20}
\DontPrintSemicolon
\caption{Simulation of Event Trajectories}
\label{alg:sim_event}
\KwIn{$\alpha$, $\delta$, $\bm\eta$, $\bm\kappa$, $\phi$, $\xi$, $br_{max}$}
\For{\forcond}{
Generate random effects:\;
    $\quad \nu_i \sim \text{Gamma}(\phi, \phi)$\;
    $\quad \omega_i \sim \text{Gamma}(\xi, \xi)$\;
Cap branching ratios:\;
    $\quad \omega_i \gets \omega_i \wedge (br_{max} \times \delta/\kappa_i$)\;
Initialize event history:\;
    $\quad j \gets 0$\;
    $\quad t_{ij} \gets 0$\;
    $\quad \lambda_1 \gets 0$\;
\While{$t_{ij} < T_i$}{
    Sample from immigrant CDF:\;
        $\quad \xsub u_0 \sim \text{U}(0, 1)$\;
        $\quad \xsub s_0 \gets \left( -\log \xsub u_0/(\nu_i\eta_i) + t_{ij}^\alpha \right)^{1/\alpha} - t_{ij}$\; 
        \vspace{1ex}
    Sample from offspring CDF:\;
        $\quad \xsub u_1 \sim \text{U}(0, 1)$\;
        $\quad d_1 \gets 1 + \delta\log u_1/\lambda_1$\;
        \vspace{1ex}
    \eIf{$d_1 > 0$}
        {$\xsub s_1 \gets -\log d_1/\delta$\;}
        {$\xsub s_1 \gets \infty$\;}
    Next interarrival time: $s \gets \xsub s_0 \wedge \xsub s_1$\;
    Offspring intensity: $\lambda_1 \gets \lambda_1 \exp(-\delta s) + \omega_i \kappa_i$\;
    \If{$t_{ij} + s < T_i$}{
        $\quad t_{i, j+1} \gets t_{ij} + s$\;
        $\quad j \gets j+1$\;    
    }
}
\KwRet{$\bm t_i$}\;
\vspace{5pt}
Aggregate events to counts at bin level.\;
\vspace{5pt}
Remove counts for any untracked bins.\;
}
\end{algorithm}

\begin{figure}[!ht]
\centering{
\includegraphics[width=1.00\textwidth]{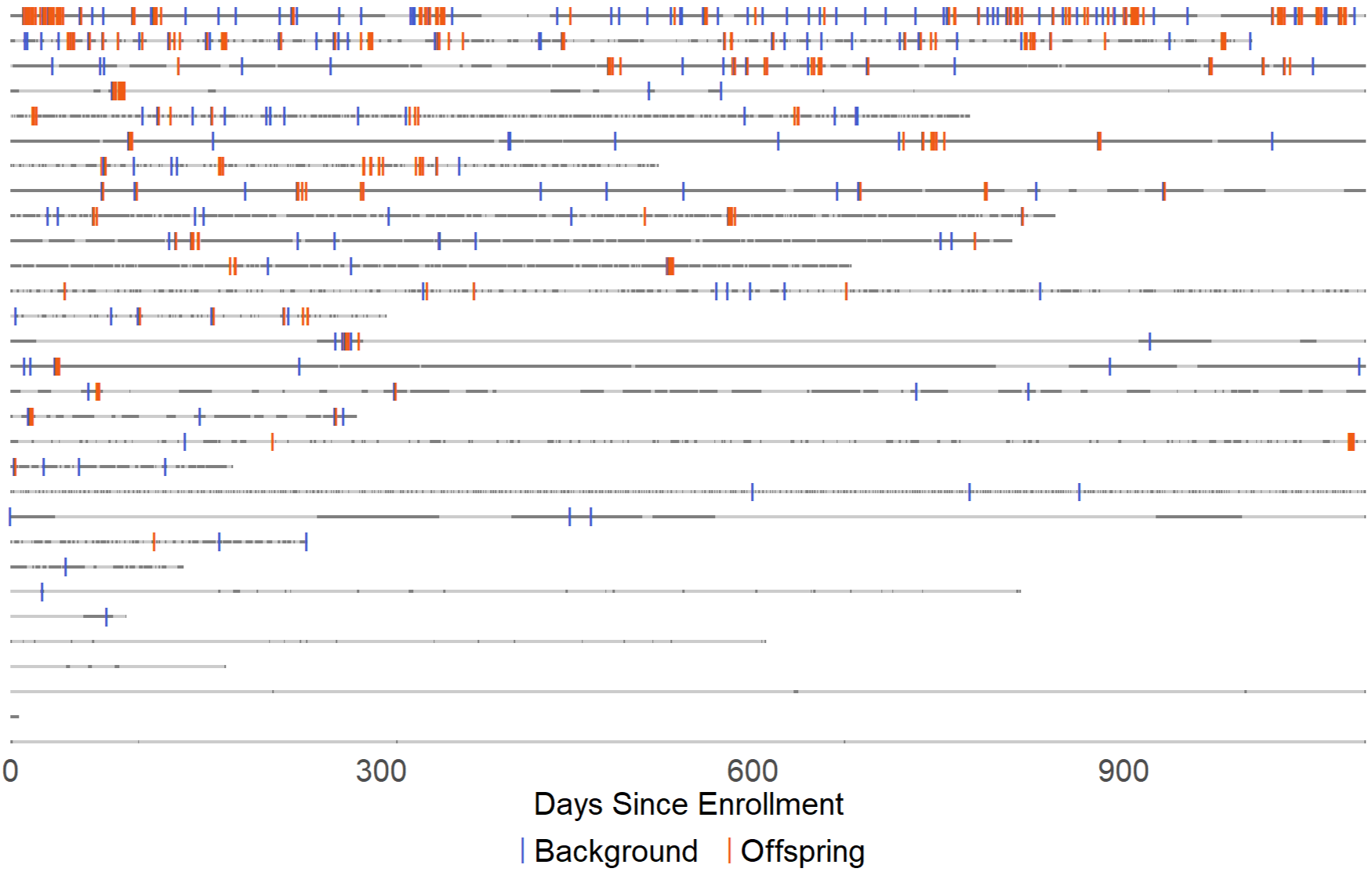}
\caption{Seizure event trajectories simulated from mixed Hawkes process model for 30 individuals with 50\% average days missing and low random effects variance ($\phi=5, \xi=50$). Tick mark shade indicates whether each event was generated from the background process or the offspring process. Each horizontal line represents the follow-up period for one participant. Darker segments represent tracked periods while lighter segments indicate unreported intervals.}
\label{fig:sim_plot_LM}
}
\end{figure}

\begin{figure}[!ht]
\centering{
\includegraphics[width=1.00\textwidth]{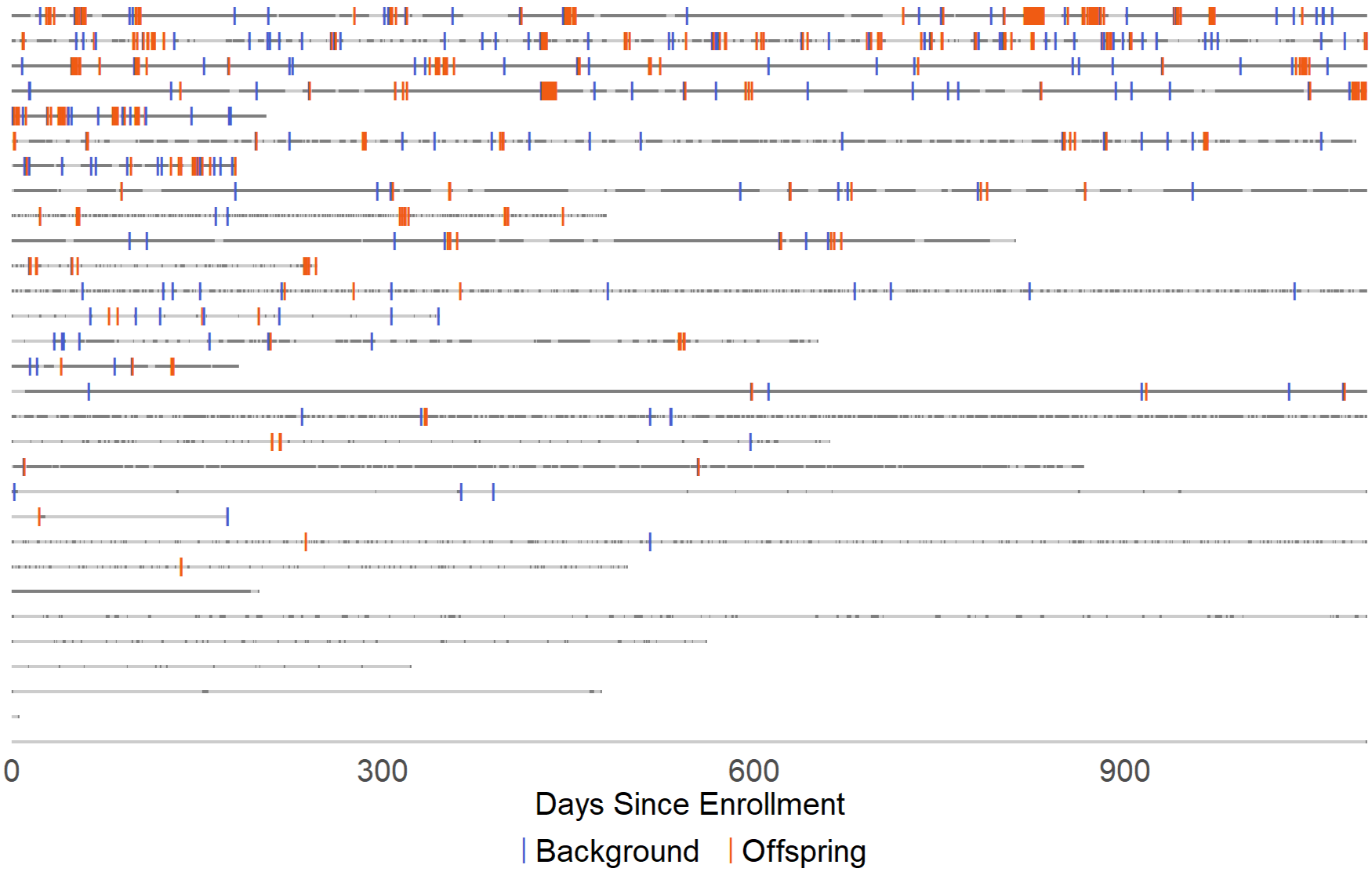}
\caption{Seizure event trajectories simulated from mixed Hawkes process model for 30 individuals with 50\% average days missing and medium random effects variance ($\phi=1, \xi=10$). Tick mark shade indicates whether each event was generated from the background process or the offspring process. Each horizontal line represents the follow-up period for one participant. Darker segments represent tracked periods while lighter segments indicate unreported intervals.}
\label{fig:sim_plot_MM}
}
\end{figure}

\begin{figure}[!ht]
\centering{
\includegraphics[width=1.00\textwidth]{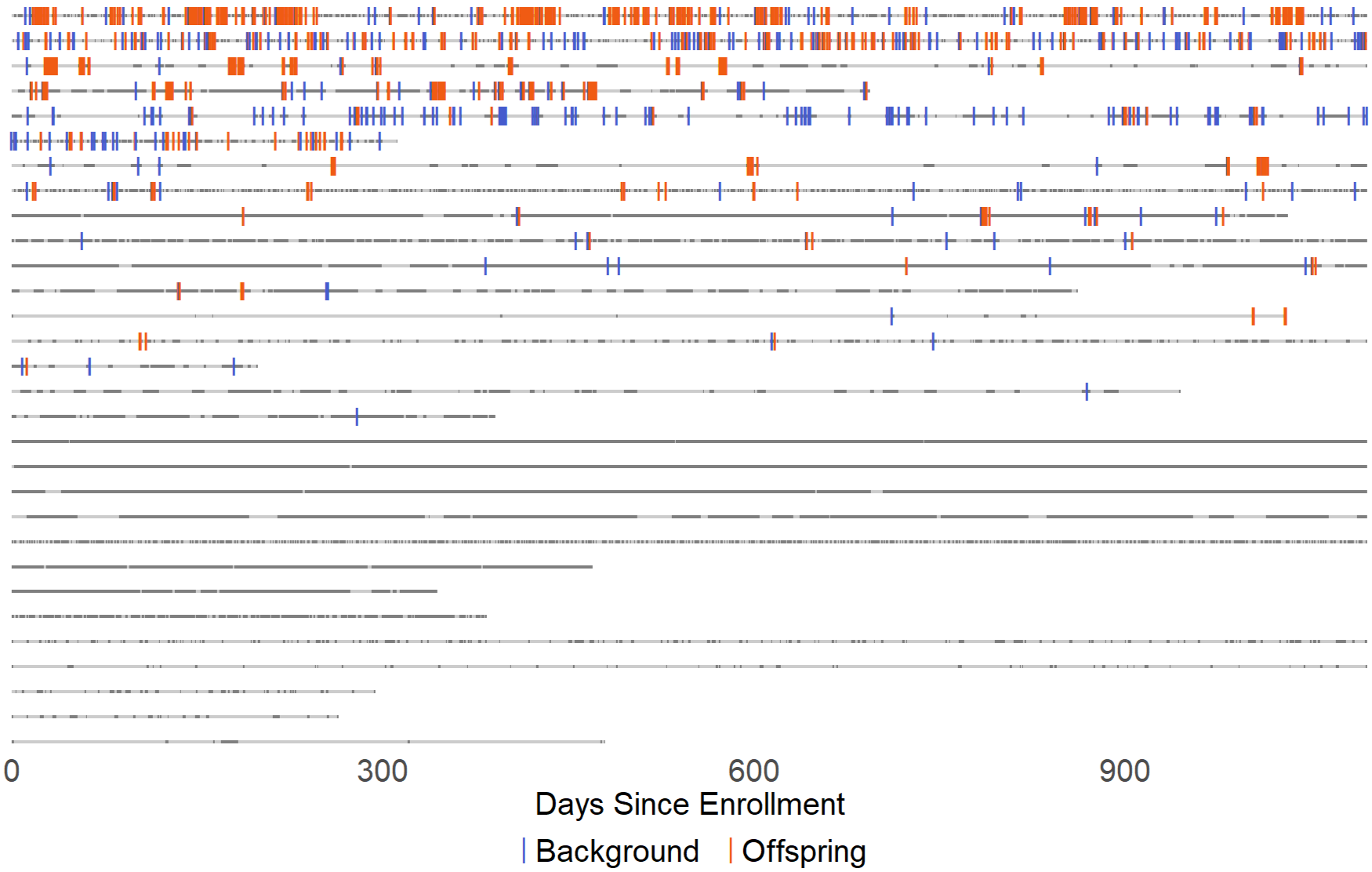}
\caption{Seizure event trajectories simulated from mixed Hawkes process model for 30 individuals with 50\% average days missing and high random effects variance ($\phi=0.2, \xi=5$). Tick mark shade indicates whether each event was generated from the background process or the offspring process. Each horizontal line represents the follow-up period for one participant. Darker segments represent tracked periods while lighter segments indicate unreported intervals.}
\label{fig:sim_plot_HM}
}
\end{figure}

\clearpage
\vspace{0pt}
\section{Additional Simulation Study Results}
\label{supp:sim}

\begin{table}[!ht]
\vspace{5ex}
\setlength{\tabcolsep}{4.75pt}
\centering{
\caption{Simulation Study I results for scenarios with 100 individuals and low random effects variance ($\phi=5, \xi=50$). Empirical means, standard deviations, and coverage of 95\% highest posterior density intervals based on 300 simulated datasets per scenario.}

\centering
\begin{tabular}{cTSSSSSSSSS}
\toprule
\multicolumn{2}{c}{ } & \multicolumn{3}{c}{25\% Missing} & \multicolumn{3}{c}{50\% Missing} & \multicolumn{3}{c}{75\% Missing} \\
\cmidrule(l{3pt}r{3pt}){3-5} \cmidrule(l{3pt}r{3pt}){6-8} \cmidrule(l{3pt}r{3pt}){9-11}
\multicolumn{1}{p{0.4cm}}{\centering } & \multicolumn{1}{p{1.2cm}}{\centering True Value} & \multicolumn{1}{p{1.2cm}}{\centering Emp Mean} & \multicolumn{1}{p{1.2cm}}{\centering Std Dev} & \multicolumn{1}{p{1.2cm}}{\centering HPD Cvg} & \multicolumn{1}{p{1.2cm}}{\centering Emp Mean} & \multicolumn{1}{p{1.2cm}}{\centering Std Dev} & \multicolumn{1}{p{1.2cm}}{\centering HPD Cvg} & \multicolumn{1}{p{1.2cm}}{\centering Emp Mean} & \multicolumn{1}{p{1.2cm}}{\centering Std Dev} & \multicolumn{1}{p{1.2cm}}{\centering HPD Cvg}\\
\midrule
$\alpha$ & 0.90 & 0.901 & 0.023 & 0.910 & 0.903 & 0.026 & 0.897 & 0.905 & 0.032 & 0.900\\
$\beta_0$ & -3.50 & -3.504 & 0.197 & 0.933 & -3.525 & 0.224 & 0.917 & -3.534 & 0.278 & 0.893\\
$\beta_1$ & -0.50 & -0.497 & 0.117 & 0.930 & -0.490 & 0.128 & 0.937 & -0.520 & 0.156 & 0.930\\
$\beta_2$ & 1.00 & 0.996 & 0.135 & 0.940 & 0.998 & 0.152 & 0.940 & 1.014 & 0.187 & 0.920\\
$\delta$ & 0.60 & 0.605 & 0.023 & 0.930 & 0.607 & 0.028 & 0.933 & 0.617 & 0.040 & 0.940\\
$\zeta_0$ & -1.10 & -1.120 & 0.089 & 0.957 & -1.129 & 0.108 & 0.933 & -1.152 & 0.141 & 0.937\\
$\zeta_1$ & -0.10 & -0.104 & 0.068 & 0.970 & -0.104 & 0.079 & 0.967 & -0.112 & 0.103 & 0.960\\
$\zeta_2$ & 0.10 & 0.110 & 0.087 & 0.947 & 0.113 & 0.103 & 0.933 & 0.132 & 0.135 & 0.920\\
$1/\phi$ & 0.20 & 0.189 & 0.042 & 0.897 & 0.187 & 0.048 & 0.893 & 0.177 & 0.060 & 0.867\\
$1/\xi$ & 0.02 & 0.029 & 0.010 & 0.957 & 0.032 & 0.011 & 0.970 & 0.037 & 0.015 & 0.970\\
\bottomrule
\end{tabular}

\label{tab:sim_LL}
}
\end{table}

\begin{table}[!ht]
\vspace{5ex}
\setlength{\tabcolsep}{4.75pt}
\centering{
\caption{Simulation Study I results for scenarios with 100 individuals and medium random effects variance ($\phi=1, \xi=10$). Empirical means, standard deviations, and coverage of 95\% highest posterior density intervals based on 300 simulated datasets per scenario.}

\centering
\begin{tabular}{cTSSSSSSSSS}
\toprule
\multicolumn{2}{c}{ } & \multicolumn{3}{c}{25\% Missing} & \multicolumn{3}{c}{50\% Missing} & \multicolumn{3}{c}{75\% Missing} \\
\cmidrule(l{3pt}r{3pt}){3-5} \cmidrule(l{3pt}r{3pt}){6-8} \cmidrule(l{3pt}r{3pt}){9-11}
\multicolumn{1}{p{0.4cm}}{\centering } & \multicolumn{1}{p{1.2cm}}{\centering True Value} & \multicolumn{1}{p{1.2cm}}{\centering Emp Mean} & \multicolumn{1}{p{1.2cm}}{\centering Std Dev} & \multicolumn{1}{p{1.2cm}}{\centering HPD Cvg} & \multicolumn{1}{p{1.2cm}}{\centering Emp Mean} & \multicolumn{1}{p{1.2cm}}{\centering Std Dev} & \multicolumn{1}{p{1.2cm}}{\centering HPD Cvg} & \multicolumn{1}{p{1.2cm}}{\centering Emp Mean} & \multicolumn{1}{p{1.2cm}}{\centering Std Dev} & \multicolumn{1}{p{1.2cm}}{\centering HPD Cvg}\\
\midrule
$\alpha$ & 0.9 & 0.901 & 0.024 & 0.907 & 0.902 & 0.028 & 0.920 & 0.904 & 0.035 & 0.907\\
$\beta_0$ & -3.5 & -3.494 & 0.283 & 0.940 & -3.516 & 0.311 & 0.900 & -3.523 & 0.367 & 0.920\\
$\beta_1$ & -0.5 & -0.502 & 0.225 & 0.927 & -0.513 & 0.236 & 0.940 & -0.528 & 0.267 & 0.930\\
$\beta_2$ & 1.0 & 0.975 & 0.250 & 0.943 & 1.002 & 0.268 & 0.947 & 0.997 & 0.307 & 0.950\\
$\delta$ & 0.6 & 0.603 & 0.022 & 0.940 & 0.607 & 0.027 & 0.927 & 0.614 & 0.038 & 0.923\\
$\zeta_0$ & -1.1 & -1.124 & 0.110 & 0.913 & -1.119 & 0.125 & 0.913 & -1.178 & 0.163 & 0.893\\
$\zeta_1$ & -0.1 & -0.092 & 0.091 & 0.940 & -0.097 & 0.099 & 0.913 & -0.093 & 0.120 & 0.907\\
$\zeta_2$ & 0.1 & 0.106 & 0.110 & 0.920 & 0.103 & 0.123 & 0.913 & 0.157 & 0.157 & 0.850\\
$1/\phi$ & 1.0 & 0.975 & 0.162 & 0.937 & 0.965 & 0.176 & 0.917 & 0.909 & 0.208 & 0.870\\
$1/\xi$ & 0.1 & 0.076 & 0.023 & 0.733 & 0.075 & 0.025 & 0.757 & 0.073 & 0.030 & 0.747\\
\bottomrule
\end{tabular}

\label{tab:sim_LM}
}
\end{table}

\clearpage
\begin{table}[!ht]
\vspace{10ex}
\setlength{\tabcolsep}{4.75pt}
\centering{
\caption{Simulation Study I results for scenarios with 100 individuals and high random effects variance ($\phi=0.2, \xi=5$). Empirical means, standard deviations, and coverage of 95\% highest posterior density intervals based on 300 simulated datasets per scenario.}

\centering
\begin{tabular}{cTSSSSSSSSS}
\toprule
\multicolumn{2}{c}{ } & \multicolumn{3}{c}{25\% Missing} & \multicolumn{3}{c}{50\% Missing} & \multicolumn{3}{c}{75\% Missing} \\
\cmidrule(l{3pt}r{3pt}){3-5} \cmidrule(l{3pt}r{3pt}){6-8} \cmidrule(l{3pt}r{3pt}){9-11}
\multicolumn{1}{p{0.4cm}}{\centering } & \multicolumn{1}{p{1.2cm}}{\centering True Value} & \multicolumn{1}{p{1.2cm}}{\centering Emp Mean} & \multicolumn{1}{p{1.2cm}}{\centering Std Dev} & \multicolumn{1}{p{1.2cm}}{\centering HPD Cvg} & \multicolumn{1}{p{1.2cm}}{\centering Emp Mean} & \multicolumn{1}{p{1.2cm}}{\centering Std Dev} & \multicolumn{1}{p{1.2cm}}{\centering HPD Cvg} & \multicolumn{1}{p{1.2cm}}{\centering Emp Mean} & \multicolumn{1}{p{1.2cm}}{\centering Std Dev} & \multicolumn{1}{p{1.2cm}}{\centering HPD Cvg}\\
\midrule
$\alpha$ & 0.9 & 0.903 & 0.026 & 0.930 & 0.901 & 0.031 & 0.890 & 0.899 & 0.038 & 0.897\\
$\beta_0$ & -3.5 & -3.509 & 0.539 & 0.940 & -3.565 & 0.565 & 0.947 & -3.510 & 0.617 & 0.920\\
$\beta_1$ & -0.5 & -0.541 & 0.488 & 0.937 & -0.489 & 0.504 & 0.957 & -0.482 & 0.543 & 0.957\\
$\beta_2$ & 1.0 & 1.005 & 0.540 & 0.927 & 1.026 & 0.559 & 0.933 & 0.977 & 0.608 & 0.947\\
$\delta$ & 0.6 & 0.603 & 0.027 & 0.950 & 0.605 & 0.033 & 0.943 & 0.613 & 0.046 & 0.930\\
$\zeta_0$ & -1.1 & -1.154 & 0.165 & 0.917 & -1.159 & 0.179 & 0.920 & -1.188 & 0.214 & 0.890\\
$\zeta_1$ & -0.1 & -0.085 & 0.140 & 0.930 & -0.076 & 0.146 & 0.913 & -0.096 & 0.171 & 0.917\\
$\zeta_2$ & 0.1 & 0.089 & 0.168 & 0.917 & 0.102 & 0.178 & 0.907 & 0.121 & 0.211 & 0.913\\
$1/\phi$ & 5.0 & 4.876 & 0.849 & 0.933 & 4.819 & 0.912 & 0.917 & 4.711 & 1.055 & 0.917\\
$1/\xi$ & 0.2 & 0.131 & 0.046 & 0.577 & 0.121 & 0.048 & 0.527 & 0.110 & 0.054 & 0.457\\
\bottomrule
\end{tabular}

\label{tab:sim_LH}
}
\end{table}

\begin{table}[!ht]
\vspace{5ex}
\setlength{\tabcolsep}{4.75pt}
\centering{
\caption{Simulation Study I results for scenarios with 400 individuals and low random effects variance ($\phi=5, \xi=50$). Empirical means, standard deviations, and coverage of 95\% highest posterior density intervals based on 300 simulated datasets per scenario.}

\centering
\begin{tabular}{cTSSSSSSSSS}
\toprule
\multicolumn{2}{c}{ } & \multicolumn{3}{c}{25\% Missing} & \multicolumn{3}{c}{50\% Missing} & \multicolumn{3}{c}{75\% Missing} \\
\cmidrule(l{3pt}r{3pt}){3-5} \cmidrule(l{3pt}r{3pt}){6-8} \cmidrule(l{3pt}r{3pt}){9-11}
\multicolumn{1}{p{0.4cm}}{\centering } & \multicolumn{1}{p{1.2cm}}{\centering True Value} & \multicolumn{1}{p{1.2cm}}{\centering Emp Mean} & \multicolumn{1}{p{1.2cm}}{\centering Std Dev} & \multicolumn{1}{p{1.2cm}}{\centering HPD Cvg} & \multicolumn{1}{p{1.2cm}}{\centering Emp Mean} & \multicolumn{1}{p{1.2cm}}{\centering Std Dev} & \multicolumn{1}{p{1.2cm}}{\centering HPD Cvg} & \multicolumn{1}{p{1.2cm}}{\centering Emp Mean} & \multicolumn{1}{p{1.2cm}}{\centering Std Dev} & \multicolumn{1}{p{1.2cm}}{\centering HPD Cvg}\\
\midrule
$\alpha$ & 0.90 & 0.900 & 0.011 & 0.913 & 0.900 & 0.013 & 0.893 & 0.902 & 0.016 & 0.883\\
$\beta_0$ & -3.50 & -3.504 & 0.096 & 0.917 & -3.502 & 0.109 & 0.913 & -3.519 & 0.135 & 0.890\\
$\beta_1$ & -0.50 & -0.504 & 0.060 & 0.943 & -0.496 & 0.066 & 0.903 & -0.499 & 0.080 & 0.947\\
$\beta_2$ & 1.00 & 1.005 & 0.068 & 0.943 & 1.006 & 0.075 & 0.937 & 1.013 & 0.093 & 0.933\\
$\delta$ & 0.60 & 0.602 & 0.011 & 0.940 & 0.604 & 0.014 & 0.940 & 0.609 & 0.019 & 0.907\\
$\zeta_0$ & -1.10 & -1.110 & 0.041 & 0.910 & -1.117 & 0.048 & 0.913 & -1.113 & 0.062 & 0.907\\
$\zeta_1$ & -0.10 & -0.100 & 0.033 & 0.940 & -0.099 & 0.037 & 0.960 & -0.109 & 0.049 & 0.943\\
$\zeta_2$ & 0.10 & 0.104 & 0.041 & 0.943 & 0.107 & 0.047 & 0.930 & 0.101 & 0.059 & 0.910\\
$1/\phi$ & 0.20 & 0.199 & 0.022 & 0.940 & 0.195 & 0.024 & 0.947 & 0.194 & 0.031 & 0.923\\
$1/\xi$ & 0.02 & 0.023 & 0.004 & 0.960 & 0.025 & 0.005 & 0.953 & 0.027 & 0.007 & 0.937\\
\bottomrule
\end{tabular}

\label{tab:sim_HL}
}
\end{table}

\clearpage
\begin{table}[!ht]
\vspace{10ex}
\setlength{\tabcolsep}{4.75pt}
\centering{
\caption{Simulation Study I results for scenarios with 400 individuals and medium random effects variance ($\phi=1, \xi=10$). Empirical means, standard deviations, and coverage of 95\% highest posterior density intervals based on 300 simulated datasets per scenario.}

\centering
\begin{tabular}{cTSSSSSSSSS}
\toprule
\multicolumn{2}{c}{ } & \multicolumn{3}{c}{25\% Missing} & \multicolumn{3}{c}{50\% Missing} & \multicolumn{3}{c}{75\% Missing} \\
\cmidrule(l{3pt}r{3pt}){3-5} \cmidrule(l{3pt}r{3pt}){6-8} \cmidrule(l{3pt}r{3pt}){9-11}
\multicolumn{1}{p{0.4cm}}{\centering } & \multicolumn{1}{p{1.2cm}}{\centering True Value} & \multicolumn{1}{p{1.2cm}}{\centering Emp Mean} & \multicolumn{1}{p{1.2cm}}{\centering Std Dev} & \multicolumn{1}{p{1.2cm}}{\centering HPD Cvg} & \multicolumn{1}{p{1.2cm}}{\centering Emp Mean} & \multicolumn{1}{p{1.2cm}}{\centering Std Dev} & \multicolumn{1}{p{1.2cm}}{\centering HPD Cvg} & \multicolumn{1}{p{1.2cm}}{\centering Emp Mean} & \multicolumn{1}{p{1.2cm}}{\centering Std Dev} & \multicolumn{1}{p{1.2cm}}{\centering HPD Cvg}\\
\midrule
$\alpha$ & 0.9 & 0.901 & 0.012 & 0.927 & 0.902 & 0.014 & 0.880 & 0.901 & 0.017 & 0.857\\
$\beta_0$ & -3.5 & -3.507 & 0.136 & 0.950 & -3.505 & 0.149 & 0.923 & -3.509 & 0.177 & 0.923\\
$\beta_1$ & -0.5 & -0.505 & 0.114 & 0.947 & -0.503 & 0.120 & 0.933 & -0.488 & 0.137 & 0.937\\
$\beta_2$ & 1.0 & 1.000 & 0.125 & 0.953 & 0.997 & 0.133 & 0.943 & 0.990 & 0.153 & 0.933\\
$\delta$ & 0.6 & 0.600 & 0.011 & 0.940 & 0.603 & 0.013 & 0.960 & 0.606 & 0.018 & 0.947\\
$\zeta_0$ & -1.1 & -1.113 & 0.052 & 0.913 & -1.107 & 0.058 & 0.947 & -1.107 & 0.071 & 0.943\\
$\zeta_1$ & -0.1 & -0.085 & 0.046 & 0.930 & -0.086 & 0.050 & 0.893 & -0.083 & 0.058 & 0.940\\
$\zeta_2$ & 0.1 & 0.090 & 0.054 & 0.923 & 0.081 & 0.058 & 0.937 & 0.083 & 0.069 & 0.930\\
$1/\phi$ & 1.0 & 0.992 & 0.081 & 0.917 & 0.991 & 0.088 & 0.933 & 0.975 & 0.106 & 0.927\\
$1/\xi$ & 0.1 & 0.078 & 0.012 & 0.513 & 0.077 & 0.013 & 0.507 & 0.071 & 0.014 & 0.450\\
\bottomrule
\end{tabular}

\label{tab:sim_HM}
}
\end{table}

\begin{table}[!ht]
\vspace{5ex}
\setlength{\tabcolsep}{4.75pt}
\centering{
\caption{Simulation Study I results for scenarios with 400 individuals and high random effects variance ($\phi=0.2, \xi=5$). Empirical means, standard deviations, and coverage of 95\% highest posterior density intervals based on 300 simulated datasets per scenario.}

\centering
\begin{tabular}{cTSSSSSSSSS}
\toprule
\multicolumn{2}{c}{ } & \multicolumn{3}{c}{25\% Missing} & \multicolumn{3}{c}{50\% Missing} & \multicolumn{3}{c}{75\% Missing} \\
\cmidrule(l{3pt}r{3pt}){3-5} \cmidrule(l{3pt}r{3pt}){6-8} \cmidrule(l{3pt}r{3pt}){9-11}
\multicolumn{1}{p{0.4cm}}{\centering } & \multicolumn{1}{p{1.2cm}}{\centering True Value} & \multicolumn{1}{p{1.2cm}}{\centering Emp Mean} & \multicolumn{1}{p{1.2cm}}{\centering Std Dev} & \multicolumn{1}{p{1.2cm}}{\centering HPD Cvg} & \multicolumn{1}{p{1.2cm}}{\centering Emp Mean} & \multicolumn{1}{p{1.2cm}}{\centering Std Dev} & \multicolumn{1}{p{1.2cm}}{\centering HPD Cvg} & \multicolumn{1}{p{1.2cm}}{\centering Emp Mean} & \multicolumn{1}{p{1.2cm}}{\centering Std Dev} & \multicolumn{1}{p{1.2cm}}{\centering HPD Cvg}\\
\midrule
$\alpha$ & 0.9 & 0.901 & 0.013 & 0.930 & 0.902 & 0.015 & 0.883 & 0.904 & 0.019 & 0.843\\
$\beta_0$ & -3.5 & -3.510 & 0.248 & 0.933 & -3.518 & 0.261 & 0.943 & -3.540 & 0.293 & 0.943\\
$\beta_1$ & -0.5 & -0.493 & 0.241 & 0.947 & -0.483 & 0.249 & 0.953 & -0.500 & 0.270 & 0.933\\
$\beta_2$ & 1.0 & 1.005 & 0.261 & 0.920 & 1.004 & 0.271 & 0.950 & 1.025 & 0.297 & 0.943\\
$\delta$ & 0.6 & 0.601 & 0.013 & 0.917 & 0.603 & 0.016 & 0.907 & 0.608 & 0.022 & 0.933\\
$\zeta_0$ & -1.1 & -1.142 & 0.077 & 0.920 & -1.144 & 0.082 & 0.897 & -1.144 & 0.097 & 0.923\\
$\zeta_1$ & -0.1 & -0.071 & 0.070 & 0.890 & -0.068 & 0.074 & 0.927 & -0.075 & 0.083 & 0.923\\
$\zeta_2$ & 0.1 & 0.082 & 0.081 & 0.950 & 0.075 & 0.085 & 0.937 & 0.083 & 0.097 & 0.933\\
$1/\phi$ & 5.0 & 4.944 & 0.419 & 0.943 & 4.951 & 0.452 & 0.943 & 4.965 & 0.536 & 0.950\\
$1/\xi$ & 0.2 & 0.134 & 0.023 & 0.223 & 0.128 & 0.024 & 0.190 & 0.114 & 0.027 & 0.167\\
\bottomrule
\end{tabular}

\label{tab:sim_HH}
}
\end{table}


\end{document}